\documentstyle[12pt]{article}
\newlength{\pubnumber} 

\catcode`\@=11
\@addtoreset{equation}{section}

\def\section{\@startsection{section}{1}{\z@}{3.5ex plus 1ex minus .2ex}
 {2.3ex plus .2ex}{\large\bf}}
\def\subsection{\@startsection{subsection}{2}{\z@}{2.3ex plus .2ex}
 {2.3ex plus .2ex}{\bf}}
\newcommand\Appendix[1]{\def\thesection{Appendix \Alph{section}}
 \section{\label{#1}}\def\thesection{\Alph{section}}}
\begin{document}

\begin{titlepage}
\samepage{
\setcounter{page}{1}
\rightline{IASSNS-HEP-95/56}
\rightline{\tt hep-th/9604112}
\rightline{published in {\it Nucl.\ Phys.}\/ {\bf B479} (1996) 113}
\rightline{April 1996}
\vfill
\begin{center}
 {\Large \bf Realizing Higher-Level Gauge Symmetries\\
       in String Theory:\\
    New Embeddings for String GUTs\\}
\vfill
\vspace{.12in}
 {\large Keith R. Dienes\footnote{
   E-mail address: dienes@sns.ias.edu}
   $\,$and$\,$ John March-Russell\footnote{
   E-mail address: jmr@sns.ias.edu}\\}
\vspace{.12in}
 {\it  School of Natural Sciences, Institute for Advanced Study\\
  Olden Lane, Princeton, N.J.~~08540~ USA\\}
\end{center}
\vfill
\begin{abstract}
  {\rm
  We consider the methods by which higher-level and non-simply laced
    gauge symmetries can be realized in free-field heterotic string theory.
  We show that all such realizations have a common underlying
  feature, namely a dimensional truncation of the charge lattice,
  and we identify such dimensional truncations with certain
  irregular embeddings of higher-level and non-simply laced
  gauge groups within level-one simply-laced gauge groups.
  This identification allows us to formulate a direct mapping
  between a given subgroup embedding, and the sorts of GSO
  constraints that are necessary in order to realize the embedding
  in string theory.  This also allows us to determine a number
  of useful constraints that generally affect string
  GUT model-building.  For example, most string GUT realizations
    of higher-level gauge symmetries $G_k$ employ the so-called diagonal
  embeddings $G_k\subset G\times G \times...\times G$.  We find that there
   exist interesting alternative
    embeddings by which such groups can
   be realized at higher levels, and we derive a complete list of all
possibilities for the
     GUT groups $SU(5)$, $SU(6)$, $SO(10)$, and $E_6$ at levels $k=2,3,4$ (and
in some
    cases up to $k=7$).
   We find that these new embeddings are always more efficient and require
     less central charge than the diagonal embeddings which have
    traditionally been employed.
   As a byproduct, we also prove that it is impossible to realize
     $SO(10)$ at levels $k>4$.
    This implies, in particular, that free-field heterotic string models
    can never give a massless {\bf 126} representation of $SO(10)$.
    }
\end{abstract}
\vspace{.10in}
\smallskip}
\end{titlepage}

\setcounter{footnote}{0}

% ========================= DEFINITIONS ===================================
\def\beq{\begin{equation}}
\def\eeq{\end{equation}}
\def\beqn{\begin{eqnarray}}
\def\eeqn{\end{eqnarray}}
\def\Tr{{\rm Tr}\,}
\def\KM{{Ka\v{c}-Moody}}

\def\ie{{\it i.e.}}
\def\etc{{\it etc}}
\def\eg{{\it e.g.}}
\def\half{{\textstyle{1\over 2}}}
\def\third{{\textstyle {1\over3}}}
\def\quarter{{\textstyle {1\over4}}}
\def\m{{\tt -}}
\def\p{{\tt +}}

\def\rep#1{{\bf {#1}}}
\def\slash#1{#1\hskip-6pt/\hskip6pt}
\def\slk{\slash{k}}
\def\GeV{\,{\rm GeV}}
\def\TeV{\,{\rm TeV}}
\def\y{\,{\rm y}}
\def\SM{Standard-Model }
\def\SUSY{supersymmetry }
\def\SSM{supersymmetric standard model}
\def\vev#1{\left\langle #1\right\rangle}
\def\l{\langle}
\def\r{\rangle}

\def\Htw{{\tilde H}}
\def\chibar{{\overline{\chi}}}
\def\qbar{{\overline{q}}}
\def\ibar{{\overline{\imath}}}
\def\jbar{{\overline{\jmath}}}
\def\Hbar{{\overline{H}}}
\def\Qbar{{\overline{Q}}}
\def\abar{{\overline{a}}}
\def\alphabar{{\overline{\alpha}}}
\def\betabar{{\overline{\beta}}}
\def\tautwo{{ \tau_2 }}
\def\calF{{\cal F}}
\def\calP{{\cal P}}
\def\calN{{\cal N}}
\def\smallmatrix#1#2#3#4{{ {{#1}~{#2}\choose{#3}~{#4}} }}
\def\bone{{\bf 1}}
\def\V{{\bf V}}
\def\b{{\bf b}}
\def\N{{\bf N}}
\def\t#1#2{{ \Theta\left\lbrack \matrix{ {#1}\cr {#2}\cr }\right\rbrack }}
\def\C#1#2{{ C\left\lbrack \matrix{ {#1}\cr {#2}\cr }\right\rbrack }}
\def\tp#1#2{{ \Theta'\left\lbrack \matrix{ {#1}\cr {#2}\cr }\right\rbrack }}
\def\tpp#1#2{{ \Theta''\left\lbrack \matrix{ {#1}\cr {#2}\cr }\right\rbrack }}

%================== BLACKBOARD BOLD CHARACTERS ==============================

\def\inbar{\,\vrule height1.5ex width.4pt depth0pt}

\def\IC{\relax\hbox{$\inbar\kern-.3em{\rm C}$}}
\def\IQ{\relax\hbox{$\inbar\kern-.3em{\rm Q}$}}
\def\IR{\relax{\rm I\kern-.18em R}}
 \font\cmss=cmss10 \font\cmsss=cmss10 at 7pt
\def\IZ{\relax\ifmmode\mathchoice
 {\hbox{\cmss Z\kern-.4em Z}}{\hbox{\cmss Z\kern-.4em Z}}
 {\lower.9pt\hbox{\cmsss Z\kern-.4em Z}}
 {\lower1.2pt\hbox{\cmsss Z\kern-.4em Z}}\else{\cmss Z\kern-.4em Z}\fi}

%========================================================================
%          MACROS FOR REFERENCES
%========================================================================
\def\AEF{A.E. Faraggi}
\def\KRD{K.R. Dienes}
\def\JMR{J. March-Russell}
\def\NPB#1#2#3{{\it Nucl.\ Phys.}\/ {\bf B#1} (19#2) #3}
\def\PLB#1#2#3{{\it Phys.\ Lett.}\/ {\bf B#1} (19#2) #3}
\def\PRD#1#2#3{{\it Phys.\ Rev.}\/ {\bf D#1} (19#2) #3}
\def\PRL#1#2#3{{\it Phys.\ Rev.\ Lett.}\/ {\bf #1} (19#2) #3}
\def\PRT#1#2#3{{\it Phys.\ Rep.}\/ {\bf#1} (19#2) #3}
\def\CMP#1#2#3{{\it Commun.\ Math.\ Phys.}\/ {\bf #1} (19#2) #3}
\def\MODA#1#2#3{{\it Mod.\ Phys.\ Lett.}\/ {\bf A#1} (19#2) #3}
\def\IJMP#1#2#3{{\it Int.\ J.\ Mod.\ Phys.}\/ {\bf A#1} (19#2) #3}
\def\nuvc#1#2#3{{\it Nuovo Cimento}\/ {\bf #1A} (#2) #3}
\def\etal{{\it et al,\/}\ }

%==============================================================================
\hyphenation{su-per-sym-met-ric non-su-per-sym-met-ric}
\hyphenation{space-time-super-sym-met-ric}
\hyphenation{mod-u-lar mod-u-lar--in-var-i-ant}
%==============================================================================

%============================== SECTION 1 ============================

%======================================================================
\setcounter{footnote}{0}
\section{Introduction and Overview}

\subsection{Motivation}

As is well-known, the simplest heterotic string constructions
lead to models for which the spacetime gauge group $G$
is both simply laced, and realized as an affine
Lie algebra with level $k_G$ equal to one.
While certain level-one simply laced models \cite{models}
have met with some phenomenological success,
higher-level and/or non-simply laced string models are important
from both phenomenological and formal perspectives.
It is therefore important to understand the nature of
such models, and the common underlying ingredients in their construction.

On the phenomenological side, for example,
we may wish to construct
string models whose low-energy spectra resemble
those of supersymmetric grand-unified theories (SUSY GUT's).
Indeed, such string GUT models represent one possible approach towards
explaining the observed unification of gauge couplings within the MSSM.
Of course, there exist other attractive string-based approaches towards
explaining this unification, such as
heavy string threshold corrections \cite{thresholds,DF},
non-standard hypercharge normalizations \cite{k1},
extra matter beyond the MSSM \cite{DF,extramatter},
and possible strong-coupling effects \cite{wittencouplings}.
For a recent general review of all of these paths to unification,
see Ref.~\cite{review}.

One advantage of the string GUT approach, however,
is that the meeting of the gauge couplings at the MSSM
scale $\approx 2 \times 10^{16}$ GeV is naturally incorporated.
Another advantage is that the GUT idea, at least in the case
of $SO(10)$, provides a compelling explanation of the fermion
matter content of the Standard Model.
One immediate observation that one faces when building string GUT models,
however,
is that if the massless spectra of such models are to contain
the adjoint Higgs representations that are required for the GUT symmetry
breaking,
then the GUT gauge group must be realized at an affine
level $k_{\rm GUT}> 1$.  Surprisingly, the construction of such
higher-level string GUT models has turned out to be
a rather complicated affair
\cite{stringguts,shygut,stringguthybrids,shynew,GxGmodels}.
It would therefore be useful to have a systematic method of
surveying the kinds of higher-level string models that can be constructed,
and for analyzing the procedures that would have to be
employed for each.

Such higher-level/non-simply laced models are also
important from a more formal perspective.
Indeed, in some sense, such string models
represent {\it generic}\/ points in the full string moduli space, and
as such they provide a crucial
arena for testing some of the predictions of
various conjectured string dualities.
For example, it has recently been demonstrated that there
exist four-dimensional heterotic string models with $N=4$
spacetime supersymmetry whose gauge symmetry groups are
non-simply laced \cite{duality}.  Therefore, unlike simply laced models,
such models cannot be self-dual under strong/weak
coupling duality ($S$-duality);
rather, the predictions of $S$-duality
for such models become highly non-trivial, relating
such models to each other in a pairwise fashion throughout
the moduli space.
Finding explicit pairs of such models, however,
is an important but as yet unsolved problem.

This paper is devoted towards understanding
the general issues that are involved when constructing
free-field heterotic string models with higher-level and/or
non-simply laced gauge symmetries.
As we shall see,
the construction of such models is substantially
more difficult than that of level-one
simply laced string models.
Thus, our goal is to develop a more abstract method
of understanding how such gauge symmetries generically arise in string theory,
and how the possibilities for their construction
can be surveyed in an efficient and general manner.

\subsection{Our Approach}

\begin{figure}[htb]
\centerline{\epsfxsize 4.5 truein \epsfbox {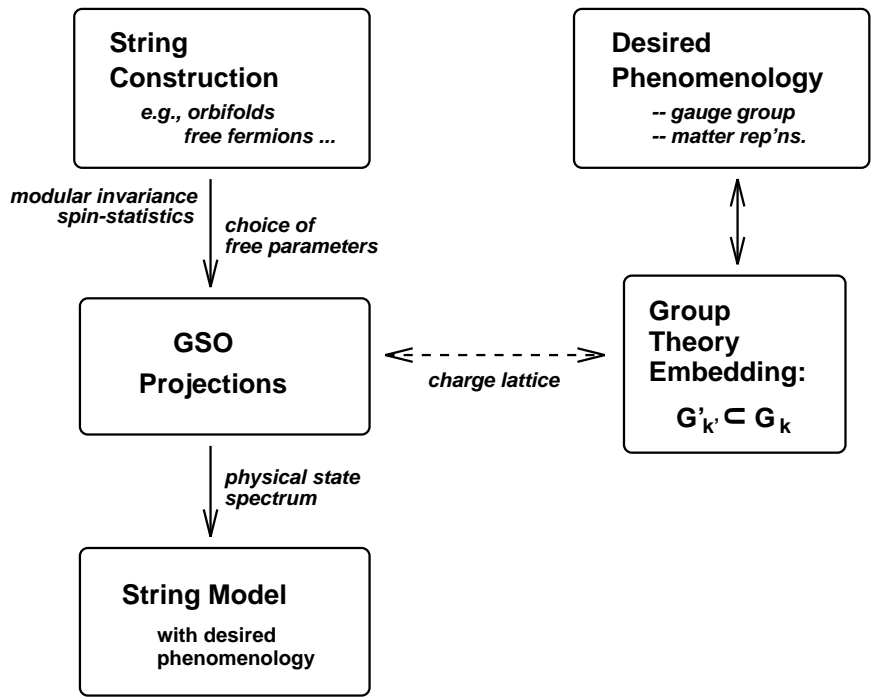}}
\caption{Procedures for string model-building, as discussed in the text.
Particular choices for the free parameters within a given string construction
determine the resulting GSO projections, which in turn define a particular
string model.  The desired phenomenology, on the other hand, dictates a
specific group-theoretic embedding of the gauge and matter representations.
The connection between the construction procedure
and the desired group-theoretic embedding occurs through the charge lattice,
at the level of the GSO projections. }
\label{flowchartfig}
\end{figure}

In this paper, we shall begin an investigation of this general issue by
approaching it from one particular, though central, direction.
In order to explain our approach, let us first
review the basic procedure which underlies the construction of
generic string models. This procedure, as we have organized it,
is schematically illustrated in Fig.~\ref{flowchartfig}.

There are basically three steps in building a heterotic string model;  these
are illustrated
along the left column of Fig.~\ref{flowchartfig}.
First, one selects a particular string construction
({\it e.g.}, orbifolds, lattices, free fermions, and so forth).
Although each of these constructions contains many free parameters,
there are also many string-theoretic self-consistency constraints
that must be satisfied:  one must maintain conformal invariance,
modular invariance, worldsheet supersymmetry,
proper spin-statistics relations,
physically sensible projections, and so forth.
Often these constraints are ``built into'' the particular
string construction
from the beginning.  For example, in the fermionic approach,
there are various rules that govern the allowed choices
of fermionic boundary conditions.  Likewise, in an orbifold
approach, only certain combinations of twists are self-consistent
and give rise to sensible models.

%================== FIGURE INSERTED HERE ==============================
%   If you do not wish to have the figure inserted, just comment
%   out the following lines:
\input epsf
%================== END OF INSERTED FIGURE ============================

However, even after these self-consistency constraints are satisfied,
there typically remain many free parameters
({\it e.g.}, additional boundary conditions, phases, twists, and so forth)
whose values are not yet fixed.
The second step in model-building therefore consists
of making specific choices for these remaining parameters.
Indeed, such specific choices essentially {\it define}\/ the string model,
and amount to selecting a self-consistent set of GSO projections that will
act on the Fock space of string states.

Finally, the third step is to enumerate the states that survive
these GSO projections, and thereby determine the physical properties
of the model thus constructed.
It is in this way that one deduces the final spectrum of the string
model, and tests its phenomenological success.

Clearly, many different string models can be generated in this way,
and only a small fraction of these will have low-energy features of
phenomenological interest.
It is therefore important to have some control or guidance
over the original choice of free parameters, so that one can efficiently
select models with the desired properties.

In the process of string model-building, this phenomenological
selection procedure typically occurs through what we can abstractly describe as
a two-step process.  We have illustrated these two steps
along the  right column of Fig.~\ref{flowchartfig}.
First, one starts with a pre-determined set of phenomenological
requirements.  For example, one may wish to realize
a particular gauge group,
along with a set of matter representations with
specific charge assignments.
Then, given these requirements, one
determines a preferred
group-theoretic {\it embedding}\/.
For example, if we wish to construct a string GUT model, we require
a GUT gauge group $G_{\rm GUT}$ such as $SU(5)$ or $SO(10)$
along with three complete chiral massless generations
and a Higgs scalar transforming in the adjoint of
the GUT group.
This latter requirement then forces us
to realize our GUT gauge symmetry
at an affine level $k_{\rm GUT}>1$, and this in turn
requires that we choose a special group-theory embedding
that is capable of realizing such a higher-level
gauge symmetry.  A choice that is typically made in the
literature is the so-called ``diagonal embedding''
\beq
        (G_{\rm GUT})_2~\subset~ (G_{\rm GUT})_1 ~\times ~(G_{\rm GUT})_1
\label{introeq}
\eeq
in which the GUT group $G_{\rm GUT}$
is realized at level $k_{\rm GUT}=2$ as
the diagonal subgroup of the two-fold tensor product of level-one group
factors.
This embedding, of course, is only one particular
way of realizing such a level-two GUT group,
and in general the desired choice of embedding
is a subtle issue which must be guided by the desired phenomenology.

Nevertheless, once a specific group-theoretic embedding is chosen,
one then makes
contact with the three-step string model-construction procedure outlined above
at the level of the GSO projections.
Specifically, one must choose the GSO projections
in such a way as to realize the desired group theory embedding.
This is illustrated via the horizontal arrow in Fig.~\ref{flowchartfig}.
Note that in free-field string constructions (such as those
based on free worldsheet bosons or fermions, thereby including all orbifold
or lattice constructions),
this relation between the GSO projections and
the desired embeddings in group theory
occurs through the {\it charge lattice}\/.
 From a string-theoretic standpoint,
this charge lattice describes
the quantum numbers that string states have under the worldsheet currents
that comprise the affine Lie algebra.  From a group-theoretic
standpoint, however, such lattices are nothing but the root systems
or weight systems of the Lie algebra representations.
It is therefore here, at the level of the GSO projections and
charge lattice, that the
connection between the desired phenomenology and a particular string
construction occurs, and through which the two can influence each other.

In this paper, we shall focus on
this latter connection.
In particular, we shall mainly concern ourselves with two
broad classes of questions:
\begin{itemize}
\item    First, given any arbitrary embedding
        $G'_{k'}\subset G_k$, what are the corresponding
       GSO projections that will be necessary in order to
       realize this embedding in string theory?
        In other words, once a particular embedding is selected,
       how can this embedding be {\it realized}\/ or incorporated
        into the string framework?
\item  Second, more generally,
         what are all of the ways of realizing a given group $G'$
        at a given level $k'$ in free-field string theory?
       In other words, what are the possible embeddings
        $G'_{k'}\subset G_k$ that one can ever hope to realize in string
       theories based on  free-field constructions, and what properties do they
share?
       For example,   the ``diagonal embedding'' presented in (\ref{introeq})
       is only one method that gives rise to
       higher-level GUT groups in string theory.  It is natural
       to wonder whether there might exist other
       options.  Can one find and classify {\it all}\/ of the embeddings
       which can realize GUT groups like $SU(5)$ or $SO(10)$
        at higher levels?
        Might such alternative, non-diagonal embeddings be more efficient than
        the
       diagonal embeddings that have been employed up until now?
\end{itemize}
In this paper, we shall provide explicit answers to all of these
questions.
Our primary focus will be on those embeddings $G'_{k'}\subset G_k$
for which $G'$ is non-simply laced and/or $k'>1$ (as these are
the more challenging cases), but our treatment will be completely
general.

Since our main interest is on this connection between the
group-theoretic embeddings and the corresponding GSO projections,
there are many other important issues
that we will {\it not}\/ be discussing in this paper.
For example, although we will determine the specific GSO projections
that are required in order to realize any arbitrary  group-theoretic
embedding, we will not approach the issue of whether such GSO
projections can be consistently realized
within a given string construction ({\it i.e.}, consistent
with modular invariance, spin-statistics constraints, and so forth).
In other words, we will not be exploring the connection
between the ``String Construction'' and ``GSO Projection'' boxes
in Fig.~\ref{flowchartfig}.  While these are important questions,
they tend to be highly dependent on the particular string construction
employed;
indeed, GSO projections that may be simple to realize in one formulation
may be impossible to realize in another.
Similarly, we will not concern ourselves
with the ways in which a particular set of phenomenological requirements
influences the selection of a group-theoretic embedding;
in terms of Fig.~\ref{flowchartfig}, this would be the connection
between the ``Desired Phenomenology'' and ``Group Theory Embedding''
boxes.  Rather, in this paper we will restrict ourselves to analyzing
the connection between
the possible group-theoretic embeddings, and the GSO projections to which
they correspond.
Our results will therefore be completely general, and relevant
for all model-construction procedures.

\subsection{Organization of this paper}

This paper is organized in three main parts.

The first part, consisting of Sects.~2--4,
develops most of our formal results.
In Sect.~2, we provide a brief review of affine Lie algebras,
and establish our notation and conventions.  Readers familiar with
affine Lie algebras are encouraged to skip this section.
Then, in Sect.~3, we identify the underlying mechanism
by which higher-level and/or non-simply laced gauge symmetries are
realized in free-field string constructions.
As we shall see, all such realizations correspond to so-called
``dimensional truncations'' of the charge lattice, and can be
analyzed purely in terms of the geometry of such truncations.
Finally, in Sect.~4, we present our general formalism, and
derive most of our formal results.
We find, in particular, that there exists a one-to-one correspondence
between consistent dimensional truncations and
so-called ``irregular'' group-theoretic embeddings, and we develop a
general procedure for determining the explicit GSO projections
that correspond to each such embedding.

The second part of this paper, consisting of Sects.~5 and 6,
works through a number of examples of such embeddings and GSO projections,
and in the process clarifies several related background issues.
Sect.~5, in particular, contains several non-trivial examples of our
general results, and Sect.~6 discusses the manner in which the
required sorts of GSO projections can be realized in a particular string
construction.  This analysis will also serve to
illustrate how string theory manages to implement an important feature
that we shall call the ``adjoint to adjoint only'' rule.
As we shall see, such a rule is ultimately necessary for the consistency
of the whole approach.

Finally, in the third part of this paper (Sects.~7 and 8),
we apply our general formalism
to the study of string GUT embeddings and their corresponding
GSO projections.
In Sect.~7, we present
a {\it complete classification of all embeddings}\/ through which the
phenomenologically interesting GUT groups
$G_{\rm GUT}= SU(5)$, $SU(6)$, $SO(10)$, and $E_6$
may be realized in free-field string theory
at affine levels $k_{\rm GUT}=2$, 3, and 4.
In the case of $SO(10)$, we also extend our results to all
levels allowed by naive central-charge constraints, {\it i.e.}, $k\leq 7$.
Since Sect.~7 is rather long and technical, we have
collected the results of this classification
together in Sect.~7.4, which may be read independently
of the rest of this paper.
Sect.~8 is then devoted to a detailed analysis of
some of the new embeddings and their associated GSO projections.

One of the main results in this part of the paper
is an identification of {\it alternative}\/
higher-level GUT embeddings
which go beyond the simple ``diagonal'' embeddings in (\ref{introeq}).
As we shall show, some of these alternative embeddings are
extraordinarily efficient when compared to the traditional diagonal embeddings,
and can be expected to give rise to entirely new classes of potential
string GUT models.  These alternative embeddings should therefore be
of direct importance for string model-builders.

Another important result is a proof that $SO(10)$ can never be
realized in free-field heterotic string models at affine levels $k>4$.
This implies, for example, that one can never realize
the potentially useful massless $\rep{126}$ representation of $SO(10)$.
This result, like all of our results, is completely general,
and holds for all free-field model-construction procedures.

Finally, in Sect.~9, we summarize our main results
and discuss various directions for future research.
We also discuss how our GUT classification can be used
to determine the {\it minimal}\/ additional central charge necessary
for the string-theoretic realization of GUT gauge symmetries,
and therefore the smallest extra chiral algebra that must be generated.
Certain technical issues pertaining to our GUT classification in Sect.~7
are then presented in an Appendix.

%======================================================================
\vfill\eject
\setcounter{footnote}{0}
\section{Technical Background:  Affine Lie Algebras}

In this section, we provide a background review of
some basic facts concerning untwisted affine Lie algebras \cite{KMalgebras},
also known as \KM\ algebras in the physics literature.
This will also serve to establish our definitions and conventions.
Readers familiar with affine Lie algebras are encouraged to skip
this section.

Affine Lie algebras $\hat G$ are infinite-dimensional
extensions of the ordinary Lie algebras $G$ that contain
this ordinary algebra $G$ as a subalgebra.
They are generated by chiral worldsheet
currents $J^a(z)$ of conformal dimension $(1,0)$ satisfying the operator
product expansions (OPE's)
\beq
       J^a(z) J^b(w) ~=~ {{\tilde k}^{ab}\over (z-w)^2} ~+~
	{i\,f^{abc}\over z-w}\,J^c(w)~+~
            {\rm regular}
\label{OPEold}
\eeq
where $f^{abc}$ are the structure constants of the Lie algebra $G$
(with $a,b,c=1,...,{\rm dim}\,G$), and where the first term on the
right side is the ``central extension'' or Schwinger term.
Equivalently, mode-expanding
the currents via $J^a(z)\equiv \sum_{n\in\IZ} J_n^a  z^{-n-1}$,
we find that the OPE (\ref{OPEold})
gives rise to the commutation relations
\beq
        [J^a_m, J^b_n] ~=~ \tilde k^{ab} \,m\,\delta_{m+n,0}~+~
i\,f^{abc}\,J^{c}_{m+n}~.
\label{commrelations}
\eeq
The subalgebra of modes with $m=n=0$
generates the ordinary Lie algebra
$G$, with vanishing central extension.

A basis of currents $J^a(z)$ may always be chosen so as to
diagonalize the coefficients of the central extension, so that
$\tilde k^{ab}=\tilde k \delta^{ab}$.
Furthermore, for a non-abelian group, one can define a unique normalization
for the currents $J^a(z)$ by fixing a particular normalization for the
structure constants.
One typically specifies the normalization of the
structure constants via
\beq
   \sum_{ab}\,f^{abc}\,f^{abd}~=~ C^{\rm (adj)}_{G} \,\delta^{cd}
\label{quadcas}
\eeq
where $C^{\rm (adj)}_G$ is the eigenvalue of the quadratic Casimir
acting on the adjoint representation.
This then fixes the normalizations of the currents, the value of
the central extension coefficient $\tilde k$, and
the lengths of the root vectors $\lbrace \vec \alpha\rbrace$ of
the corresponding Lie algebra.
Alternatively, one can define the normalization-independent
quantities
\beq
    \tilde h_G ~\equiv~ {C^{\rm (adj)}_G \over \vec{\alpha_{h}}^2} ~,~~~~~
    k_G ~\equiv~ {2\,\tilde k \over \vec{\alpha_{h}}^2}
\label{coxeter}
\eeq
where $\vec \alpha_h$ is the longest root.
Here $\tilde h_G$ is the so-called ``dual Coxeter number'' of
the group $G$,
and $k_G$ is the so-called ``level'' of the affine Lie algebra.
Thus, the level $k_G$ of an affine Lie algebra has invariant meaning
only for a non-abelian group $G$.
In general the dual Coxeter number $\tilde h_G$ can be calculated
for any Lie algebra as
\beq
   \tilde h_G ~=~ {1\over {\rm rank}(G)}\, \left\lbrack
       n_L ~+~  \left({L\over S}\right)^{-2} n_S\right\rbrack
\eeq
where $n_{L,S}$ are the numbers of long and short non-zero roots in the root
system $\lbrace \vec \alpha \rbrace$, and where $L/S$ is the corresponding
ratio of their lengths.
It turns out that $\tilde h_G$ is always an integer;
likewise, unitarity requires $k_G$ to be integral as well.
The {\it central charge}\/ of the corresponding
conformal field theory is:
\beq
        c_G ~=~ {k\, {\rm dim}(G)\over k+\tilde h_G}~.
\label{centralcharge}
\eeq
For each of the classical Lie algebras $G$, the corresponding rank, dimension,
root-length ratio, dual Coxeter number, and central charges are tabulated
below:
\beq
\begin{tabular}{c||c|c|c|c|c}
 $G$ &  rank($G$) & dim($G$) & $L/S$ & $\tilde h_G$ & $c_G$ \\
\hline
\hline
 $A_n \equiv SU(n+1)$ & $n$ &  $n(n+2)$ & $1$ & $n+1$ & $n(n+2)k/(n+k+1)$ \\
 $B_n \equiv SO(2n+1)$ & $n$ &  $n(2n+1)$ & $\sqrt{2}$ & $2n-1$ &
$n(2n+1)k/(2n+k-1)$ \\
 $C_n \equiv Sp(2n)$ & $n$ &  $n(2n+1)$ & $\sqrt{2}$ & $n+1$ &
$n(2n+1)k/(n+k+1)$ \\
 $D_n \equiv SO(2n)$ & $n$ &  $n(2n-1)$ & $1$ & $2n-2$ & $n(2n-1)k/(2n+k-2)$ \\
 $E_6$ & $6$ &  $78$ & $1$ & $12$ & $78k/(k+12)$ \\
 $E_7$ & $7$ &  $133$ & $1$ & $18$ & $133k/(k+18)$ \\
 $E_8$ & $8$ &  $248$ & $1$ & $30$ & $248k/(k+30)$ \\
 $F_4$ & $4$ &  $52$ & $\sqrt{2}$ & $9$ & $52k/(k+9)$ \\
 $G_2$ & $2$ &  $14$ & $\sqrt{3}$ & $4$ & $14k/(k+4)$ \\
\end{tabular}
\eeq

%===========================================================================
\vfill\eject
\setcounter{footnote}{0}
\section{Dimensional Truncations of the Charge Lattice}

In this section, we begin by reviewing the subtleties
encountered when attempting to realize higher-level
or non-simply laced gauge symmetries in string theories based
on free-field constructions.
We shall then discuss, in a model-independent manner,
how these difficulties are ultimately resolved.  As we shall see,
the resolution involves a special type of GSO projection.

\subsection{The Subtleties}

It in order to fully appreciate the subtleties that enter
the construction of string models with higher-level
or non-simply laced gauge symmetries, let us first recall
the simplest string constructions ---
 {\it e.g.}, those based on free worldsheet bosons, or complex
worldsheet fermions.  In a four-dimensional heterotic string,
the conformal anomaly
on the left-moving side
can be saturated by having $22$ internal bosons
$\Phi^I$, $I=1,...,22$, or equivalently $22$ complex
fermions $\psi^I$.
If we treat these bosons or fermions
indistinguishably, this generates an internal symmetry group
$SO(44)$, and we can obtain other internal symmetry
groups by distinguishing  between these different worldsheet fields
({\it e.g.}, by giving different toroidal boundary conditions to
different fermions $\psi^I$).
 Such internal symmetry groups are then interpreted
as the gauge symmetry groups of the effective low-energy theory.
Note that in this paper, we are focusing on the gauge symmetries
that arise from the left-moving ({\it i.e.}, internal)
degrees of freedom of the heterotic string.
If any gauge group arises from the compactified right-moving
degrees of freedom, it will appear only via a tensor product
with the left-moving gauge group, and will not affect
our subsequent analysis.

In general, the spacetime gauge bosons of such left-moving symmetry groups
fall into two classes:  those of the form
$\psi^\mu |0\rangle_R \otimes
            i\partial\phi^I  |0\rangle_L$
give rise to the 22 {\it Cartan}\/ elements of the gauge symmetry,
and those of the form
$\psi^\mu |0\rangle_R \otimes
  e^{i \alpha \phi_I}
  e^{i \beta \phi_J} |0\rangle_L$
with $\alpha^2+\beta^2=2$
give rise to the {\it non-Cartan}\/ elements.
In the language of complex fermions, both groups of gauge bosons
take the simple form
 $\psi^\mu |0\rangle_R \otimes \psi^I \psi^J |0\rangle_L$;
if $I= J$, we obtain the Cartan elements,
whereas if $I\not =J$ we obtain the non-Cartan elements.
Together these fill out the adjoint
representation of some Lie group.
The important point to notice here, however, is the fact
that in the fermionic formulation, {\it two}\/ fermionic excitations
are required on the left-moving side (or equivalently that
$\alpha^2+\beta^2=2$ in the bosonic formulation).  Not only is this
required in order to produce the two-index tensor
representation that contains the adjoint representation
(as is particularly evident in the fermionic construction),
but precisely this many excitations are
also necessary in order for the resulting gauge boson state
to be {\it massless}.

The next step is to consider the corresponding charge lattice.
Each of the left-moving worldsheet bosons $\phi^I$ has, associated
with it, a left-moving current $J_I\equiv i\partial \phi_I$
(or in a fermionic
formulation, $J_I\equiv \overline{\psi}_I \psi_I$).
The eigenvalues $Q_I$
of this current when acting on a given state yield the charge
of that state.  The complete left-moving charge of a given state
is a 22-dimensional vector ${\bf Q}$, and the charges of the above
gauge boson states together comprise the root system
of a rank-22 gauge group (which can be simple or non-simple).
However, the properties of this gauge group are highly constrained.
For example, the fact that we require two fundamental excitations in order
to produce the gauge boson state (or equivalently that $\alpha^2+\beta^2=2$ in
bosonic language) implies that
each non-zero root must have (length)$^2=2$.
Thus, we see that we can obtain only {\it simply laced}\/ gauge groups
in such constructions!  Moreover,
it turns out that in such constructions,
the GSO projections only have the power to project a
given {\it non}\/-Cartan root into or out of the spectrum.
Thus, while we are free to potentially alter
the particular gauge group in question via GSO projections, we cannot go beyond
the set of rank-22 simply laced gauge groups.

As the final step, let us now consider the affine level at which such
groups are ultimately realized.  Indeed, this is another property of
the gauge group that cannot be altered in such constructions.
With fixed normalizations for the currents $J_a$ and structure
constants $f^{abc}$, we see from (\ref{OPEold}) that
\beq
    k_G\cdot |\vec \alpha_h|^2 ~  =  ~{\rm constant} ~=~2~.
\label{lengthlevel}
\eeq
Thus, with roots of (length)$^2=2$, we see that our gauge symmetries
are realized at level $k_G=1$.
Indeed, if we wish to realize our gauge group at a higher level
({\it e.g.}, $k_G=2$), then we must somehow devise a special mechanism
for obtaining roots of smaller length ({\it e.g.}, length $=1$).
However, as discussed above, this would naively appear to
conflict with the masslessness requirement.
Thus, on the face of it, it would seem to be impossible to realize
higher-level gauge symmetries in string theory.

Note that this problem arises in {\it all}\/ constructions based on
free worldsheet fields.  Indeed, this is because all such constructions
automatically give rise to a charge lattice
for which the conformal dimensions of the
non-Cartan gauge boson states
can be identified as $h={\bf Q}^2/2$.

\subsection{The Resolution}

Fortunately, even within free-field constructions,
there do exist various methods which
are capable of yielding higher-level gauge symmetries.
Indeed, although such methods
are fairly complicated,
they all share certain simple underlying features.
We shall now describe, in the language of the above discussion,
the general underlying mechanism which enables such symmetries to be
realized.
It is this general mechanism which ultimately
forms the foundation for the rest of this paper.

As we have seen, the fundamental problem that we face is
that we need to realize our gauge boson string states
as massless states, but with smaller corresponding charge vectors (roots).
To do this, let us for the moment
imagine that we could somehow {\it project}\/ or {\it truncate}\/
these roots onto
a certain hyperplane in the 22-dimensional charge space,
and consider only the surviving components of these roots.
Clearly, thanks to this projection, the ``effective length''
of our roots would be shortened.  Indeed, if we cleverly
choose the orientation of this hyperplane of projection,
we can imagine that our shortened, projected roots could
either
\begin{itemize}
\item combine with other longer, unprojected roots ({\it i.e.},
     roots which originally lay in the projection hyperplane)
      to fill out the root system of a {\it non}\/-simply laced
      gauge group, or
\item combine with other similarly shortened roots to
    fill out the root system of a {\it higher-level}\/ gauge symmetry.
\end{itemize}
Thus, such a projection would be exactly what is required.

The question then arises:  how can we achieve or interpret
such a hyperplane projection in charge space?  Clearly, such a
projection would imply that one or more dimensions of the charge lattice
should no longer be ``counted'' towards building the gauge group, or
equivalently that one or more of the gauge quantum numbers should be lost.
Indeed, such a projection would entail a {\it loss of rank}\/
(commonly called ``rank-cutting''), which corresponds to a loss
of Cartan generators.
Thus, we see that we can achieve the required projection if and only
if we can somehow construct a special GSO projection which,
unlike those described above, is capable of projecting out
a {\it Cartan root}.
This corresponds to a dimensional truncation of the charge lattice.

Indeed, from the above discussions, it is also easy to see this result
by proving the reverse statement:  {\it without}\/ rank-cutting,
the only gauge groups
that can be realized in free-field string constructions
are at level one and must be simply laced.
This follows directly as follows.
Conformal invariance and masslessness constraints, as we have seen,
require that the left-moving vertex operators of gauge-boson
states must have conformal dimensions equal to one.
In theories without rank-cutting, however, the conformal dimension
of a non-Cartan gauge-boson state is related to its 22-dimensional charge
vector ${\bf Q}$ via $h={\bf Q}^2/2$.
We therefore find that
such gauge-boson states must always have ${\bf Q}^2=2$,
which implies that the gauge groups that they produce are necessarily
simply laced and realized at level one.
This observation is completely general (since it relies
on only conformal symmetry and the masslessness constraint), and applies to all
free-field heterotic string constructions.
Note, in particular, that this argument applies
regardless of whether
such level-one simply laced groups
are realized directly (such as the case of $SO(44)$, for which all
gauge bosons arise in the same Neveu-Schwarz sector), or as an
enhanced gauge symmetry (such as $E_8\times E_8\times...$, for which
some gauge bosons arise in additional twisted sectors).

Thus, in order to realize a higher-level and/or non-simply laced
gauge symmetry in free-field string models, we must
 {\it start with a level-one simply laced gauge group, and then
perform a dimensional truncation of the charge lattice corresponding
to that group}\/.
As we have said, such dimensional truncations correspond to projecting
out {\it Cartan roots}\/ from the string spectrum.

What kinds of string constructions can give rise to such
unusual GSO projections?
As we indicated, such GSO projections cannot arise
in simple free-field constructions based on free bosons or complex fermions.
Instead, we require highly ``twisted'' orbifolds (typically
asymmetric, non-abelian orbifolds \cite{asymmorbifolds}), or
constructions based on so-called ``necessarily
real fermions'' \cite{KLST}.  The
technology for constructing such string theories is
still being developed \cite{stringguts,shygut,stringguthybrids,shynew,KT}.
For the purposes of this paper, however,
the basic point is that such ``dimensional truncations''
of the charge lattice are the common underlying feature in
 {\it all}\/ free-field constructions of higher-level or non-simply laced
string models.

Finally, we comment again on the possibility of gauge symmetries
arising from the {\it right-moving}\/ ({\it i.e.}, supersymmetric)
worldsheet degrees of freedom of the heterotic string.
A consideration of the possible realizations of the worldsheet supercurrent
enables one to show that the maximal right-moving
gauge symmetry that can arise in this case is $[SU(2)]^6$.
Furthermore, because the superconformal algebra yields a right-moving
vacuum energy of $-1/2$ rather than $-1$, all such states must
have ${\bf Q}^2_{\rm right}=1$ rather than $2$.  Thus, such
right-moving gauge symmetries are always realized at affine level two.
This is particularly evident in the free-fermionic construction
\cite{freefermions},
where each potential $SU(2)_2$ right-moving gauge-group factor
is realized purely in terms of three Majorana-Weyl fermions.
Note that such right-moving gauge symmetries are therefore
the {\it sole}\/ case in which a higher-level gauge symmetry
can be realized without a dimensional truncation.\footnote{
    In certain limits of moduli space, it has been shown \cite{witten}
    that extra $Sp(2n)$ gauge symmetries can arise due to
    the effects of small worldsheet instantons.
    However, like the right-moving gauge symmetries,
    these extra non-perturbative gauge symmetries appear
    as an extra tensor-product factor in the total gauge group.
    Furthermore, such gauge symmetries are not realized as affine
    Lie algebras, and cannot be understood through an ordinary
    conformal-field-theoretic analysis of the classical string degrees of
freedom.
    They therefore do not have affine ``levels'' in the usual sense,
    and are beyond the scope of this paper.  }
In any case,
such right-moving gauge symmetries are too small to be of phenomenological
interest
in the case of string GUT models,
and are often entirely absent.
They will therefore not concern us further.

\subsection{An Example}

\begin{figure}[thb]
\centerline{\epsfxsize 3.5 truein \epsfbox {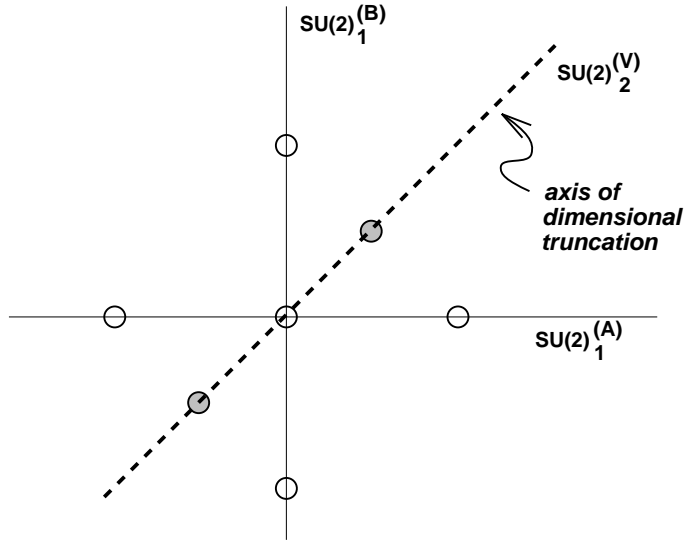}}
\caption{ The root system of $SU(2)_1^{(A)}\times SU(2)_1^{(B)}$ (denoted
by open circles), and its
dimensional truncation onto the diagonal subgroup $SU(2)_2^{(V)}$
(with new non-zero roots denoted by shaded circles).}
\label{su2su2}
\end{figure}

Before proceeding to a general analysis of such ``dimensional
truncations'', it is useful
to have
an explicit example of how they work.
Let us therefore consider the well-known
method of achieving a level-two symmetry algebra which consists of
tensoring together two copies of any group $G$ at level one,
and then modding out by the interchange symmetry.
This leaves behind the diagonal subgroup $G$ at level two.
This construction is well-known in the case $G=E_8$, where it serves
as the underlying mechanism responsible
for the (level-two) $E_8$ string model in ten dimensions \cite{Eeightmodel}.
For simplicity, let us analyze this construction
for the case $G=SU(2)$. If we start with an
$SU(2)_1^{(A)}\times SU(2)_1^{(B)}$ gauge symmetry,
as illustrated in Fig.~\ref{su2su2}, then modding
out by the interchange symmetry corresponds
to projecting the roots onto the diagonal axis corresponding to
the diagonal Cartan generator $J_z^{(V)}=J_z^{(A)}+J_z^{(B)}$.
As we can see from Fig.~\ref{su2su2},
this reproduces the $SU(2)$ root system, but scaled
so that roots which formerly had length $\sqrt{2}$ now
have length $1$. Thus we realize $SU(2)_2$
as the diagonal survivor of the dimensional truncation.
In terms of the Cartan
generators $U_1\equiv J_z^{(A)}$ and $U_2\equiv J_z^{(B)}$
of the original $SU(2)$ factors, it is clear that
we simply need project out the linear combination $U_1-U_2$,
retaining the orthogonal combination $U_1+U_2$.
Therefore if we want to realize
this particular embedding in string theory, we must
construct GSO projections
that remove the $|U_1\rangle -|U_2\rangle$ state,
but preserve $|U_1\rangle + |U_2\rangle$.

By analyzing the root systems in this way --- {\it i.e.}, as
dimensional truncations of the charge lattice --- we now have a powerful
tool at our disposal for determining the possibilities
for realizing higher-level and/or non-simply laced gauge
symmetries in string theory, and for determining the particular
GSO projections to which they correspond.
For example, given this simple geometrical interpretation, it is immediately
clear  that this ``diagonal'' construction generalizes to {\it any}\/ group
$G$, for
the roots of each level-one gauge factor $G$ always project onto the diagonal
hyperplane with
a reduction in length by a factor $\cos\, 45^\circ=1/\sqrt{2}$,
causing a doubling of the resulting affine level.
Moreover, we also see that this procedure even generalizes to
 {\it any}\/ number $n$ of identical group factors tensored together,
$G_1\times G_1\times...\times G_1$,
leaving the completely diagonal subgroup $G$ at level $n$.

%================== FIGURE INSERTED HERE ==============================
%   If you do not wish to have the figure inserted, just comment
%   out the following lines:
%================== END OF INSERTED FIGURE ============================

However, the vast majority of cases are not this simple.
For example, we will see that there exists a method of realizing
$SU(2)$ at level $k=4$
which does {\it not}\/ involve
taking the diagonal tensor product of four copies of $SU(2)$,
but which instead realizes $SU(2)_4$ as a special subgroup of $SU(3)$ at
level $k=1$:  $SU(2)_4\subset SU(3)_1$.  An even more dramatic example
of a higher-level $SU(2)$ realization
is $SU(2)_{10} \subset SO(5)_1$.
In fact, for even higher levels,
the possible embeddings become more and more unexpected.
At level 28, for example,
it turns out that one can realize $SU(2)_{28}$
not through 28 copies of $SU(2)$,
but rather through the single rank-two group $G_2$:
$SU(2)_{28} \subset G_2$.
Note that if we wished to realize $SU(2)_{28}$
in string theory, this latter embedding would be crucial,
for the ``diagonal method'' would require that we start with
a group of rank at least 28 --- too high to be realized
in a classical heterotic string.

Clearly, then, for the purposes of string model-building,
it is important to have a general way of surveying the possibilities
for alternative embeddings,
and for determining the GSO projections to which they correspond.
For example, if we are interested in building an $SU(5)$ or $SO(10)$
string GUT model, we would like to survey all of the possible methods
of realizing $SU(5)_2$ or $SO(10)_2$ in free-field string constructions.
Each different embedding, although yielding the same gauge symmetry,
will nevertheless have a drastically different
stringy realization, and consequently will correspond to a
different low-energy phenomenology.
For example, an embedding of the form $SU(10)_2\subset SO(10)_1\times SO(10)_1$
will involve more dimensions of the string charge lattice than an embedding
into a smaller-rank group, and will consequently give rise to not only
a different ``hidden sector'' gauge symmetry,
but also a different embedding for the {\it matter}\/ fields.
Such alternate embeddings might be especially useful for tackling
the important question of obtaining higher-level GUT gauge symmetries
while simultaneously producing three generations.

Furthermore, even after a suitable embedding is chosen,
there then remains the separate question of
determining the particular GSO projections that will {\it realize}\/ this
embedding in string theory. For example, let us suppose that we wish
to realize the $SU(2)_{10}\subset SO(5)_1$ embedding mentioned above.
What linear combination of the two Cartan roots $U_1$ and $U_2$ of the
$SO(5)_1$ gauge group  must then be projected out of the spectrum?
(We will answer this question explicitly in Sect.~5.)

Thus, we see that there are two general questions that we would like to
answer:
\begin{itemize}
\item  What are all of the ways of realizing a given group $G$
    at a given level $k$ in free-field string constructions?
\item  In order to realize a particular embedding ${G}'_{k'}\subset
	{G}_k$, what linear combination of Cartan roots of ${G}_k$
     must be projected out of the string spectrum?
\end{itemize}
In the next section, we shall provide general answers to these questions.

%======================================================================
\vfill\eject
\setcounter{footnote}{0}
\section{Irregular Embeddings and GSO Projections}

In the previous section, we considered
the general process of ``dimensional truncation'' of the
charge lattice, and demonstrated that this
is the general underlying mechanism by
which higher-level and non-simply laced
gauge symmetries are realized in free-field string constructions.
In this section, we shall analyze this procedure
in a general way.  First, given an arbitrary affine Lie algebra $G$,
we shall determine the consistent dimensional truncations
that may be performed.
Then, armed with this knowledge, we will develop a systematic
method of determining
which Cartan generators of $G$ must be GSO-projected out of the
string spectrum in order to explicitly realize
each such truncation.

\subsection{Regular vs. Irregular Embeddings}

We first address the question of determining, for an arbitrary affine
Lie algebra $G$, which dimensional truncations are consistent.
As we have discovered in the previous section,
higher-level and/or non-simply laced gauge symmetries
in free-field string constructions are realized only through {\it dimensional
truncations of the charge lattices of level-one simply laced groups}\/.
Such dimensional truncations correspond
to GSO projections which remove one or more Cartan roots from
the string spectrum.
By contrast, the ``ordinary'' GSO projections
in string theory can delete only various {\it non}\/-Cartan roots,
and are therefore capable of producing only
level-one, simply laced algebras.

Of course, starting with a level-one simply laced gauge symmetry $G$,
it is not necessarily consistent
to GSO-project an arbitrary linear combination of Cartan generators
out of the spectrum.
For example, it is easy to imagine
that by removing a poorly-chosen combination of Cartan generators
(or equivalently, by projecting onto an improperly oriented
hyperplane in charge space),
we would wind up with a nonsensical collection of surviving
roots that does not correspond to the root system of any
Lie algebra at any level, whether simply or non-simply laced.
Thus, we must require that any potential dimensional truncation
of the adjoint representation of $G$
should yield at least the adjoint representation of some other Lie algebra
$G'$.
However, this is not sufficient.
Indeed, we must also demand that the corresponding dimensional truncation of
 {\it every representation}\/ of $G$
produce a representation (not necessarily irreducible) of
$G'$.
Otherwise, the dimensional truncation will produce non-sensical
results in other sectors of the theory.
Obviously, taken together, these conditions
amount to the simple requirement that the final Lie algebra $G'$ be
a {\it subalgebra}\/ of the original Lie algebra $G$.

This much is trivial.
However,
thanks to our dimensional truncation results
of the previous section,
we can make some further distinctions.
Recall that a group $G$ can have two types of subgroups $G'$, often called
 {\it regular}\/ and {\it irregular} (or ``special'').  A subgroup $G'$ is
called `regular' if its roots are a subset of the roots of the original
group $G$.
By contrast, an `irregular' subgroup $G'$ must necessarily
contain some roots which were {\it not present}\/ in the original group $G$;
in general, such new roots of $G'$ are non-trivial linear combinations
of the roots of $G$.

This distinction is crucial, because we have seen that
dimensional truncations not only reduce
rank, but must also truncate at least some of the roots
in such a way that these roots are projected onto hyperplanes.
Indeed, it was this exactly this projection which
gave rise to the required shorter roots with (length)$^2<2$.
However, this projection implies that the final (shorter projected)
roots that are obtained are necessarily {\it different}\/ from the
original (longer, unprojected) roots that we started with ---
 {\it i.e.}, the final set of
roots that are obtained are necessarily not a subset of
original roots. Thus, we conclude that dimensional truncations necessarily
correspond to {\it irregular}\/ embeddings.
It is therefore only
through irregular embeddings
that higher-level or non-simply laced gauge symmetries can be realized in
free-field string constructions.

This is an important point.
Note, in particular, that this does not mean  that
only irregular embeddings are rank-reducing;  indeed,
in general a given group will typically contain both
regular and irregular subgroups of smaller rank.
Thus, both regular and irregular embeddings can
correspond to rank-reduction.
However, by definition,
the roots of regular subgroups always retain the lengths
found in the original group.
Thus, since the masslessness condition
forces us to begin with roots of (length)$^2=2$,
we see that regular embeddings are incapable
of producing higher-level or non-simply laced gauge groups
in such constructions.

It is also an important point that we are making these claims
in the physical context of {\it string theory}.
In particular, there do exist regular embeddings which
can give rise to higher-level or non-simply laced
gauge groups;  two examples of such embeddings are
\beqn
        SU(2)_1 \times SU(2)_3 ~&\subset& ~ G_2~\nonumber\\
        SU(5)_2 \times U(1)~&\subset& ~ Sp(10)_1~.
\label{regembedding}
\eeqn
In the first of these embeddings,
the non-zero roots of the $SU(2)_1$ gauge factor are taken from the long
roots of $G_2$, while the
non-zero roots of the higher-level $SU(2)_3$
gauge factor are taken from the (orthogonal set of) short roots of $G_2$.
Similarly, in the second case,
the 20 non-zero roots of $SU(5)_2$ are taken from the
short roots of $Sp(10)_1$.
However, note that both of these embeddings require
the presence of non-simply laced gauge groups such as $G_2$ or $Sp(10)$
in order to provide us with the short roots we require.
As we have seen, such groups can be realized in string theory only through
a dimensional truncation procedure of the sort we discussed in Sect.~3.
Thus, the realization of higher-level
group factors via regular embeddings of the sort (\ref{regembedding})
still requires the existence of an {\it irregular}\/ embedding
at some prior symmetry-breaking step in the string construction.

\subsection{Connecting Irregular Embeddings to Dimensional Truncations:
     General Formalism}

Having made these observations, we now wish to sharpen
our connection between irregular embeddings $G'_{k'}\subset G_k$
and dimensional truncations.  Note, for example,
that our identification of dimensional truncations with irregular
embeddings implies that irregular embeddings must somehow
be associated
with projections in root space.
This is indeed the case, and developing the exact connection
between the two will enable us to determine which linear
combinations of Cartan roots of $G$
must be projected out of the spectrum in order to realize $G'$.
This will also enable us to
geometrically determine the affine level $k'$
at which the subalgebra $G'$ is realized.

To do this, let us first recall some elementary facts about
Lie algebras and their representations.
In general, the roots $\lbrace \vec \alpha \rbrace$ of a Lie
algebra of rank $r$ are vectors in an $r$-dimensional
vector space, and among these roots there always exists
a special basis of $r$ ``simple roots''.
Likewise, to each representation of the Lie algebra there
corresponds a set of $r$-dimensional vectors (the so-called ``weights'') which
fill out the representation, and among which there exists a highest weight
$\vec\Lambda$ from which all others may be obtained by repeated
subtractions of the simple roots.
In general, a given root or weight $\vec \Lambda$ may be specified
in a coordinate-independent manner by specifying its
Dynkin indices or labels $a_i$ with respect to each of the
simple roots $\vec\alpha_i$.  These Dynkin indices are defined as
\beq
        a_i ~\equiv~ { 2\,(\vec\Lambda,\vec \alpha_i)\over
                 (\vec\alpha_i, \vec\alpha_i) }~
\label{Dynkinlabels}
\eeq
where the inner products between any two roots
or weights are evaluated in the (Euclidean) root/weight space.
Once these Dynkin indices are known,
inner products between any roots or weights can then be determined directly
from these indices via the metric tensor $G_{ij}$:
\beq
         (\vec\Lambda,\vec\Lambda')~=~ \sum_{ij}\,a_i\,G_{ij}\,a'_j~.
\label{innerproduct}
\eeq
The metric tensors, highest weights, and Dynkin indices
for all of the Lie algebras and their representations can
be found, {\it e.g.}, in Ref.~\cite{slansky}.
Finally, we also recall the definition of the
so-called {\it quadratic index}\/ of a representation of a group $G$
with highest weight $\vec \Lambda$:
\beq
      \ell_G(\vec \Lambda)~\equiv~
       {{\rm dim}\,R \over {\rm dim}\,G}~
	(\vec \Lambda,\vec \Lambda+2\vec \delta)
       ~=~ {1\over {\rm rank}\,G}\, \sum_{i=1}^{{\rm dim}\,R}
	\,(\vec \lambda_i,\vec \lambda_i)~.
\label{indexdef}
\eeq
Here $\vec \delta$ is defined
as half of the sum of the positive roots of $G$, and the $\vec \lambda_i$ are
all of the weights of the representation.  In general,
$\vec \delta$ has Dynkin indices $(1,1,\dots,1)$.
The inner products $(\vec \Lambda,\vec \Lambda+2\vec \delta)$
or $(\vec \lambda_i,\vec \lambda_i)$
are then evaluated as in (\ref{innerproduct}).

Given an embedding $G'\subset G$, it turns out that
there is a simple method for determining the
ratio of the corresponding affine levels $k'/k$.
First, recall that
the ratio of the affine levels
is given by the ratio of the $({\rm length})^2$ of the roots of
$G'$ and $G$.
Next, note that the
quadratic index $\ell_G$ in (\ref{indexdef}) is directly
proportional to this $({\rm length})^2$,
for the quadratic index scales with the normalization of the root system
of $G$.
Now, it turns out to be a general feature of
irregular embeddings $G'\subset G$ that
there always exists at least one representation $R^\ast$ of $G$ which,
under the decomposition $G\to G'$, has the simple branching rule
$R^\ast\to R^\ast$.  (This will be discussed more fully in Sect.~7.)
Thus, for the $R^\ast$ representation of $G$, we find that
\beq
     {k'\over k}~=~ {\ell_{G'}(R^\ast) \over \ell_{G}(R^\ast)}~.
\eeq
Indeed, this result generalizes to {\it any}\/ representation $R$
of $G$ as follows.
Let us assume that under the decomposition $G\to G'$, we have
the branching rule $R\to \sum_i R_i$ where $R_i$ are irreducible
representations of $G'$.
Then the so-called {\it embedding index}\/
$j(G'\subset G)$ of the corresponding
irregular embedding $G'\subset G$ is defined as \cite{dynkin2}
\beq
          j(G'\subset G) ~\equiv ~{  \sum_i \ell_{G'}(R_i) \over
                           \ell_G(R)}~,
\label{embeddingindex}
\eeq
and we have the simple identification
\beq
                k'_{G'}~= ~ j(G'\subset G)\, k_{G}~.
\label{kjj}
\eeq
Of course, in (\ref{embeddingindex}), each of the indices $\ell_G(R)$ and
$\ell_{G'}(R_i)$
must be computed using the {\it same}\/ normalization for
the root systems of $G$ and $G'$.
Note that the identification (\ref{kjj}) has been previously exploited
in the physics literature (see, {\it e.g.}, \cite{BB}).

These results are useful and elegant, but for our purposes in string theory
we wish to identify each irregular embedding $G'_{k'}\subset G_k$  with a
particular dimensional truncation, for it is only in this
geometric way that we will be able to determine which linear combinations of
Cartan roots of $G$ need to be GSO-projected out of the spectrum
in order to realize the embedding.
Thus, we shall now take a slightly different approach, and consider
instead the so-called {\it embedding matrix}\/ that corresponds
to a given subgroup embedding.

Such embedding matrices may be defined as follows.
For any given embedding of
a subgroup $G'$ of rank $r'$
within a group $G$ of rank $r$,
the corresponding embedding matrix ${\cal P}(G'\subset G)$ is a matrix of
dimensionality $r'\times r$ which
maps the Dynkin labels $(a_1,...,a_r)$ of the highest
weight of a given representation of $G$
onto the Dynkin labels $(a'_1,...,a'_{r'})$ of the
highest weight of the corresponding (not necessarily irreducible)
representation of $G'$:
\beq
    {\vec a'} ~=~ \calP(G'\subset G)\,{\vec a}~.
\eeq
The reduction in rank which necessarily accompanies irregular
embeddings implies that the embedding matrix $\cal P$ will not be square
in such cases.
However, by providing a complete mapping between the representations
of $G$ and the corresponding representations of $G'$, the embedding matrix
$\cal P$ thus succinctly contains all information about the particular
embedding of $G'$ within $G$, and no further information is required.
Thus, given only this embedding matrix $\calP(G'\subset G)$, will be
able to determine not only  which linear combinations of
Cartan generators of $G$
must be projected out of the spectrum
in order to produce $G'$, but also the affine level $k'$ at which
$G'$ will be realized.
We shall now give an explicit procedure for carrying out both tasks.

Since we have already determined that irregular embeddings
must somehow correspond to dimensional truncations of the charge
lattice, our first
step must be to determine the orientation of the hyperplane
of truncation onto which the roots of the original
group must be projected.
Now, in general, the orientation of an $r'$-dimensional
hyperplane passing through the origin
within an $r$-dimensional space may be given by
specifying $(r-r')$ independent $r$-dimensional vectors $\vec\beta_i$,
$i=1,...,r-r'$, each of which is perpendicular to the hyperplane.
Because such vectors $\vec\beta_i$ are orthogonal
to the hyperplane of truncation, they must have
vanishing projection onto this hyperplane.
Hence, they satisfy
\beq
      \vec 0 ~=~ \calP(G' \subset G)\,\vec \beta_i~.
\label{eigenvectors}
\eeq
In other words, from (\ref{eigenvectors}), we see that such
vectors $\vec \beta_i$ are simply a set of linearly independent
vectors which span the nullspace of $\calP$.
If desired, we can subject these vectors $\vec\beta_i$
to a Gram-Schmidt orthogonalization
procedure in order to make them mutually orthogonal.

Given these orthogonal vectors $\vec\beta_i$,
it is now straightforward, following the
general procedure outlined in Sect.~4,
to determine  the affine level $k'$ of the subgroup.
We simply choose an arbitrary weight $\vec \Lambda$
in the weight space of $G$, and determine its projection
$\vec \Lambda'$ onto the corresponding hyperplane:
\beq
     \vec \Lambda' ~=~
         \vec\Lambda ~-~ \sum_{i=1}^{r - r'}
      \, {(\vec\Lambda,\vec\beta_i)\over
        (\vec\beta_i,\vec\beta_i)}\,\vec\beta_i~.
\label{alphaprojected}
\eeq
Thus $(\vec\Lambda',\vec\beta_i)=0$ for all $\vec\beta_i$.
The squared length of this projected vector $\vec \Lambda'$
is then simply
\beq
       |\vec\Lambda'|^2~=~ (\vec\Lambda',\vec\Lambda') ~=~
      (\vec\Lambda,\vec\Lambda) ~-~ \sum_{i=0}^{r-r'}\,
       {(\vec\Lambda,\vec\beta_i)^2\over
        (\vec\beta_i,\vec\beta_i)}~.
\eeq
Thus, in order to determine the level $k'$ of the subgroup
$G'$, we
simply compare $|\vec\Lambda'|^2$ against the expected length
of the weight in the $G'$ system to which $\vec\Lambda$ corresponds.
It is clear, however, that this $G'$-weight is nothing but
$\calP \vec\Lambda$.
We thus find that the level $k'$ of the subgroup $G'$ is
related to the level $k$ of the original group $G$ via
\beq
    {k'\over k}~=~ {(\calP\vec\Lambda,\calP\vec\Lambda) \over
      (\vec\Lambda,\vec\Lambda) ~-~ \sum_{i=0}^{r-r'}\,
       \lbrack (\vec\Lambda,\vec\beta_i)^2/ (\vec\beta_i,\vec\beta_i)\rbrack}~.
\label{kresult}
\eeq
Note that this result is independent of our choice for
$\vec\Lambda$, as well as the normalizations of any of the vectors
$\vec \beta_i$.
This result does depend, however, upon
the overall normalization for $\calP$ being chosen appropriately
for similarly normalized roots.
In other words,
the embedding matrix $\calP$ should map the Dynkin labels $\lbrace a_i\rbrace$
of any $G$-weight onto the Dynkin labels $\lbrace a'_i\rbrace$ of the
corresponding $G'$-weight, where all of these Dynkin labels $a_i$ are to
be evaluated, as in (\ref{Dynkinlabels}), relative to the
identically normalized roots $\vec\alpha_i$ of the corresponding groups.

We therefore now have two (ultimately equivalent) methods
of calculating the affine level of an irregularly
embedded subgroup:  we can calculate either the ``embedding indices''
of (\ref{embeddingindex}), or the general expression in (\ref{kresult}).
Which is easier depends on the information available.
In particular, while
the embedding indices are often tabulated in mathematical references,
the formulation (\ref{kresult}) is more geometrical, and relies
only on the embedding matrix $\calP(G\subset G')$.
Such an embedding matrix $\calP(G\subset G')$ not only contains {\it all}\/
information concerning the embedding in question, but is also intimately
related
to the GSO projections that must be performed in order to realize the
embedding.

We now turn, therefore, to the remaining task:  the determination of the
appropriate
linear combinations of Cartan generators of $G$ that must be projected
out of the spectrum in order to realize
the general embedding $G'_{k'}\subset G_1$.
Once again, however, the
vectors $\vec\beta_i$ are precisely what we
want, for each corresponds uniquely to a different linear combination
of Cartan generators of $G$ that must be GSO projected out of the string
spectrum in order to realize $G'$.
In particular, let $G$ and $G'$ have rank $r$ and $r'$ respectively, so that
the nullspace of ${\cal P}$ is $(r-r')$-dimensional, and is spanned by $r-r'$
different vectors $\vec \beta_i$.
Also let $U_\ell$ ($\ell=1,...,R$)  be the Cartan generators of $G$, such that
each different generator $U_\ell$ corresponds to a different (orthogonal)
lattice direction $\hat e_\ell$
in the root space of $G$.
Of course, strictly speaking we have $R=r$, but at this point we leave
open the possibility that $R$ may exceed $r$ due to the presence of extra
group factors (such as extra $U(1)$ factors) that may arise along with $G$.
In other words, we allow for the possibility
that the $r$-dimensional root space of $G$
may ultimately be non-trivially embedded within
an even larger $R$-dimensional lattice
whose directions correspond to an increased
number of Cartan generators $U_\ell$ ($\ell=1,...,R)$.
Then, if the Cartesian (lattice) components of
each vector $\vec\beta_i$  are given by $b^{(i)}_\ell$,
it follows that the $r-r'$ different linear combinations of $U_\ell$ which
must be projected from the string spectrum are simply given by
\beq
     \sum_{\ell=1}^R\, b^{(i)}_\ell \, U_\ell~,~~~~~       i=1,...,r-r'~.
\label{lincombs}
\eeq

Thus, given the $\vec\beta_i$ vectors as specified by their Dynkin indices,
we must first determine their Cartesian coordinates $b^{(i)}_\ell$.
 {\it A priori}\/, this can be done straightforwardly:
if a given vector $\vec\beta$ in the root space of $G$
has Dynkin labels $a_j$ ($j=1,...,r$), then
its corresponding Cartesian coordinates $b_\ell$ ($\ell=1,...,R$) are given by:
\beq
    b_\ell~=~   (\vec \beta,\hat e_\ell) ~=~
           \sum_{j,k=1}^r \,
         {  4 \,(\vec\beta,\vec\alpha_j)\,G_{jk}\,(\vec\alpha_k,\hat e_\ell)
\over
                    |\vec\alpha_j |^2 \,|\vec \alpha_k|^2 }
         ~=~
       \sum_{j,k=1}^r \, a_j\, {2\,G_{jk}\over |\vec\alpha_k|^2}
\,(\vec\alpha_k,\hat e_\ell)
\label{cartesian}
\eeq
where $G_{jk}$ is the metric tensor in root space,
and where we have used the definition (\ref{Dynkinlabels}).
Thus, we see that we must first determine the inner products
     $(\vec\alpha_k,\hat e_\ell)$.
It is here, however, that an important subtlety arises,
for we see that we must first determine the relative {\it orientation}\/
of the simple roots with respect to the underlying Cartesian
coordinate system.
In other words, we must determine
the ultimate orientation of the charge lattice in terms of the underlying
string
degrees of freedom.  For example, in a string formulation
based upon complex worldsheet bosons or fermions,
each lattice direction $\hat e_\ell$ --- and consequently each
generator $U_\ell$ --- will correspond to a different boson
or fermion:  $U_\ell\equiv i\partial \phi_\ell=\overline{\psi}_\ell \psi_\ell$.
Given such a construction,
it is therefore necessary to determine the orientation or embedding
of the simple roots of the gauge group $G$ with respect
to these lattice directions.

Fortunately,
this can be determined by recalling how such (simply laced) gauge groups $G$
are ultimately realized in free-field string constructions.
In particular, the gauge boson states in string theory
are always realized in terms of simple particle and anti-particle excitations
of the underlying worldsheet fields.
We shall see an explicit example of this in Sect.~6.2.
Thus, we see that the appropriate realization of the roots of
$G$ is essentially fixed (up to irrelevant overall lattice permutations and
inversions)
in terms of the lattice directions $\hat e_\ell$.
For example, in the case of $SO(2r)$, the roots $\lbrace \vec \alpha\rbrace$
are
given as $\lbrace \pm \hat e_i \pm \hat e_j\rbrace$, with the simple roots
given by $\vec\alpha_i = \hat e_i - \hat e_{i+1}$ for $i\leq r-1$,
and $\vec \alpha_r = \hat e_{r-1}+ \hat e_{r}$.
Given such an explicit realization, the relative orientation of the simple
roots
with respect to the Cartesian coordinate system is then fixed.
For convenience, we now list the appropriate
inner products $(\vec \alpha_i,\hat e_\ell)$
for each of the classical Lie groups:
\beqn
    SU(r+1):&~~~~~~~~~& (\vec\alpha_k,\hat e_\ell)~=~\phantom{\bigl\lbrace}
          \delta_{k,\ell}-\delta_{k+1,\ell} ~~~~{\rm for~all}~  1\leq k \leq r
\nonumber\\
    SO(2r+1):&~~~~~~~~~& (\vec\alpha_k,\hat e_\ell)~=~\cases{
          \delta_{k,\ell}-\delta_{k+1,\ell} & if $1\leq k < r$\cr
          \delta_{r,\ell} & if $k = r$\cr}\nonumber\\
    Sp(2r):&~~~~~~~~~& (\vec\alpha_k,\hat e_\ell)~=~\cases{
          (\delta_{k,\ell}-\delta_{k+1,\ell})/\sqrt{2} & if $1\leq k < r$\cr
          \sqrt{2}\,\delta_{r,\ell} & if $k = r$\cr}\nonumber\\
    SO(2r):&~~~~~~~~~& (\vec\alpha_k,\hat e_\ell)~=~\cases{
          \delta_{k,\ell}-\delta_{k+1,\ell} & if $1\leq k < r$\cr
          \delta_{r-1,\ell} +\delta_{r,\ell} & if $k = r$\cr}
\label{innerprods}
\eeqn
As required, in each case we have normalized the simple roots so that
the long roots have length $\sqrt{2}$.
The inner products in (\ref{innerprods}) can then be
substituted into (\ref{cartesian}) and (\ref{lincombs}) in order to
determine which linear combinations of Cartan generators must be projected
out of the string spectrum.

Note that, in general, the $SU(n)$ groups are realized in string theory
by first realizing $U(n)\equiv SU(n)\times U(1)$ in an $n$-dimensional
lattice.  In this $n$-dimensional lattice,
the $U(1)$ group factor amounts to the trace
of the $U(n)$ symmetry, and corresponds to the lattice
direction ${\bf E}\equiv \sum_{\ell=1}^n \hat e_\ell$.
The $(n-1)$-dimensional hyperplane
orthogonal to ${\bf E}$ then corresponds to the $SU(n)$ gauge group.
This explains why,
as in (\ref{innerprods}),
the number of required lattice directions $\hat e_\ell$
for the $SU(r+1)$ case is larger than the rank of the group.
Thus, if $r$ is the rank of the group $G$
in (\ref{lincombs}) and if $R$ is the corresponding number of required
string lattice directions,
we find that we can have $R=r$ for the $G=SO$ and $Sp$ groups,
but require $R=r+1$ for the $SU$ groups.

Thus, through the general procedure outlined in this section,
we see that we have succeeded in identifying every irregular embedding
with a particular geometric dimensional truncation in root space.
Indeed, for every irregular embedding $G'_{k'}\subset G_k$, we now
know precisely the dimensional truncation to which it corresponds,
and the particular GSO projections of Cartan generators
that are required to achieve it.
Thus, in a general fashion, we find that have completely
described the underlying mechanism
which is responsible for the generation of higher-level and/or non-simply
laced gauge symmetries in free-field string constructions.
In the remainder of this paper, we shall consider various extensions
and applications of these general results.

%======================================================================
\vfill\eject
\setcounter{footnote}{0}
\section{Examples}

In this section we shall give two
explicit examples of the general results of Sect.~4.
We shall choose two particular
irregular embeddings, and show how each corresponds
to precisely
a dimensional truncation of the sort that arises
in an actual string-theoretic
construction.  We will also determine the
affine levels of the subgroups, and
deduce the GSO projections that are required
in order to realize these particular embeddings.

\subsection{First Example:   $SU(2)_4 \subset SU(3)_1$}

\begin{figure}[htb]
\centerline{\epsfxsize 3.5 truein \epsfbox {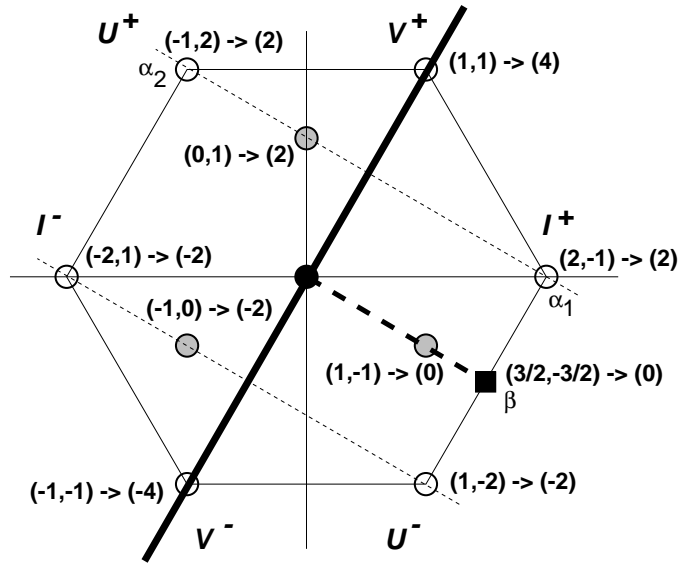}}
\caption{The irregular embedding of $SU(2)_4$ within $SU(3)_1$.
The non-zero roots of $SU(3)_1$ lie along the outer hexagon,
as denoted by empty circles.
The roots labelled $\vec\alpha_1$ and $\vec\alpha_2$ are the simple
roots, and the weights of the fundamental \rep{3} representation of
$SU(3)$ are also shown (shaded circles).
Next to each weight we have listed its Dynkin labels, along
with the Dynkin label of the $SU(2)$ weight
to which it corresponds according to the embedding matrix in
Eq.~(\protect\ref{Psu3su2}).  The axis of $SU(2)$ truncation is also
shown (dark line), with the vector $\vec\beta$ (black square)
chosen perpendicular to this axis.}
\label{su3roots}
\end{figure}

Perhaps the simplest case to consider is the irregular embedding
of $SU(2)$ within $SU(3)$.
Since this is an irregular embedding, the roots
of $SU(2)$ are not a subset of the roots of $SU(3)$, and in
particular the Cartan generator
of $SU(2)$ is neither of the Cartan generators of $SU(3)$.
Rather, the three generators $J_{x,y,z}$ of this
irregularly-embedded $SU(2)$ are realized as {\it linear combinations}\/ of
the non-Cartan generators of $SU(3)$:
\beqn
      J_x ~\equiv& -\sqrt{2}\,i\,(V_+ - V_-)&=~2\,F^7\nonumber\\
      J_y ~\equiv& -\sqrt{2}\,i\,(I_+ - I_-)&=~2\,F^2\nonumber\\
      J_z ~\equiv& -\sqrt{2}\,i\,(U_+ - U_-)&=~2\,F^5
\label{gellmannbasis}
\eeqn
Here $\lbrace I_\pm, U_\pm, V_\pm\rbrace$
are the non-Cartan generators of $SU(3)$ as labelled in Fig.~\ref{su3roots},
and the $F^i$ refer to the $SU(3)$ generators in the Gell-Mann basis.
One then easily finds, given the $SU(3)$ commutation relations for
$\lbrace I_\pm, U_\pm, V_\pm\rbrace$, that $\lbrace J_x,J_y,J_z\rbrace$ satisfy
the $SU(2)$ commutation relations.
Indeed, we can determine the level of the $SU(2)$ affine Lie algebra
by calculating the affinized commutation relations as follows.
The $SU(3)$ structure constants $f^{ijk}$ in the Gell-Mann
basis are normalized as
$\sum_{jk} f^{ijk}f^{\ell jk}=3\delta^{i\ell}=\tilde
h_{SU(3)}\delta^{i\ell}$,
which corresponds to the highest root having length $1$.
In particular,
$f^{257}=1/2$, and there are no
other non-zero structure constants
involving any two of these indices.
Thus, if the $SU(3)$ group is realized at any arbitrary
level $k$, then the corresponding commutation relations between
the $F^{2,5,7}$ generators are given by
\beq
   [F^i_m,F^j_n] ~=~ {i}\,f^{ijk}\,F^{k}_{m+n} ~+~
              {k\over 2}\,m\,\delta^{ij}\,\delta_{m+n,0}~
\label{commsu3}
\eeq
and cyclic permutations.
Among these generators, the structure constant $f^{ijk}$ will
take only the values $\lbrace 0,\pm 1/2\rbrace$.  From
(\ref{commsu3}) we therefore see that
\beq
   [J^i_m,J^j_n] ~=~ {i}\,(2f^{ijk})\,J^{k}_{m+n} ~+~
        {k'\over 2}\,m\,\delta^{ij}\,\delta_{m+n,0}~
\label{doublecommsu3}
\eeq
where $k'=4k$.  Furthermore, the new structure constants $2f^{ijk}$
with $i,j,k=2,5,7$ are nothing but $\epsilon^{ijk}$, the structure
constants for $SU(2)$ in a normalization with $SU(2)$ roots
having length $1$.
We thus identify $k'=4k$ as the level of the $SU(2)$ subgroup, so that
$SU(2)_4\subset SU(3)_1$ for this irregular embedding.

%================== FIGURE INSERTED HERE ==============================
%   If you do not wish to have the figure inserted, just comment
%   out the following lines:
%================== END OF INSERTED FIGURE ============================

This much is standard.
However, in order to relate this to the geometrical
dimensional truncations of Sect.~3,
let us now follow the procedure outlined in Sect.~4 and
consider the decompositions of the various $SU(3)$ representations.
This irregular embedding is defined by the $\rep{3} \to\rep{3}$ and
$\rep{8} \to\rep{5}+\rep{3}$ branching rules.
Now, in general, the Dynkin label of the highest weight
of the $\rep{2j+1}$ representation of $SU(2)$ is simply $(2j)$,
while the Dynkin labels for highest weights of the fundamental and
adjoint representations of $SU(3)$ are respectively $(0,1)$ and $(1,1)$.
Thus the embedding matrix $\calP$ must map $(0,1)$ to (2), and $(1,1)$ to
$(4)$,
implying
\beq
      \calP(SU(2)\subset SU(3)) ~=~ \pmatrix{ 2 & 2 \cr}~.
\label{Psu3su2}
\eeq

Given this, we see from Fig.~\ref{su3roots} that for each $SU(3)$ weight,
the corresponding $SU(2)$ Dynkin label is proportional to the length of
its {\it projection}\/ onto the axis indicated.
Thus, the irregular $SU(2)\subset SU(3)$ embedding
corresponds precisely to the geometrical process of dimensional truncation
that we discussed in Sect.~3.
Indeed,
such an identification would have seemed somewhat mysterious,
given only the explicit realization of the subgroup
as listed in (\ref{gellmannbasis}).
Moreover, just as outlined in Sect.~4, the proportionality factor between the
actual length of the projection and the length anticipated from the $SU(2)$
Dynkin label allows us to deduce the level of the subalgebra.
Since the $SU(2)$ root system realized through this projection is scaled down
by a
factor of two, the affine level increases by a factor of $4$, in agreement with
the explicit calculation above.
Of course, given the six-fold Weyl symmetry of the $SU(3)$ root system,
the irregular embedding $SU(2)_4\subset SU(3)_1$
actually corresponds in general
to a dimensional truncation onto {\it any}\/ axis which
is related to that
in Fig.~\ref{su3roots} by an element of the Weyl group.

Finally, we now determine which GSO projection
is required in order to realize this $SU(2)_4\subset SU(3)_1$
embedding.
The nullspace of the $\calP$-matrix in (\ref{Psu3su2}) is spanned
by the single vector $\vec\beta$ whose Dynkin labels are given
by $(1,-1)$.
Thus, we must first convert these to Cartesian coordinates.
Since the original group in this case is an $SU$ group, we cannot
simply use the graphical representation given in the figure,
for we recall from the previous section that such an $SU(3)$ group,
along with an additional orthogonal $U(1)$ group factor,
will be realized together in a {\it three}\/-dimensional lattice.
Indeed, following the formalism given the previous section,
we find that the
two simple roots $\vec\alpha_1,\vec\alpha_2$ of $SU(3)$,
as well as the nullspace vector $\vec \beta$,
will have the following Cartesian coordinates:
\beq
             \vec\alpha_1 ~=~ (1,-1,0)~,~~~~~~~
             \vec\alpha_2 ~=~ (0, 1,-1)~,~~~~~~~
             \vec\beta ~=~ (1, -2,1)~.
\eeq
Note that all of these coordinates are {\it rational}\/ in this
three-dimensional realization;
indeed, this is one of the reasons that string theory requires
such a three-dimensional space in order to realize $SU(3)_1$.
Thus, we conclude that if $U_i$ are the three Cartan generators that correspond
respectively to these three lattice directions, the subsequent
$SU(2)_4\subset SU(3)_1$ embedding can be realized by projecting
out the linear combination
\beq
        SU(2)_4\subset SU(3)_1:~~~~~~ {\rm project~out}~~~U_1 - 2\, U_2 + U_3~.
\eeq

Note that although we required {\it three}\/ Cartan generators in
order to realize the $SU(3)$ group factor and to express this
dimensional truncation, this does {\it not}\/ imply that $SU(2)_4$
is really embedded in $SU(3)_1\times U(1)$.
Indeed, the entire dimensional truncation occurs within the
two-dimensional
subspace that corresponds to $SU(3)$ alone (as illustrated in the figure), and
the extra $U(1)$ factor, which corresponds to the Cartan generator
$U_1+U_2+U_3$,
is truly orthogonal to the entire process.
Thus, as claimed, we have truly realized an $SU(2)_4\subset SU(3)_1$ embedding.
Such an embedding is an extremely efficient way of realizing $SU(2)_4$ in
string
theory, for we see that only one lattice dimension must be sacrificed in the
process.
By contrast,
the diagonal embedding $SU(2)_4\subset [SU(2)_1]^4$
would have required the sacrifice of {\it three}\/ lattice dimensions.

\vfill\eject

\subsection{Second Example:  $SU(2)_{10} \subset SO(5)_1$}

\begin{figure}[htb]
\centerline{\epsfxsize 4.5 truein \epsfbox {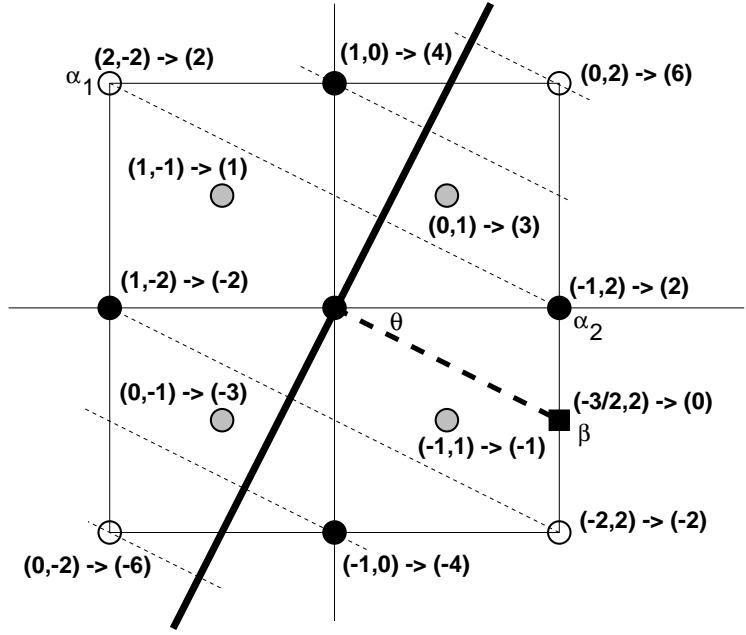}}
\vskip -0.5truein
\caption{The irregular embedding of $SU(2)_{10}$ within $SO(5)_1$.
The non-zero roots of $SO(5)_1$
lie along the outer square, with the empty circles denoting the
long roots and the black circles denoting the short roots.
The simple roots are $\vec\alpha_1$ and $\vec\alpha_2$.
The non-zero weights of the \rep{5} representation of $SO(5)$
comprise the short roots alone, and the weights of the \rep{4}
representation are also superimposed (shaded circles).
Next to each weight we have listed its Dynkin labels,
along with the Dynkin label of the $SU(2)$ weight
to which it corresponds according to
the embedding matrix in Eq.~(\protect\ref{Pso5su2}).
The axis of $SU(2)$ projection is also shown (dark line), with the
vector  $\vec\beta$ (black square) chosen perpendicular to
this axis.  The Dynkin labels of $\vec\beta$ thus uniquely define
this axis, and specify the orientation angle
to be $\theta=\protect\cos^{-1}(2/\protect\sqrt{5})$.}
\label{so5su2embedding}
\end{figure}

As a less-trivial example (indeed, one for which
the axis of truncation does not correspond to any
symmetry axis of the weight diagram),
let us examine the irregular embedding of $SU(2)$ within $SO(5)$.
This is illustrated in Fig.~\ref{so5su2embedding}.
In this embedding, the \rep{5} and \rep{4}
representations [for which the highest
weights have Dynkin labels (1,0) and (0,1) respectively]
map directly onto the \rep{5} and \rep{4} representations of $SU(2)$
[with respective Dynkin labels (4) and (3) in the $SU(2)$
normalization with roots of $({\rm length})^2=2$].
Thus the embedding matrix in this case is given by
\beq
   \calP(SU(2)\subset SO(5))  ~=~ \pmatrix{4 & 3 \cr}~.
\label{Pso5su2}
\eeq
The effects of this embedding matrix
on the Dynkin labels of the
\rep{4}, \rep{5}, and \rep{10} representations
of $SO(5)$ are shown in the figure.

Given these Dynkin label mappings, it is straightforward to deduce
the corresponding orientation of the $SU(2)$ axis of truncation (also
shown in the figure). As indicated in
the figure, the vector $\vec\beta$ with Dynkin labels $(-3/2,2)$
defines the nullspace of $\calP$, and
hence defines the orientation of the axis of projection.
Given this orientation, we immediately see
that the \rep{5} representation of $SO(5)$ fills out
the $j=2$ representation of $SU(2)$, and that the \rep{4} representation
fills out the $j=3/2$ representation.
The full adjoint representation of $SO(5)$ likewise decomposes
into the $\rep{7} +\rep{3}$ representations of $SU(2)$.

To determine the level of the $SU(2)$ subgroup,
we can calculate, for example, the length of the projection
of the simple root $\vec\alpha_2$ onto the $SU(2)$ axis.
This is most easily done by first determining the angle
$\theta$ between $\vec\alpha_2$ and $\vec\beta$, as follows.
Clearly the length $|\vec\alpha_2|$ is $1$, and we can
determine the length of $\vec\beta$ directly from its Dynkin labels
using the $SO(5)$ metric tensor, yielding
$(\vec\beta,\vec\beta)= 5/4$.
The angle $\theta$ can then be determined by
evaluating the inner product $1=(\vec\alpha_2,\vec\beta)
 =|\vec\alpha_2||\vec\beta| \cos\,\theta$
via the metric tensor, yielding
\beq
      \cos\,\theta~=~ 2/\sqrt{5}~.
\eeq
It then follows that the projection of $\vec\alpha_2$
onto the $SU(2)$ axis has length $\sin \theta=1/\sqrt{5}$.
Since this projected root corresponds to the highest
weight of the adjoint representation of $SU(2)$ (which at
level 1 would have length $\sqrt{2}$ in our normalization), we conclude
that the $SU(2)$ is here realized at level $k=10$.
This is of course the same result as we would have obtained
by straightforward use of (\ref{kresult}).

%================== FIGURE INSERTED HERE ==============================
%   If you do not wish to have the figure inserted, just comment
%   out the following lines:
%================== END OF INSERTED FIGURE ============================

Finally,
it is also straightforward to determine the GSO projection that
corresponds to this $SU(2)_{10}\subset SO(5)$ embedding.
Using the formalism discussed in the previous section,
we know that this $SO(5)$ group
can be realized directly in a two-dimensional lattice,
and indeed we find that these two lattice
dimensions correspond to the orthogonal
Cartesian axes indicated in Fig.~\ref{so5su2embedding}, with $U_1$ and
$U_2$ corresponding to the vertical and horizontal directions respectively.
The nullspace vector $\vec\beta$, with Dynkin indices $(-3/2,2)$, then has
Cartesian coordinates $(-1/2,1)$.
Hence, in order to realize this embedding, we find that we must
project out the linear combination of Cartan generators
\beq
   SU(2)_{10}\subset SO(5)_1:~~~~~~ {\rm project~out}~~~U_1 - 2\, U_2
    ~=~\sqrt{5}\,\biggl\lbrack
         (\sin\,\theta)\,U_1 - (\cos\,\theta)\,U_2\biggr\rbrack~.
\eeq
Of course, in string theory,
this original $SO(5)$ group factor must itself be realized from
a level-one simply laced group via a prior dimensional
truncation.

%======================================================================
\vfill\eject
\setcounter{footnote}{0}
\section{Satisfying the ``Adjoint to Adjoint Only'' Rule:
      A Pedagogical Example}

In this section we shall take a brief detour,
and discuss how the required sorts of dimensional truncations
are actually realized in a particular string construction ---
one based upon real fermions \cite{freefermions,KLST}.  We shall
assume no prior familiarity with this construction,
and keep our presentation as non-technical as possible.
However, by studying this example, we shall also
be able to discuss
how string theory manages to satisfy what we shall call
the ``adjoint to adjoint only'' rule.
As we have seen, a dimensional truncation is consistent
if and only if it corresponds to a {\it bona-fide}\/
irregular embedding.  However,
this is only part of the story.  In particular,
in the {\it physical context of string theory}\/,
there is an additional constraint that comes
into play.

\subsection{The ``adjoint to adjoint only'' rule}

In order to discuss this additional constraint, let us begin
by recalling that under an irregular embedding of $G'$ into  $G$,
the adjoint representation of $G$ generally decomposes into a sum of
irreducible
representations of $G'$, one of which includes
the adjoint of $G'$:
\beq
        {\bf R}_{\rm adj} ~\to~
        {\bf R'}_{\rm adj} ~+~ {\rm other}~G'~{\rm representations}~.
\label{adjtomess}
\eeq
This causes no inconsistency as far as the mathematical embedding of the $G'$
subgroup is concerned.  However, within the physical context of string theory,
a decomposition of the form (\ref{adjtomess}) leads to serious problems.
To see this,
let us imagine the particular string model {\it before}\/ introducing
the final GSO projections that induce the dimensional truncation
and  break $G$ to $G'$.
In this parent string model, all of the gauge-boson states that fill out the
adjoint representation of $G$ will be massless, and
carry a spacetime vector index.
However, according to (\ref{adjtomess}), {\it after}\/ the
final GSO projections are performed,
the states that survive will in general fill out not only
the adjoint representation of $G'$, but also ``other'' non-adjoint
representations
as well.  However, all of these states will continue
to be massless, and moreover they will all continue
to carry a spacetime vector index.
Thus, {\it a priori}\/, all of these states will give rise
to gauge bosons.  This is impossible, however,
since such states will not fill out
the adjoint of any Lie group!
In other words, within the physical context of string theory, such a
decomposition (\ref{adjtomess}) would still not be consistent.

In general, string theory manages to avoid this problem in a very
elegant fashion:  the very same GSO projections that
effect the dimensional truncation in the first place also
simultaneously project the ``other'' unwanted $G'$ representations
out of the spectrum.  In other words, string theory manages
to enforce an ``adjoint to adjoint only'' rule.  Clearly,
this property goes beyond mere group theory, and is ultimately
guaranteed by the self-consistency of the underlying string construction.
It is therefore instructive to see how this arises in practice.

\subsection{The ``adjoint to adjoint only'' rule:  An example}

The pedagogical example we shall consider demonstrates
not only how the required sorts of GSO projections
can be achieved in a particular string construction,
but also how the ``adjoint to adjoint only'' rule is
simultaneously and automatically enforced.

The example we shall consider
consists of a simple worldsheet theory
containing ten Majorana-Weyl fermions.
In an actual string model, these two-dimensional fermions (which we
shall label $\lambda_{1,...,10}$) might be part of the internal
worldsheet degrees of freedom.  Now, if these fermions are treated
symmetrically
(meaning that their excitations are all interchangeable and
commute with the GSO projections in each sector), then
the internal symmetry corresponding to these ten fermions is $SO(10)$.
Typically the GSO constraints for such gauge boson
states take the form
\beq
     \sum_{i=1}^{10} N_i ~=~ 2
\label{GSOone}
\eeq
where the real-fermion
number operators $N_i$ are $1$ if the lowest mode of the $i^{\rm th}$
fermion is excited, and $0$ otherwise.
This gives rise to $45$ different states, which comprise
the adjoint of $SO(10)$.  That this group is realized at level
one is easily determined by calculating the central charge
of this fermionic representation, $c_{\rm tot}=10\times 1/2=5$,
and comparing with (\ref{centralcharge}).

It is also possible to obtain these results by
considering the charge lattice corresponding to the gauge bosons.
This is done as follows.  Because these ten
real fermions are treated completely symmetrically
by the single GSO constraint equation (\ref{GSOone}), it is possible
to pair these fermions to form five complex fermions via
$\psi_k\equiv  (\lambda_{2k-1} +i \lambda_{2k})/\sqrt{2}$
for $k=1,...,5$.
Through this relation, we can directly relate the particle and anti-particle
excitation mode operators $b^{(a)}_r, \overline{b}^{(a)}_r$
of the complex fermions $\psi_a$ to the particle excitation mode operators
$d^{(i)}_r$ of the real fermions $\lambda_i$:
\beqn
        b^{(k)}_r &=& {1\over \sqrt{2}}\, \left( d_r^{(2k-1)} +
                    i\,\overline{d}_r^{(2k)} \right)\nonumber\\
        \overline{b}^{(k)}_r &=&
      {1\over \sqrt{2}}\, \left( d_r^{(2k-1)} -
	i\,\overline{d}_r^{(2k)} \right)~.
\label{moderelations}
\eeqn
Here the index $r$ signifies the energy of the excitation, odd half-integer
for the Neveu-Schwarz sector (such as we encounter in this toy model,
with the lowest mode $r=1/2$ producing the gauge bosons), and integer
otherwise.  Likewise, we can define the number operators\footnote{
        In giving this form for the real-fermion number operators, we are
        omitting a number of subtleties which are important for
        the case of Ramond boundary conditions.  These are discussed in
        Ref.~\cite{KLST}, and will not be needed for what follows.}
for the individual
real fermions,
$N^{(i)}\equiv \sum_r\, d_r^{\dagger (i)} d_r^{(i)}$,
as well as
the complex-fermion number operators ${\cal N}_a$ for each complex fermion
$\psi_a$:  ${\cal N}^{(a)}\equiv \sum_r\,( b_r^{\dagger (i)} b_r^{(i)} -
\overline{b}_r^{\dagger (i)} \overline{b}_r^{(i)})$.
Thus ${\cal N}^{(a)}$ yields
$+1$ for a single particle excitation, $-1$ for a single
anti-particle excitation, and $0$ if both or neither are excited.
Note that excitations of both of the real fermions
$(\lambda_{2i-1},\lambda_{2i})$
in a single pair amount to a joint particle/anti-particle excitation in the
corresponding complex fermion $\psi_i$.
Hence, of the 45 gauge boson states above, five give rise to
states with all ${\cal N}_{a}=0$,
while the remaining 40 states have different configurations of
non-zero $\vec \calN$.  Now, in the conventional normalization, the
five-dimensional charge lattice ${\bf Q}$ corresponding to these
gauge-boson states is simply the set of allowed $\vec {\cal N}$.
Thus the five states with $\vec {\cal N}=0$ correspond to the
generators of the Cartan subalgebra, and the 40 remaining states
fill out the five-dimensional root lattice of $SO(10)$.
This much is of course simply the standard realization
of $SO(10)_1$ in terms of five complex fermions.

Let us now consider what happens if, along with the single
GSO constraint equation (\ref{GSOone}),
we impose two additional constraint equations of the form
\beqn
          N_1+N_2+N_3 +N_4 &\in& 2\,\IZ \nonumber\\
          N_1+N_2+N_3 +N_5 &\in& 2\,\IZ~.
\label{GSOtwo}
\eeqn
Such extra constraint equations can be
realized, for example, in string models built out of so-called
``necessarily real fermions'' \cite{KLST}.
It is clear that this has a number of consequences,
among them a decrease in the number of surviving states.

In this example, it is straightforward
to determine the residual subgroup that survives.
Considering the three constraint equations (\ref{GSOone}) and
(\ref{GSOtwo}) simultaneously, we see that no
states are allowed in which either $\lambda_4$ or $\lambda_5$ are excited.
Indeed, our set of
allowed excitations splits into two disjoint groups,
the first consisting of any two excitations from the set
$\lbrace \lambda_{1,2,3}\rbrace$ (thereby giving rise to three
possible states), and the second
consisting of any two excitations from the
set $\lbrace \lambda_{6,7,...,10}\rbrace$ (giving
rise to ten states).  These correspond
to the adjoint representations of $SU(2)$ and $SO(5)$ respectively.
It is also trivial to verify, at least in this case, that the $SU(2)$
symmetry is in fact realized at level two, since
this $SU(2)$ gauge factor is now essentially represented in
terms of the three real fermions $\lambda_{1,2,3}$, with total
central charge $c=3/2$.

In this simple example, it was straightforward
to deduce that the level of the $SU(2)$ gauge factor was increased
thanks to an obvious representation in terms of three real
fermions, and a quick comparison of the central charges
involved.
However, the same results can be obtained by considering
the effects on the original $SO(10)$ charge lattice
induced by the additional constraints in (\ref{GSOtwo}).
Recall that the first step in determining the charge lattice
was to determine a pairing or complexification of
the real fermions, for it is only in terms of such complex
fermions that $U(1)$ charges can be defined.
However, with the new constraints
(\ref{GSOtwo}) adjoined, we now see that
no consistent complexifications are possible for all ten
real fermions.  The maximal number of
complex fermions that can be formed is three ({\it i.e.},
$\psi_1$, $\psi_4$, and $\psi_5$),
corresponding to the rank of the resulting gauge group.
Hence, in our former five-dimensional charge-vector space,
we see that two dimensions have simply been extinguished.
This is of course nothing but a dimensional truncation of the charge lattice,
now explicitly realized through the sets of GSO projections
(\ref{GSOone}) and (\ref{GSOtwo}).

In order to analyze this remaining charge lattice,
let us first consider the three states which form
the adjoint representation of the $SU(2)$ gauge group factor.
In the notation $|N_1,N_2,N_3,...\rangle$, these three states
are
$|1\rangle \equiv |1,1,0,...\rangle$,
$|2\rangle \equiv |1,0,1,...\rangle$,
$|3\rangle \equiv |0,1,1,...\rangle$.
Let us describe these states in terms of the
full five-dimensional space corresponding to the
five complex fermions $\psi_{1,...,5}$.
The state $|1\rangle$, in the operator language of the complex fermions,
is nothing but $\psi_1^\dagger \psi_1$.
This is therefore the Cartan generator of the $SU(2)$ group.
For the remaining states, we may, for convenience,
switch to a new basis defined by
$|2'\rangle\equiv (|2\rangle + i |3\rangle)/\sqrt{2}$
and $|3'\rangle\equiv (|2\rangle - i |3\rangle)/\sqrt{2}$.
Using the mode relations (\ref{moderelations}),
we then find that these new states can be expressed in
terms of the number operators $\calN_{a=1,2}$
corresponding to the complex fermions $\psi_1$ and $\psi_2$ as
\beqn
     |2'\rangle &=& {1\over\sqrt{2}}\,
      \biggl(|\calN_1=1,\calN_2=1\rangle + |\calN_1=1,\calN_2=
-1\rangle\biggr)\nonumber\\
     |3'\rangle &=&
      {1\over\sqrt{2}}\,
      \biggl(|\calN_1= -1,\calN_2=1\rangle + |\calN_1= -1,\calN_2=
-1\rangle\biggr)~.
\label{su2rootsfullspace}
\eeqn
This is the description that would be appropriate {\it if}\/
there truly existed a full five-dimensional charge space.
However, the second and third dimensions of this lattice
have actually been truncated.
Thus, concentrating on only the eigenvalue of $\calN_1$
(or equivalently, applying the above Cartan generator $\psi_1^\dagger \psi_1$
to determine the quantum numbers of our non-zero roots),
we find that the $|2'\rangle$ state corresponds to the
positive root at the point $+1$ in the remaining
one-dimensional $SU(2)$ root lattice, and
that the $|3'\rangle$ state corresponds to the lattice site at
$-1$. These three shortened roots at lattice
sites $\lbrace 0,\pm 1\rbrace$
comprise the root system of $SU(2)_2$.

It is also instructive to consider how the non-simply
laced group $SO(5)$ is realized from the remaining
ten gauge bosons in this example --- {\it i.e.}, those
which are constructed via any two excitations from the fermion
set $\lbrace \lambda_{6,7,...,10}\rbrace$.
Since we can consistently form the two complex fermions
$\psi_{4,5}$ from the four real fermions $\lambda_{7,8,9,10}$,
the six excitations involving only $\lambda_{7,8,9,10}$
give rise to points in the charge lattice of lengths
zero or $\sqrt{2}$, as expected.  These correspond to the
two zero roots and the four long roots in the $SO(5)$
root system.  However, because the four states
involving excitations of both $\lambda_6$ and one of the
remaining fermions all have components in the truncated
directions, they suffer dimensional projections
and are reduced in length from $\sqrt{2}$
to $1$. Their projections onto the surviving directions then form
the four short roots of $SO(5)$, thereby
completing the root system of this non-simply laced algebra.

Thus, to summarize the dimensional truncations in this ten-fermion
example, we see that
imposing the constraint (\ref{GSOone}) alone yields
the gauge group $SO(10)_1$, and
then additionally imposing {\it only the first}\/ of
the constraints in (\ref{GSOtwo}) breaks this gauge group
to $SO(4)_1\times SO(6)_1$.
However, imposing the final constraint in (\ref{GSOtwo})
effects the dimensional truncation, removing
two dimensions from the total charge lattice.
One of these dimensions is removed from the $SO(4)= SU(2)\times SU(2)$
lattice, and produces $SU(2)$ at level two in the manner that we have
already outlined in Fig.~\ref{su2su2}.
The other dimension is removed from the $SO(6)= SU(4)$ lattice,
and produces the non-simply laced group $SO(5)_1$.

Moreover, this ten-fermion example also shows precisely
how the ``adjoint to adjoint only'' rule is automatically satisfied.
In this example, the \rep{15} representation of $SO(6)$
decomposes into the \rep{10} and \rep{5} representations of $SO(5)$.
However, the third GSO constraint, which  not only effects the
dimensional truncation of the charge lattice,
also projects out the \rep{5} representation.
Thus, only the adjoint \rep{10} representation of $SO(5)$ survives.
Similarly, of the original gauge bosons of $SO(4)=SU(2)\times SU(2)$, only one
copy
of the $SU(2)$ gauge bosons survives.
Thus, as required, we see that the GSO projections that effect the dimensional
truncation also simultaneously project out the non-adjoint representations,
so that only the adjoint representation of the
final subgroup survives.

%==============================================================================
\vfill\eject
\setcounter{footnote}{0}
\section{Classification of String GUT Group Embeddings}

As we have shown in Sects.~3 and 4, higher-level and/or non-simply laced
gauge symmetries can arise in free-field
heterotic string constructions only through dimensional
truncations
of the charge lattice, which correspond uniquely to irregular embeddings.
Irregular embeddings, however, have been completely classified by
mathematicians.
Thus, by virtue of our identification,
we now have at our disposal the means for a powerful classification
of the possible embeddings through which such gauge symmetries can be realized
in free-field string constructions.
This will enable us to answer questions of direct relevance
to string GUT model-builders, such as classifying all possible ways
of realizing, {\it e.g.}, $SU(5)_2$ or $SO(10)_2$ gauge groups
in free-field string models.

In this section, we shall perform such a classification.
We shall begin in Sect.~7.1 by recalling the reasons that higher-level GUT
groups are of interest in string theory, and then we shall proceed in Sect.~7.2
to
describe the mathematical classification of irregular embeddings.
In Sect.~7.3 we will then use these results to completely classify
all methods of obtaining $G_{\rm GUT}$ at levels $k=2,3,4$, for
the cases $G_{\rm GUT}=SU(5)$, $SU(6)$, $SO(10)$, and $E_6$;
furthermore, we shall prove in Sect.~7.3.5 that it is impossible
to realize $SO(10)_{k>4}$ or $(E_6)_{k>3}$ in free-field string theory.
The material in Sects.~7.2 and 7.3 is fairly technical,
and is not necessary for understanding the final results of
our classification.
We have therefore collected together and summarized the results of our GUT
classification in Sect.~7.4, which can be read independently
of the other sections.

\subsection{Why higher-level GUT groups?}

Affine Lie algebras with levels $k>1$ are of particular interest
in string theory because
their unitary representations include
various phenomenologically desired representations which
are otherwise precluded at levels $k=1$.
These include, for example, the adjoint representations which
are necessary for Higgs scalars
in order to realize
the standard symmetry-breaking scenarios of
most conventional GUT theories, such as those of $SU(5)$ or $SO(10)$.
In general, the unitary irreducible representations
of a given affine Lie algebra at level $k$
(and consequently, the only representations
that can appear in a consistent string model realizing such
an algebra) are those for which
\beq
            k~\in~\IZ~~~~~{\rm and}~~~~~~
        0~\leq~  \sum_{i=1}^{{\rm rank}(G)} a_i\,m_i ~\leq~ k~
\label{unitary}
\eeq
where $m_i$ are the so-called ``co-marks'' corresponding
to each simple root $\vec \alpha_i$, and where $a_i$ are the Dynkin
labels of the highest weight of the representation.
The conformal dimension of such a representation
is then given by
\beq
           h_{(R)} ~=~ {C^{(R)}_G /\vec \alpha_h^2 \over k+\tilde h_G}
\label{confdim}
\eeq
where $C^{(R)}_G$ is the eigenvalue of the quadratic Casimir
acting on the representation $R$.  This eigenvalue is
defined analogously to (\ref{quadcas}), via
\beq
       \sum_{a=1}^{{\rm dim}(G)}\, (T^a T^a)_{ij}
            ~=~ C^{(R)}_G\,\delta^{ij}
\label{quadcasR}
\eeq
where $T^a$ are the group generators in the representation $R$.
In general, $C^{(R)}_G=(\vec \Lambda,\vec \Lambda+2\vec \delta)$
where $\vec \Lambda$ and $2\vec\delta$ are respectively
the highest root and the sum of the positive roots of $G$.
Thus, the conformal dimension $h^{(R)}$ is directly related
to the {\it quadratic index}\/ $\ell_G(\vec\Lambda)$ of the representation,
as defined in (\ref{indexdef}), via
\beq
         h_{(R)} ~=~ { {\rm dim}\, G \over {\rm dim}\,R}~
         {1\over k+ \tilde h_G} ~
           {\ell_G(\vec \Lambda) \over \vec\alpha_h^2}~.
\eeq

In heterotic string theory, a particular representation $R$ can
appear in the massless spectrum if and only if its conformal dimension
satisfies $h_{(R)}\leq 1$.
This, along with the unitarity constraint (\ref{unitary}),
then limits the allowed representations that may appear
for a given gauge group realized at a given affine level.
Below, we have tabulated the complete set of unitary representations
that can appear in the massless string spectrum
for the phenomenologically
interesting gauge groups $SU(5)$, $SU(6)$,
$SO(10)$, and $E_6$  at levels $1\leq k\leq 4$.  In each
case, we have listed the values of
$(\rep{dim(R)},h_{(R)})$ for each such representation,
and have also listed the central charge
corresponding to the relevant group factor.
Note that for each group $G_k$, the maximum affine level $k_{\rm max}$
that is {\it a priori}\/ allowed
is determined by requiring that $c(G_k)\leq 22$.

\def\bbox#1{\parbox[t]{1.15 in}{{#1}}}
\def\bbreak{\vfill\break}

\beq
\begin{tabular}{c|c|c|c|c}
  ~& $SU(5)$ & $SU(6)$ & $SO(10)$ & $E_6$ \\
%  \begin{tabular}{c|c|c|c|c|c|c|c}
  %  ~& $SU(2)$ & $SU(3)$ & $SU(4)$ & $SU(5)$ & $SU(6)$ & $SO(10)$ & $E_6$ \\
 & ($k_{\rm max}=55$) & ($k_{\rm max}=10$) & ($k_{\rm max}=7$) & ($k_{\rm
max}=4$) \\
\hline
\hline
$k=1$ &
        %   \bbox{
        %   $c=1:$\bbreak
        %   ~(\rep{2},~1/4)
        %   }
        %   &
        %   \bbox{
        %   $c=2:$\bbreak
        %   ~(\rep{3},~1/3)
        %   }
        %   &
        %   \bbox{
         %   $c=3:$\bbreak
          %   ~(\rep{4},~3/8)\bbreak
          %   ~(\rep{6},~1/2)
          %   }
        %   &
        \bbox{
         $c=4:$\bbreak
         ~(\rep{5},~2/5)\bbreak
           ~(\rep{10},~3/5)
           }
        &
        \bbox{
         $c=5:$\bbreak
         ~(\rep{6},~5/12)\bbreak
         ~(\rep{15},~2/3)\bbreak
           ~(\rep{20},~3/4)
           }
        &
        \bbox{
         $c=5:$\bbreak
           ~(\rep{10},~1/2)\bbreak
           ~(\rep{16},~5/8)
           }
        &
        \bbox{
         $c=6:$\bbreak
           ~(\rep{27},~2/3)
           }
        \\
\hline
$k=2$ &
        %   \bbox{
         %   $c=3/2:$\bbreak
         %   ~(\rep{2},~3/16)\bbreak
           %   ~(\rep{3},~1/2)
           %   }
        %   &
        %   \bbox{
         %   $c=16/5:$\bbreak
         %   ~(\rep{3},~4/15)\bbreak
           %   ~(\rep{6},~2/3)\bbreak
           %   ~(\rep{8},~3/5)
           %   }
        %   &
        %   \bbox{
         %   $c=5:$\bbreak
           %   ~(\rep{4},~5/16)\bbreak
           %   ~(\rep{6},~5/12)\bbreak
           %   ~(\rep{10},~3/4)\bbreak
           %   ~(\rep{15},~2/3)\bbreak
           %   ~(\rep{20},~13/16)\bbreak
           %   ~(\rep{20$^\prime$},~1)
           %   }
        %   &
        \bbox{
         $c=48/7:$\bbreak
           ~(\rep{5},~12/35)\bbreak
           ~(\rep{10},~18/35)\bbreak
           ~(\rep{15},~4/5)\bbreak
           ~(\rep{24},~5/7)\bbreak
           ~(\rep{40},~33/35)\bbreak
           ~(\rep{45},~32/35)
           }
         &
        \bbox{
         $c=35/4:$\bbreak
           ~(\rep{6},~35/96)\bbreak
           ~(\rep{15},~7/12)\bbreak
           ~(\rep{20},~21/32)\bbreak
           ~(\rep{21},~5/6)\bbreak
           ~(\rep{35},~3/4)\bbreak
           ~(\rep{84},~95/96)
           }
         &
        \bbox{
         $c=9:$\bbreak
           ~(\rep{10},~9/20)\bbreak
           ~(\rep{16},~9/16)\bbreak
           ~(\rep{45},~4/5)\bbreak
           ~(\rep{54},~1)
           }
         &
        \bbox{
         $c=78/7:$\bbreak
           ~(\rep{27},~13/21)\bbreak
           ~(\rep{78},~6/7)
           } \\
\hline
$k=3$ &
        %   \bbox{
         %   $c=9/5:$\bbreak
           %   ~(\rep{2},~3/20)\bbreak
           %   ~(\rep{3},~2/5)\bbreak
           %   ~(\rep{4},~3/4)
           %   }
          %   &
        %   \bbox{
         %   $c=4:$\bbreak
           %   ~(\rep{3},~2/9)\bbreak
           %   ~(\rep{6},~5/9)\bbreak
           %   ~(\rep{8},~1/2)\bbreak
           %   ~(\rep{10},~1)\bbreak
           %   ~(\rep{15},~8/9)
           %   }
          %   &
        %   \bbox{
         %   $c=45/7:$\bbreak
           %   ~(\rep{4},~15/56)\bbreak
           %   ~(\rep{6},~5/14)\bbreak
           %   ~(\rep{10},~9/14)\bbreak
           %   ~(\rep{15},~4/7)\bbreak
           %   ~(\rep{20},~39/56)\bbreak
           %   ~(\rep{20$^\prime$},~6/7)\bbreak
           %   ~(\rep{36},~55/56)
           %   }
          %   &
        \bbox{
         $c=9:$\bbreak
           ~(\rep{5},~3/10)\bbreak
           ~(\rep{10},~9/20)\bbreak
           ~(\rep{15},~7/10)\bbreak
           ~(\rep{24},~5/8)\bbreak
           ~(\rep{40},~33/40)\bbreak
           ~(\rep{45},~4/5)\bbreak
           ~(\rep{75},~1)
           }
          &
        \bbox{
         $c=35/3:$\bbreak
           ~(\rep{6},~35/108)\bbreak
           ~(\rep{15},~14/27)\bbreak
           ~(\rep{20},~7/12)\bbreak
           ~(\rep{21},~20/27)\bbreak
           ~(\rep{35},~2/3)\bbreak
           ~(\rep{70},~11/12)\bbreak
           ~(\rep{84},~95/108)\bbreak
           ~(\rep{105},~26/27)
           }
          &
        \bbox{
         $c=135/11:$\bbreak
           ~(\rep{10},~9/22)\bbreak
           ~(\rep{16},~45/88)\bbreak
           ~(\rep{45},~8/11)\bbreak
           ~(\rep{54},~10/11)\bbreak
           ~(\rep{120},~21/22)\bbreak
           ~(\rep{144},~85/88)
           }
          &
        \bbox{
         $c=78/5:$\bbreak
           ~(\rep{27},~26/45)\bbreak
           ~(\rep{78},~4/5)
           }
         \\
\hline
$k=4$ &
        %   \bbox{
         %   $c=2:$\bbreak
           %   ~(\rep{2},~1/8)\bbreak
           %   ~(\rep{3},~1/3)\bbreak
           %   ~(\rep{4},~5/8)\bbreak
           %   ~(\rep{5},~1)
           %   }
         %   &
        %   \bbox{
         %   $c=32/7:$\bbreak
           %   ~(\rep{3},~4/21)\bbreak
           %   ~(\rep{6},~10/21)\bbreak
           %   ~(\rep{8},~3/7)\bbreak
           %   ~(\rep{10},~6/7)\bbreak
           %   ~(\rep{15},~16/21)
           %   }
         %   &
        %   \bbox{
         %   $c=15/2:$\bbreak
           %   ~(\rep{4},~15/64)\bbreak
           %   ~(\rep{6},~5/16)\bbreak
           %   ~(\rep{10},~9/16)\bbreak
           %   ~(\rep{15},~1/2)\bbreak
           %   ~(\rep{20},~39/64)\bbreak
           %   ~(\rep{20$^\prime$},~3/4)\bbreak
           %   ~(\rep{20$^{\prime\prime}$},~63/64)\bbreak
           %   ~(\rep{36},~55/64)\bbreak
           %   ~(\rep{45},~1)\bbreak
           %   ~(\rep{64},~15/16)
           %   }
         %   &
        \bbox{
         $c=32/3:$\bbreak
           ~(\rep{5},~4/15)\bbreak
           ~(\rep{10},~2/5)\bbreak
           ~(\rep{15},~28/45)\bbreak
           ~(\rep{24},~5/9)\bbreak
           ~(\rep{40},~11/15)\bbreak
           ~(\rep{45},~32/45)\bbreak
           ~(\rep{50},~14/15)\bbreak
           ~(\rep{70},~14/15)\bbreak
           ~(\rep{75},~8/9)
           }
         &
        \bbox{
         $c=14:$\bbreak
           ~(\rep{6},~7/24)\bbreak
           ~(\rep{15},~7/15)\bbreak
           ~(\rep{20},~21/40)\bbreak
           ~(\rep{21},~2/3)\bbreak
           ~(\rep{35},~3/5)\bbreak
           ~(\rep{70},~33/40)\bbreak
           ~(\rep{84},~19/24)\bbreak
           ~(\rep{105},~13/15)\bbreak
           ~(\rep{120},~119/120)\bbreak
           ~(\rep{189},~1)
           }
         &
        \bbox{
         $c=15:$\bbreak
           ~(\rep{10},~3/8)\bbreak
           ~(\rep{16},~15/32)\bbreak
           ~(\rep{45},~2/3)\bbreak
           ~(\rep{54},~5/6)\bbreak
           ~(\rep{120},~7/8)\bbreak
           ~(\rep{144},~85/96)\bbreak
           ~(\rep{210},~1)
           }
         &
        \bbox{
         %$c=39/2:$\bbreak
         ~~\bbreak
          ~{\rm cannot be}\bbreak
          %~{\rm be}\bbreak
          ~{\rm realized}\bbreak
          ~{\rm in}\bbreak
          ~{\rm free-field}\bbreak
          ~{\rm string}\bbreak
          ~{\rm theory}
           }
         \\
\hline
\end{tabular}
\label{bigtable}
\eeq

Note that in the above table we have not listed singlet representations
 [for which the corresponding entry is (\rep{1},0) for all groups and levels].
We have also omitted
the complex-conjugate representations,
and representations obtained from those listed above
via exchange of
the simple roots when such exchanges are symmetries of the Dynkin
diagram.  In the cases of distinct representations having the
same dimension [such as the
\rep{70} representations of $SU(5)$
and the \rep{210} representations of $SO(10)$],
we have adopted the primed labelling conventions of
Ref.~\cite{slansky}.
Furthermore, note that in
our calculations of the conformal dimensions, we have assumed
that each representation of a given group $G$ transforms as
a singlet under all {\it other}\/ group factors.
Otherwise, for representations
that are charged with respect to several groups simultaneously,
one need only add the corresponding conformal dimensions.

There are several things to note from this table.
First, note that $E_6$ at affine levels $k\geq 4$ cannot be
realized in free-field string models.
Although the central charge of such a group factor at $k=4$
would only be $39/2$ --- which is less than 22 and hence potentially
realizable --- we will prove later in this section
(as the result of our classification)
that there are ultimately no string-theoretic {\it embeddings}\/
for $E_6$ at such levels $k\geq 4$.
We thereby conclude that the {\it only}\/ massless representations of $E_6$
that can {\it ever}\/ be realized in free-field string models are the
$\rep{27}$ and $\rep{78}$ representations.

Second, note that the $\rep{126}$ representation of $SO(10)$
does not appear for $SO(10)$ realizations at levels $1\leq k\leq 4$.
Indeed, the first level for which the $\rep{126}$ can potentially appear
in the massless string spectrum is $k_{SO(10)}=5$.
Like the $E_6$ case, however,
we will prove later in this section (again as a
result of our classification) that the maximum affine level
at which $SO(10)$ can ever be realized is in fact $k=4$.
Thus, it is {\it impossible}\/ to realize the $\rep{126}$ representation
of $SO(10)$ in free-field heterotic string theory.
This result is completely general, and applies for all
free-field string constructions.

Most importantly, however, we see from this table that regardless of the GUT
gauge group in question, we cannot achieve the desired massless adjoint
representation required for the GUT Higgs scalar
unless the GUT group is realized at an affine level $k\geq 2$.
Moreover, in order to avoid too many unwanted representations appearing in the
massless string spectrum, one usually wishes to keep the affine level
from being {\it too}\/ large.  Typically, the choices $2\leq k \leq 4$ are
of direct phenomenological interest, especially for the standard
GUT groups such as $SU(5)$, $SU(6)$, $SO(10)$, and $E_6$.

We therefore seek to classify the ways in which such groups
and levels can be realized in free-field string constructions.

\subsection{Classification of irregular embeddings}

We begin by listing the mathematical results for the complete classification
of irregular embeddings.
As far as we are aware, the results of a complete classification do not appear
in any one place in the mathematics or physics literature.  We have therefore
gleaned and compiled the following classification from various
different sources \cite{cahn,slansky},
most notably the original papers of Dynkin \cite{dynkin2,dynkin3}.

It turns out that there are only three classes of embeddings that
we need to consider,
each with its own set of rules.  These three classes correspond to the
following
irregular embeddings:  embeddings of simple algebras into
classical algebras (which we shall call Class I),
embeddings of non-simple algebras into classical algebras
(Class II),
and embeddings of all algebras into exceptional algebras (Class III).
Note that, as in previous sections, we use a notation in
which $SU(r+1)$, $SO(2r)$, $Sp(2r)$, and $SO(2r+1)$ each have rank $r$.

\underbar{Class I}:
Of the three classes we shall consider,
the irregular embeddings of simple algebras into classical algebras
are the most complicated.
Nevertheless,
the general property that
governs such embeddings is most easily stated as follows.
For each simple group $G'$ and for each $n$-dimensional irreducible
representation $\rep{n}$
of $G'$, there always exists at least one irregular embedding of $G'$ into
a larger-rank classical group $G''$
which also has an $n$-dimensional irreducible representation.
In each case, such $G'\subset G''$ embeddings then give rise to the branching
rule $\rep{n}\to\rep{n}$.
If $G$ is the smallest of such groups $G''$
(so that there exists no proper subgroup of $G$ which contains $G'$),
then the embedding $G'\subset G$ is called {\it maximal}.
In order to achieve a classification of embeddings, one therefore
focuses on maximal embeddings, since these can be iterated in order
to reproduce all other embeddings.

The sets of maximal embeddings $G'\subset G$ generally come
in two distinct patterns, which we shall label Class IA and Class IB.
Class IA maximal embeddings are those that arise from the following
``theorem'':
for {\it almost}\/ every $n$-dimensional representation $\rep{n}$ of $G'$,
the following is a maximal irregular embedding:
\beqn
      {\rm if~\rep{n}~ is~real}~&~~\Longrightarrow&~~~ G'_{k'}\subset
                 SO(n)_2\nonumber\\
      {\rm if~\rep{n}~ is~pseudo{-}real}~&~~\Longrightarrow&~~~ G'_{k'}\subset
                 Sp(n)_1
           \nonumber\\
      {\rm if~\rep{n}~ is~complex}~&~~\Longrightarrow&~~~ G'_{k'}\subset
                 SU(n)_1~.
\label{rule1}
\eeqn
In (\ref{rule1}), the subscripts indicate the affine levels of the embeddings,
with
$k'$ in each case given by $k'=\ell_{G'}(\rep{n})$ where
$\ell_{G'}(\rep{n})$ is the index (\ref{indexdef})
of the $\rep{n}$ representation of $G'$.
These affine levels have been determined as follows.
Thanks to the simple branching rules $\rep{n}\to\rep{n}$ that
arise in such embeddings, we see that
we can easily determine the affine
level of the irregular $G'$ subgroup in each case
by comparing, as in (\ref{embeddingindex}),
the index of the $\rep{n}$ representation of $G'$ with the
index of the fundamental representation of the group in which
it is embedded according to (\ref{rule1}).
In general, the index of the fundamental representation for these latter
groups (in our present normalization with longest roots of length $\sqrt{2}$
for all groups) is
\beqn
             \ell_{SU(n)}\, ({\rm fundamental}) &=& 1~\nonumber\\
             \ell_{SO(2n+1)}\, ({\rm fundamental}) &=& 2~\nonumber\\
             \ell_{Sp(2n)}\, ({\rm fundamental}) &=& 1~\nonumber\\
             \ell_{SO(2n)}\, ({\rm fundamental}) &=& 2~.
\label{fundindices}
\eeqn
Thus, the level of the $G'$ subgroup in each
case becomes simply the index $\ell_{G'}(\rep{n})$ of the $\rep{n}$
representation of $G'$.
Of course, for consistency,
these indices must be calculated in a normalization for which
the longest roots of $G'$ have length $\sqrt{2}$.

As an example of the rule (\ref{rule1}), let us consider the
irregular embeddings of $SU(3)$.
Recall that the representations of $SU(3)$ can be labelled by the
two Dynkin indices $(a_1,a_2)$ of their highest weights;
the corresponding representation is
then real if $a_1=a_2$, and complex otherwise.
Thus, by scanning the irreducible representations of $SU(3)$,
we easily find that $G'\equiv SU(3)$ can be maximally irregularly embedded in
all of the following groups at level one:  $SU(3)$ (with $k_{G'}=1$),
$SU(6)$ (with $k_{G'}=5$),
$SO(8)$ (with $k_{G'}=3$), $SU(10)$ (with $k_{G'}=15$),
$SU(21)$ (with $k_{G'}=70$), and so forth.
(The group $SU(15)$ is omitted for reasons to be explained.)
Of course, the first of these embeddings is that of $SU(3)$ into
itself, and does not correspond to a change in the affine level.

Unfortunately,
there are several special cases in which the embeddings (\ref{rule1}),
though valid, are not maximal.
An example of such a special case is the $\rep{15}$ representation of $SU(3)$:
the above rule would assert that the embedding into $G=SU(15)$ is maximal,
whereas in fact the true maximal embedding satisfying the $\rep{15}\to\rep{15}$
branching rule is the same one that also simultaneously satisfies the
$\rep{6}\to\rep{6}$ branching rule, namely $SU(3)_5\subset SU(6)_1$.
The complete set of special cases of (\ref{rule1}) has been enumerated by
Dynkin,\footnote{See, {\it e.g.}, Table I of Ref.~\cite{dynkin3}.}
but for the purposes of analyzing embeddings involving the string GUT
groups, we shall require only those special cases
for which $G'$ has rank $\geq 5$
and $G$ has rank $\leq 22$.
The complete set of such special cases is then the following:
\beq
\begin{tabular}{c|c||c}
 $\vec\Lambda'$ &  $G'$ & $G$  \\
\hline
\hline
 $a_1=a_3=1 $ & $SU(n+1)_{n-1}$, $n\geq 4$  &  $SU(n(n+1)/2)_1$ \\
 $a_1=2$, $a_2=1$  & $SU(n+1)_{n+3}$, $n\geq 2$  &  $SU((n+1)(n+2)/2)_1$ \\
 $a_n=k$, $k\geq 2$ &  $SO(2n+1)_1$, $n\geq 3$, $n$ odd &  $SO(2n+2)_1$ \\
 $a_2=a_4=1$ &  $SU(6)_6$ &  $Sp(20)_1$ \\
 $a_2=a_4=1$ &  $SO(10)_4$ & $SU(16)_1$ \\
 $a_4=1$; $a_3=a_5=1$ &  $SO(12)_8$ & $Sp(32)_1$ \\
\end{tabular}
\label{exceptions}
\eeq
For each case listed in (\ref{exceptions}),
we have indicated the relevant representation(s)
of $G'$ by specifying the non-zero Dynkin labels of its
highest weight(s) $\vec \Lambda'$ in $G'$.
Thus, except for the special cases (\ref{exceptions}),
the complete set of Class IA maximal irregular embeddings is given by
(\ref{rule1}).
We have calculated the affine levels in (\ref{exceptions}) following
the procedure discussed above.

The maximal irregular embeddings in Class IB follow a different set of
embedding patterns.\footnote{
   Note that these Class IB embeddings are often overlooked in
   the mathematical literature.  We have gleaned these embeddings
   from Table 5 of Ref.~\cite{dynkin3};  they are among the embeddings that
   Dynkin ``rejected'' as not being among the exceptions to the Class IA rule.
   However, for our purposes, we must include these embeddings because,
   like the others, they give rise to higher-level and non-simply laced
subgroups.  }
Those for which $G'$ has rank $\geq 4$ and
$G$ has rank $\leq 22$, along with some of the relevant representation(s)
satisfying
the $\rep{n}\to\rep{n}$ branching rule in each case, are as follows:
\beqn
         \rep{n}:& ~~SO(n)_2~&\subset~ SU(n)_1 \nonumber\\
         \rep{n}:& ~~Sp(n)_1~&\subset~ SU(n)_1 \nonumber\\
          a_n=k, ~k\geq 2:& ~~SO(2n+1)_1~&\subset~ SO(2n+2)_1 ~.
\label{rule1b}
\eeqn
Note that the third special case in (\ref{exceptions})
therefore reproduces a pattern in Class IB rather than a pattern in Class IA.
As before, we have determined the affine levels in (\ref{rule1b})
via (\ref{fundindices}).
Note that the last two embeddings in (\ref{rule1b})
give rise to subgroups which are not at higher levels,
but which instead are non-simply laced.

\underbar{Class II}:
Irregular embeddings of {\it non-simple}\/ algebras
into classical algebras also fall
into simple patterns which can be grouped into two subclasses.
For Class IIA, a complete list of rules is the following:
\beqn
        SU(m)_n~\times~ SU(n)_m ~&\subset&~ SU(mn)_1 \nonumber\\
        SO(m)_n~\times~ SO(n)_m ~&\subset&~ SO(mn)_1 \nonumber\\
        SO(m)_{2n}~\times~ Sp(n)_m ~&\subset&~ Sp(mn)_1 \nonumber\\
        Sp(m)_{n/2}~\times~ Sp(n)_{m/2} ~&\subset&~ SO(mn)_1~.
\label{rule2}
\eeqn
Once again, these embeddings are maximal, and
have a simple branching rule relating
their respective fundamental representations:  $\rep{mn}\to(\rep{m},\rep{n})$.
It is this branching rule
which has allowed us to determine the affine levels listed
in (\ref{rule2}).  To see this, let us consider the first embedding in
(\ref{rule2}),
namely $SU(m)\times SU(n)\subset SU(mn)$,
and let us assume that the $SU(mn)$ factor is realized at level one.
Then, since the branching rule for the fundamental representations
is $\rep{mn}\to(\rep{m},\rep{n})$, we see that as far as the $SU(m)$ subgroup
factor is concerned, this branching rule
essentially amounts to $\rep{mn}\to \rep{m}+ \rep{m}+... ~ \rep{m}$ (where
the $\rep{m}$ representation appears $n$ times).
Thus, given (\ref{fundindices}), we see that
the affine level of the $SU(m)$ factor
is simply $k_{SU(m)}= n$, and likewise $k_{SU(n)}=m$.
All of the affine levels given in (\ref{rule2}) are determined in this way.
Note, in particular, that the results for the last embedding listed in
(\ref{rule2})
are consistent since $n/2$ and $m/2$ are necessarily integers.

As examples of the Class IIA embeddings, let us once again consider
potential realizations of $SU(3)$ at higher levels.
In addition to those found above (which follow from the Class IA embeddings),
we now have the embeddings
$SU(3)_n\subset SU(3n)$ for all $n\geq 1$.  These follow from the Class IIA
embedding pattern
\beq
       SU(3)_n\times SU(n)_3 ~\subset~ SU(3n)_1~.
\label{diagsu3}
\eeq
As explained in the Appendix, however,
these embeddings ultimately
turn out to be
equivalent to the diagonal embeddings
\beq
  SU(3)_n~\subset~ SU(3)_1\times SU(3)_1 \times SU(3)_1 \times .... ~\subset~
SU(3n)_1~.
\eeq
Thus, we do not obtain new non-diagonal embeddings in this case.

Indeed, as we shall discuss in the Appendix,
it turns out to be a general property that
 {\it all}\/ of the embeddings in (\ref{rule2})
are ultimately equivalent to diagonal embeddings
once they are ``turned around'' and expressed
as a simple group embedded in a non-simple group.
This includes not only the $SU(n)$ embeddings, as discussed above,
but also the $SO(n)$ and $Sp(n)$ embeddings.
We conclude, therefore, that the Class IIA embeddings are all equivalent to the
diagonal
embedding of any level-$n$ group $G$ within an $n$-fold tensor product of
level-one factors
of $G$:
\beq
       G_n~\subset~ G_1 \times G_1 \times ... \times G_1~=~
       \mathop{\bigotimes}_{i=1}^{n}\,G^{(i)}_1~.
\label{diagrule}
\eeq
Of course, as we know, this diagonal embedding pattern extends
even to the cases when $G$ is one of the exceptional groups.

In this context,
it is worth observing that the existence of such diagonal embeddings
(\ref{diagrule})
in turn implies the existence of more general embeddings of the form
\beq
     G_{K}~\subset~ \mathop{\bigotimes}_{i=1}^{n} \, G_{k_i}^{(i)}
         ~~~~{\rm where}~K\equiv \sum_{i=1}^{n} k_i~
\label{diagextended}
\eeq
for all groups $G$.
This embedding is more subtle than it may at first appear, for while it is
clear
from (\ref{diagrule}) that
\beq
      \mathop{\bigotimes}_{i=1}^{n} \, G_{k_i}^{(i)}
      ~\subset~ \mathop{\bigotimes}_{i=1}^{K} \,G_1^{(i)}~
         ~~~~{\rm where}~K\equiv \sum_{i=1}^{n} k_i~,
\eeq
it is not obvious that the relation (\ref{diagextended}) actually follows
as a result.  In other words,
while both $G_K$ and $\otimes_i G_{k_i}^{(i)}$
are subgroups of
$G_1\times G_1 \times...\times G_1$, it is not clear that these subgroups are
actually related to each other as in (\ref{diagextended}).  However, it turns
out
that (\ref{diagextended}) is indeed valid.
This is most easily demonstrated via the dimensional truncation procedure
described in Sect.~3, in particular via a generalization of Fig.~\ref{su2su2}
to the case when two identical group factors $G_{k_1}\times G_{k_2}$ are
realized at different levels.
In such cases, the axis of truncation does not have an orientation angle
$\theta= 45^\circ$
as in Fig.~\ref{su2su2},
but rather an orientation angle $\theta=\tan^{-1}(k_2/k_1)$.
Induction then demonstrates (\ref{diagextended}) for all $n$.

We now turn to the Class IIB embeddings.
These embeddings consist of the single general rule
\beq
        SO(s)_1~\times~ SO(t)_1 ~\subset~ SO(s+t)_1 ~~~~{\rm for}~s,t~{\rm odd}
   ~,
\label{rule2b}
\eeq
with the branching rule for the fundamental representation
given by $(\rep{s+t})\to(\rep{s},\bone)+(\bone,\rep{t})$.
Once again, we have determined the affine levels in (\ref{rule2b})
via this branching rule and (\ref{embeddingindex}).
For the $SO(s)$ group factor,
the relevant branching rule is essentially
$(\rep{s+t})\to\rep{s}+\bone+\bone+...+\bone$
where there are a total of $t$ singlets (each of which has vanishing index).
Thus, we find that $k_{SO(s)}=1$, and likewise $k_{SO(t)}=1$.
Note that like the last two Class IB embeddings,
these Class IIB embeddings
also give rise to subgroups which are non-simply laced rather
than at higher levels.

\underbar{Class III}:
Finally, we consider the case of irregular embeddings
into the {\it exceptional}\/ groups.  In this case,
there does not exist a general
rule.  Nevertheless, all such embeddings have been compiled,
yielding the following complete list:\footnote{
       Note that in many of the standard mathematical references,
       the first embedding within $E_6$
        is often erroneously listed as $SU(2)_9$, rather than
        the correct $SU(3)_9$.  Likewise, the sixth embedding
        into $E_7$ is often erroneously listed as $SU(3)_3\times (F_4)_1$,
        rather than the correct $SU(2)_3\times (F_4)_1$.}
\beqn
     SU(2)_{28}~&\subset&~(G_2)_1\nonumber\\
     SU(2)_{156},\,
          SU(2)_8\times (G_2)_1 ~&\subset&~(F_4)_1\nonumber\\
     SU(3)_{9},\, Sp(8)_1,\, (F_4)_1,\, (G_2)_3,\,
          SU(3)_2\times (G_2)_1~&\subset&~(E_6)_1\nonumber\\
     SU(2)_{231},\, SU(2)_{399},\, SU(3)_{21},\, SU(2)_{15}\times
SU(2)_{24},~~~~&&\nonumber\\
     SU(2)_{7}\times (G_2)_{2},\, SU(2)_3\times (F_4)_1,\, Sp(6)_1 \times
(G_2)_1 ~&\subset&~
               (E_7)_1 \nonumber\\
     SU(2)_{520},\, SU(2)_{760},\, SU(2)_{1240},\,SO(5)_{12},~~~~&&\nonumber\\
          SU(2)_{16}\times SU(3)_{6},\,
       (F_4)_1\times (G_2)_1 ~&\subset&~(E_8)_1~.
\label{rule3}
\eeqn
In writing this list of Class III embeddings, we have explicitly included
the affine levels of the subgroups, as originally calculated by Dynkin.
These affine levels are calculated directly from (\ref{embeddingindex}).
All of these embeddings are maximal.

\subsection{Realizing the GUT groups $SU(5)$, $SU(6)$, $SO(10)$, and $E_6$ at
higher levels}

Given the above results, we now seek to classify the
methods by which phenomenologically interesting GUT groups
such as $G_{\rm GUT}=SU(5)$, $SU(6)$, $SO(10)$, and $E_6$ can be realized at
higher levels, {\it e.g.}, $k=2,3,...$.
As we have already seen, such higher-level realizations of these
groups are typically obtained in the literature as the diagonal
component within a tensor product of multiple copies of the level-one group.
Such diagonal embeddings, however, tend to make it difficult (though not
impossible)
to simultaneously obtain three generations.
We therefore seek to know whether there might exist other potentially
useful embeddings which are also capable of yielding $(G_{\rm GUT})_k$
for, {\it e.g.}, $k=2$, $k=3$, and $k=4$.
In this section,
we shall give a {\it complete}\/ classification
of all embeddings that realize
these groups at levels $k=2$, $k= 3$, and $k=4$.
In the case of $SO(10)$, we will actually extend this classification
to $k=7$ (which is the maximum level allowed by central-charge constraints).
Of course, the methods that we shall use are
applicable to all groups and levels.

\subsubsection{\it Classifying GUT Embeddings:
      ``initially irregular'' vs.\ ``initially regular''}

Recall that free-field string constructions
require that a higher-level gauge group ultimately be
realized as a subgroup of a level-one simply laced group (henceforth
denoted ``L1SL'') of rank $\leq 22$.
In general, there are various ways in which this can be done,
through sequences of both regular and irregular embeddings
involving an entire chain of intermediate groups $G_i$.
It is therefore first necessary to find an efficient way of
organizing our classification procedure.

In our case, we are interested in specifying
a particular smallest subgroup ({\it e.g.}, $SU(5)$ or $SO(10)$ at
a particular affine level).  We are less interested in specifying
the large group into which these subgroups are embedded,
provided it is an L1SL of appropriate rank.  (Indeed,
building our embeddings from the ground up,
we see that once we obtain an L1SL of appropriate rank,
we can always append purely
regular embeddings in order to embed this group
into the expected $SO(44)$ gauge group or into any of its enhanced L1SL's
such as $E_8\times E_8\times...$.)
Therefore, for the purposes of our investigation, the
primary distinction that naturally arises will involve
the sorts of {\it smallest}\/ or {\it initial}\/ embeddings that we shall use.

To this end, we shall first invent some terminology.
``Initially irregular embeddings'' will correspond to
embeddings in which the $SU(5)_k$ or $SO(10)_k$
subgroup factor is realized directly as the result of a maximal irregular
embedding into some intermediate group $G^\ast$.
In other words, the first embedding in the sequence of
maximal embeddings $G_{\rm GUT}\subset G^\ast
\subset ... \subset$ L1SL  must be irregular.
In some cases, this group
$G^\ast$ will be an L1SL itself, in which case we are done.
Otherwise, if the group $G^\ast$ is non-L1SL,
we then iterate the procedure.
Note that in this subsequent iteration that embeds $G^\ast$ into an L1SL,
we do not limit ourselves to only sequences of maximal irregular embeddings;
regular embeddings are also permitted.
Of course, in order to change the non-L1SL group $G^\ast$ into an L1SL,
at least one irregular embedding must appear in this subsequent iteration.
Nevertheless, we refer to such combined embeddings of $G_{\rm GUT}\subset$ L1SL
as ``initially irregular'' because the lowest initial embedding is irregular.

The remaining possible route for achieving higher-level GUT groups
will therefore involve the ``initially regular embeddings''.
Such initially regular embeddings
first realize $(G_{\rm GUT})_k$
as a {\it regular} subgroup of the first intermediate group $G^\ast$.
Note that we do not require that this initial regular embedding
of $G_{\rm GUT}$ into $G^\ast$ be maximal, since sequences of maximal
regular embeddings are also regular.
Thus, in general $G^\ast$ is defined as the highest group $G_i$ in the sequence
$G_{\rm GUT}\subset G_1 \subset G_2 \subset ... \subset$ L1SL
that can be reached from $G_{\rm GUT}$ through only regular embeddings.
We then subsequently seek to realize $G^\ast$ via an
embedding (necessarily irregular, but not necessarily maximal)
into an L1SL.
Note that this subsequent embedding of $G^\ast\subset$ L1SL is
itself initially irregular.

These ``initially regular'' embeddings of $(G_{\rm GUT})_k$ into an L1SL
generally proceed through two types
of intermediate group $G^\ast$.  The first type is a level-$k$
simply or non-simply laced group;  in such cases the regular embedding
of $SU(5)_k$ or $SO(10)_k$ into $G^\ast$ is constructed using only the long
roots of $G^\ast$.  We shall call these ``long-root embeddings''.
If $k\in 2\IZ$,
then a second possible type of intermediate group $G^\ast$ is a level-$(k/2)$
 {\it non}\/-simply laced intermediate group $G^\ast$;
in such cases, $(G_{\rm GUT})_k$
is realized through regular embeddings constructed
from the short roots of $G^\ast$.
We shall call these ``short-root embeddings''.
We have already seen examples of such short-root regular embeddings
in (\ref{regembedding}).

Thus, in order to realize a particular higher-level gauge group,
we must consider both ``initially irregular'' embeddings,
and long- and short-root ``initially regular'' embeddings.
These classifications exhaust all possibilities.
We shall therefore examine each case in turn.

\subsubsection{\it Initially irregular embeddings of the GUT groups}

We begin by studying the initially irregular embeddings of $SU(5)_k$,
$SU(6)_k$,
and $SO(10)_k$ into an L1SL.
We shall first focus on the cases in which such embeddings are maximal,
and then proceed to the non-maximal cases.

\underbar{Maximal embeddings}:~~  We begin with the possibilities provided by
the maximal
Class IA embeddings listed in (\ref{rule1}).
In general, the representations of $SU(5)$ can be described by
the four Dynkin indices $(a_1,a_2,a_3,a_4)$ of their highest weights;
the representation is real if $a_1=a_4$ and $a_2=a_3$, and
complex otherwise.  Thus, we find that
there exist non-trivial irregular embeddings of $SU(5)$ into
only the following groups with rank $\leq 22$:
$SU(10)$, $SU(15)$, and $SO(24)$.  Calculating the affine levels of the
$SU(5)$ subgroup in each case, however, we find that only the embedding
into $SU(10)$ realizes $SU(5)$ with an affine level less than or equal to four:
$SU(5)_3\subset SU(10)_1$.
(The $SU(5)$ levels in the other two cases are seven and five respectively.)
Thus, only the $SU(5)_3\subset SU(10)_1$ embedding is suitable for the purposes
of
our classification.
This is indeed an efficient way of realizing $SU(5)_3$, for it requires only
the rank-nine group $SU(10)_1$ rather than the
rank-12 group $[SU(5)_1]^3$ that would have been required for
the straightforward diagonal embedding.

Likewise, for the $SU(6)$ GUT group, we find that only the embedding
$SU(6)_4\subset SU(15)_1$ satisfies our criteria of having $k_{\rm GUT}\leq 4$
and rank $G \leq 22$.

In the case of the $SO(10)$ GUT group,
recall that the representations of $SO(10)$ are described by five Dynkin labels
$(a_1,a_2,a_3,a_4,a_5)$, with an ordering for which
$a_4$ and $a_5$ correspond to the spinor labels.
In general, these representations are real if $a_4=a_5$, and complex
otherwise.
Hence, we find that (\ref{rule1}) yields only one non-trivial
embedding possibility into a group of rank and central charge less than 22,
namely
$SO(10)\subset SU(16)$.  Comparing the indices of these representations,
we then find that the level of the $SO(10)$ factor in this embedding is $k=4$.
Thus $SO(10)_4\subset SU(16)_1$.  Note that such an embedding is particularly
useful
if we wish to realize $SO(10)_4$, especially since the
diagonal method would have required the rank-20 group
$[SO(10)_1]^4$.

This concludes the possibilities that arise from Class IA embeddings.
We now examine the possible embeddings from the other classes.
 {\it A priori}\/, it is possible that some of the Class IA special cases
listed
in (\ref{exceptions}) can yield embeddings for our GUT groups $G_{\rm GUT}$ at
at affine levels $2\leq k\leq 4$.
The special cases involving $SU(5)$ pertain to
its $\rep{45}$ and $\rep{105}$ representations;
likewise,
the special cases involving $SU(6)$ pertain
to its $\rep{105}$ and $\rep{210}$ representations,
and the special case involving $SO(10)$ pertains
to its $\rep{560}$ representation.
Comparing the indices of these representations, however, we find that
these special cases merely reproduce the above Class IA embeddings.
Thus, we do not obtain any new embeddings from the Class IA
special cases.

Turning to Class IB embeddings,
however, we see that while no new Class IB possibility exists for realizing
$SU(5)_k$ or $SU(6)_k$ with $k>1$, we now have the single new possible Class IB
embedding for $SO(10)_2$:  $SO(10)_2\subset SU(10)_1$.

Finally, as we have already discussed, the Class IIA embeddings are equivalent
to the diagonal embeddings, and the Class IIB embeddings do not embed any of
our
GUT groups of interest.
Likewise, scanning the Class III embeddings, we see
that in no case are the subgroups of sufficient rank to be of interest to us.

We thus conclude that the following is the complete list of maximal initially
irregular embeddings which are of interest to us:
\beqn
          \#1:&~~~~SU(5)_3 ~&\subset~ SU(10)_1\nonumber\\
          \#2:&~~~~SU(6)_4 ~&\subset~ SU(15)_1\nonumber\\
          \#3:&~~~~SO(10)_2 ~&\subset~ SU(10)_1\nonumber\\
          \#4:&~~~~SO(10)_4 ~&\subset~ SU(16)_1~.
\label{beginlist}
\eeqn
Note that we do not list the corresponding diagonal embeddings,
which in principle are also possible.
We have numbered the embeddings in (\ref{beginlist}) for future ease of
reference.

\underbar{Non-maximal embeddings}:~~
We now consider the {\it non-maximal}\/ initially irregular embeddings.
Note that if we are to embed $SU(5)_k$, $SU(6)_k$, or $SO(10)_k$
via a non-maximal initially irregular embedding into an L1SL,
then for the $k=2$ and $k=3$ cases,
the initial maximal irregular embedding
in the series must embed the GUT group into a group $G^\ast$ which
is necessarily {\it non}\/-simply laced,
and realized either at level $k$ or at level one.
This is because any embedding into a simply laced group at level $k$
would be a regular embedding (which will be considered later, as
an ``initially regular'' embedding);  likewise, any such embedding into
a simply laced group at level one would have already been sufficient
for our purposes, and would therefore have already been
classified above as a maximal ``initially irregular'' embedding.
A quick examination of the possible maximal irregular embeddings then
shows that there do not exist any
maximal irregular embeddings of $(G_{\rm GUT})_{k=2,3}$ into
non-simply laced groups.

For the $k=4$ case, however, the situation is more
complicated:  the first maximal irregular embedding
can be into a non-simply laced group at level $1$ or $4$,
or into {\it any}\/ group (simply or non-simply laced) at level $2$.
Indeed, such possibilities for irregular embeddings into a group at
a level not
equal to $1$ or $k$ arise whenever $k$ itself is not prime, and represent
the primary complication for the $k=4$ case.
Now, as above, it is simple to check that there do not exist any maximal
irregular
embeddings of $(G_{\rm GUT})_{4}$ into
non-simply laced groups.
Thus we are left with the single remaining option,
an embedding into a level-two simply laced group.
This ultimately yields several possibilities, since
$SO(10)_4\subset SU(10)_2$
(in addition to the diagonal embeddings
$SU(5)_4\subset SU(5)_2\times SU(5)_2$ and
$SO(10)_4\subset SO(10)_2\times SO(10)_2$).
However, $SO(10)_2\subset SU(10)_1$
(in addition to the diagonal embeddings
$SU(5)_2\subset SU(5)_1\times SU(5)_1$ and
$SU(10)_2 \subset SU(10)_1\times SU(10)_1$).
Thus, combining these embeddings, we find that we obtain the
following two new {\it non}\/-diagonal irregular embeddings:
\beqn
          \#5:&~~~~ SO(10)_4 ~&\subset~ SU(10)_1 \times SU(10)_1 \nonumber\\
          \#6:&~~~~ SO(10)_4 ~&\subset~ SU(10)_1 \times SO(10)_1 \times
SO(10)_1 ~.
\label{pseudodiag}
\eeqn
These two embeddings are marginally more efficient than the purely
diagonal embedding $SO(10)_4\subset [SO(10)_1]^4$.
No other non-diagonal non-maximal initially irregular embeddings
in this $k=4$ case exist.

\subsubsection{\it Initially regular embeddings of the GUT groups}

\underbar{Long-root embeddings}:~~
Next we examine the cases in which our GUT groups $(G_{\rm GUT})_k$
are realized through {\it regular}\/ embeddings
into level-$k$ simply laced or non-simply laced groups $G^\ast$.
As discussed above, these are the ``initially regular'' embeddings that
employ initial embeddings through the long roots.
For the sake of generality and clarity of presentation,
at this stage of the analysis
we shall ignore the question of whether our particular GUT groups of
interest can actually be realized as regular subgroups of $G^\ast$,
and instead we shall simply classify the possible
embeddings that follow from such groups $G^\ast$ themselves.
We will then discuss
the regular embeddings $(G_{\rm GUT})_k\subset G^\ast_k$ at the end.

In the $k=2$ case, the possible groups $G^\ast$ are:
$SU(n)_2$, $SO(n)_2$, $Sp(2n)_2$,
$(E_6)_2$, $(E_7)_2$, and $(E_8)_2$.
We therefore consider each of these possibilities in turn, testing whether
each such possible group $G^\ast_2$
can then be embedded
through an initially irregular (maximal or non-maximal) embedding into an L1SL.
For the exceptional groups, there exist no
subsequent initially irregular embeddings other than the diagonal
embeddings.  Since a regular embedding followed by a diagonal irregular
embedding is equivalent to a diagonal irregular embedding followed by
a regular embedding,  we disregard these cases.
Likewise, for the $SU(n)_2$ and $Sp(2n)_2$ groups with ranks $\geq 5$,
there also exist no non-diagonal embeddings.
Thus we are left with the single potential embedding pattern that
employs the $G^\ast=SO(n)_2$ option:
\beq
          \#7:~~~~ (G_{\rm GUT})_2 ~\subset~ SO(n)_2 ~\subset~ SU(n)_1~.
\eeq
In the $k=3$ case, none of the potential level-three groups $G^\ast$
can be realized by non-diagonal embeddings into an L1SL.
Finally, in the $k=4$ case, by scanning the previous cases we
see that there are
several groups $G^\ast$ for which subsequent non-diagonal embeddings exist:
\beqn
          \#8:&~~~~ (G_{\rm GUT})_4 ~&\subset~ SU(6)_4 ~\subset~ SU(15)_1
\nonumber\\
          \#9:&~~~~ (G_{\rm GUT})_4 ~&\subset~ SO(10)_4 ~\subset~ SU(16)_1
\nonumber\\
          \#10:&~~~~ (G_{\rm GUT})_4 ~&\subset~ SO(10)_4 ~\subset~
SU(10)_1\times SU(10)_1 \nonumber\\
          \#11:&~~~~ (G_{\rm GUT})_4 ~&\subset~ SO(10)_4 ~\subset~
SU(10)_1\times SO(10)_1\times SO(10)_1~.
\eeqn

This then exhausts all possibilities for long-root initially regular
embeddings.

\underbar{Short-root embeddings}:~~
Finally, we consider the case of short-root initially regular embeddings.
This case amounts to finding regular embeddings of $(G_{\rm GUT})_k$
into non-simply laced groups $G^\ast$ at levels $k/2$
in such a way that $G_{\rm GUT}$ is realized from the short roots of $G^\ast$.
We then seek to realize $G^\ast$ via initially irregular embeddings into an
L1SL.
Of course,
for $k=3$ (and more generally, for all cases when $k$ is odd),
it is clear that there are no possibilities for such groups $G^\ast$.
This is because all non-simply laced groups of rank $\geq 5$
have short roots which are reduced in length relative to the long roots
by a factor of $\sqrt{2}$, implying that any short-root regular subgroup
must be realized at a level which is twice the level of the original group.
Therefore, we shall only need to focus on the $k=2$ and $k=4$ cases when
considering short-root initially regular embeddings.

For $k=2$, the only possible groups $G^\ast$ are $SO(2n+1)_1$ and $Sp(2n)_1$.
Each of these, of course, has a subsequent non-trivial irregular embedding
into an L1SL.  We thus have the potential new short-root GUT embedding
patterns:
\beqn
          \#12:&~~~~ (G_{\rm GUT})_2 ~&\subset~ SO(2n+1)_1 ~\subset~ SO(2n+2)_1
\nonumber\\
          \#13:&~~~~ (G_{\rm GUT})_2 ~&\subset~ Sp(2n)_1 ~\subset~ SU(2n)_1 ~.
\eeqn
For $k=4$, the only possible groups $G^\ast$ are $SO(2n+1)_2$ and $Sp(2n)_2$.
In this case, we then have a variety of subsequent potential irregular
embedding
patterns at our disposal:
\beqn
          \#14:&~~~~ (G_{\rm GUT})_4 ~&\subset~ SO(2n+1)_2
                         ~\subset~ SO(2n+2)_2
                         ~\subset~ [SO(2n+2)_1]^2 \nonumber\\
          \#14':&~~~~ (G_{\rm GUT})_4 ~&\subset~ SO(2n+1)_2
                         ~\subset~ [SO(2n+1)_1]^2
                         ~\subset~ [SO(2n+2)_1]^2 \nonumber\\
          \#15:&~~~~ (G_{\rm GUT})_4 ~&\subset~ SO(2n+1)_2
                         ~\subset~ SO(2n+2)_2
                         ~\subset~ SU(2n+2)_1 \nonumber\\
          \#16:&~~~~ (G_{\rm GUT})_4 ~&\subset~ Sp(2n)_2
                         ~\subset~ SU(2n)_2
                         ~\subset~ [SU(2n)_1]^2~.
\eeqn
This exhausts all possibilities.

\subsubsection{\it Simplifications, Redundancies, and Impossibilities}

We have thus far generated 16 embedding patterns
which represent an exhaustive classification of all of the methods
by which a general GUT group can be realized at levels $2\leq k\leq 4$
through irregular embeddings into level-one simply laced groups (L1SL's).
However, not all of these 16 patterns are independent of each other, and
moreover several of them are ultimately equivalent to the diagonal embedding.
Furthermore, many of these embedding patterns cannot be realized for the
specific GUT group choices for $G_{\rm GUT}$ that we have in mind, namely
$G_{\rm GUT}=SU(5), SU(6)$, and $SO(10)$.
In this section we shall eliminate all of these redundant and unrealizable
possibilities, and obtain our final complete list of independent, non-diagonal
embeddings for these GUT groups of interest.

It is clear that embeddings \#1 through \#4 are independent of each other,
since they each satisfy different branching rules;  moreover, it is also
clear (by comparing, {\it e.g.}, the ranks of the groups on both sides of
the embeddings) that they cannot be equivalent to the diagonal embeddings.
Thus we retain each of these embeddings.  These are, in some sense,
the ``core'' embeddings that we have used to generate the others.

Similarly, embeddings \#5 and \#6 are independent of each other and
distinct from the diagonal;  we therefore retain these as well.

We now turn to embedding pattern \#7.
Unfortunately, for the GUT groups of interest,
this pattern reduces either to the diagonal embedding or to previous cases.
To see this, let us consider this embedding for the case
$G_{\rm GUT}=SU(5)$,
 {\it i.e.},   $SU(5)_2\subset SO(n)_2 \subset SU(n)_1$, with $n\geq 10$.
It is easy to verify that the $n=10$ case amounts to the diagonal embedding
$SU(5)_2\subset SU(5)_1 \times SU(5)_1$,
supplemented with the final regular embedding
$SU(5)_1 \times SU(5)_1\subset SU(10)$.
This $n=10$ case is therefore essentially equivalent to the diagonal embedding,
and the cases with $n\geq 11$ amount merely to enlarging the final
regular embedding.  A similar conclusion holds for $G_{\rm GUT}=SU(6)$.
We therefore reject embedding pattern \#7
for the cases  $G_{\rm GUT}=SU(5)$
and $SU(6)$.
By contrast,
for the case of $SO(10)$, embedding pattern \#7 gives us
$SO(10)_2\subset SO(n)_2\subset SU(n)_1$ for $n\geq 10$.
For $n=10$, of course, this reduces to the case of embedding \#3,
and for $n\geq 11$ this amounts to embedding \#3 supplemented with
final regular embeddings of $SU(10)_1$ into $SU(n)_1$.
Thus we reject embedding pattern \#7 in all cases.

Turning now to embedding pattern \#8, we see that there is only
one GUT group of interest for which this pattern is non-trivial:
$G_{\rm GUT}=SU(5)$.  We therefore retain this pattern in this case only.
The same conclusion holds for embedding pattern \#9.
For embedding patterns \#10 and \#11, we see that there is likewise
only one GUT group, namely $G_{\rm GUT}=SU(5)$, for which these
embeddings are non-trivial.  However, in these cases, we then obtain
embeddings which are ultimately equivalent to the diagonal embeddings,
supplemented with final regular embeddings.  We therefore obtain no
new non-trivial embeddings from patterns \#10 and \#11.

We now focus on the short-root embedding patterns \#12 through \#16.
In these cases, the primary concern is whether the initial
embeddings $(G_{\rm GUT})_k\subset (G^\ast)_{k/2}$ can actually
be realized via short-root regular embeddings in the cases of interest.
Towards this end, the following observations are useful.
First, recall that in each of these cases, $G^\ast$ is non-simply laced, and
is either $SO(2n+1)$ or $Sp(2n)$.
Next, recall that in general, $SO(2n+1)$ does not have any short-root
regular subgroups except $SU(2)$.
This is most easily seen by deleting simple roots from the extended Dynkin
diagrams
of $SO(2n+1)$.  This can also be anticipated from the fact
that although both $SO(2n+1)$ and $Sp(2n)$ have
$n(2n+1)$ non-zero roots, only $2n$ of these
roots are short for $SO(2n+1)$, whereas $2n(n-1)$ of these roots are short
for $Sp(2n)$.  Thus we immediately eliminate embeddings \#12, \#14, and \#15
for each of the GUT groups of interest.
Moreover, focusing our attention on the remaining embedding patterns with
$G^\ast=Sp(2n)$,
we recall that the only regular short-root subgroups of $Sp(2n)$ are $SU(n')$,
$n'\leq n$.
This is once again apparent by deleting roots from the extended Dynkin diagram
of $Sp(2n)$.  Thus, we conclude that $SO(10)$ can never be realized as a
short-root regular subgroup, and we need only focus on the cases with $G_{\rm
GUT}=SU(5)$
or $SU(6)$.
In such cases, embedding pattern \#13 then becomes
$SU(5)_2\subset Sp(2n)_1\subset SU(2n)_1$ for $n\geq 5$, and
$SU(6)_2\subset Sp(2n)_1\subset SU(2n)_1$ for $n\geq 6$, but
in each case these embeddings then reduce
to the embeddings \#7 (which we have already determined to be equivalent
to the diagonal embeddings supplemented with final regular embeddings).
We therefore eliminate embedding pattern \#13.
The same equivalences arise for embedding patterns \#16.

We therefore find that only the following embeddings are achievable,
and distinct
from the diagonal:  \#1 through \#6, \#8, and \#9.

\subsubsection{\it More Impossibilities:  $SO(10)$ at levels $k>4$, and $E_6$
at levels $k>3$}

We now consider two further classifications:
those of $SO(10)$ at levels $k=5,6,7$,
and those of the GUT group $E_6$ at all levels.
Note, {\it a priori}\/,
that $SO(10)$ can be realized
only at levels $k\leq 7$ (since its central charge exceeds
22 for higher levels);  similarly,
$E_6$ can be realized only at levels $k\leq 4$.
In this section, however, we shall show
that $SO(10)$ can in fact be realized only at levels $k\leq 4$,
and that $E_6$ can
be realized only at levels $k\leq 3$.
These stronger bounds arise because of the absence
of suitable {\it embeddings}\/ at these higher affine levels.

First,
we now classify the embeddings of $SO(10)_k$ for $k=5,6,7$,
extending the classification of the $2\leq k \leq 4$ cases that
was performed in the previous subsection.
Note, first of all,
that no diagonal embeddings for these cases are possible,
for they would require tensor-product groups of rank exceeding 22.
Thus, if there are to be {\it any}\/ embeddings for $SO(10)_k$ at levels
$k=5,6,7$, they must be non-diagonal.
To determine if non-diagonal embeddings exist, we follow the same procedures as
before.
For $k=5,6,7$, we find that there are no maximal initially
irregular embeddings of $SO(10)_k$;  similarly, for $k=5,7$,
there are no non-maximal initially irregular embeddings.
Turning to the initially regular embeddings, we note that $SO(10)$
can never be realized as a short-root subgroup;  we therefore
investigate only the long-root embeddings.  The only possible
intermediate groups $G^\ast$, each realized at level $k$,  are $SO(n)$ for
$n\geq 11$,
$E_6$, $E_7$, and $E_8$.
The exceptional groups can be disregarded
because their only subsequent embeddings are through the diagonal.
These are ultimately equivalent to the diagonal embedding
for $SO(10)_k$ itself, and are therefore ruled out.
We therefore focus on the cases with $G^\ast= SO(n)$ for $n\geq 11$.
For levels $k=5$ and $7$, there are no subsequent embeddings
that realize such $G^\ast$ subgroups.
Thus, we conclude that for $k=5$ and $7$, there are no embeddings
for $SO(10)$, either diagonal or non-diagonal.

For $k=6$, the only
remaining cases that we have not yet ruled out are
the non-maximal initially irregular embeddings
and the long-root initially regular embeddings.
As explained above,
the $k=6$ case is more complex because $k$
is not prime.
Thus, when examining the non-maximal initially irregular
embeddings, we must consider the additional possibilities
in which $SO(10)_6$ is first irregularly embedded
into groups $G^\ast$ at levels two or three.
However, such embeddings
are equivalent to the embeddings of $SO(10)$
at levels $k=2,3$ into level-one groups $G^\ast$, and we have
already classified the embeddings of $SO(10)_{2,3}$.
Recall that we found that $SO(10)_3$ can only be diagonally embedded
into $[SO(10)_1]^3$,
while $SO(10)_2$ can be diagonally embedded into $[SO(10)_1]^2$
or non-diagonally embedded into $SU(10)_1$.
Thus, for $SO(10)_6$, we see that there are only three possibilities for
$G^\ast$:
$[SO(10)_3]^2 $,
$[SO(10)_2]^3 $, and $SU(10)_3$.
The first and second choices for $G^\ast$, however, have no subsequent
non-diagonal embeddings
into groups of rank $\leq 22$,
and the third choice, $SU(10)_3$, has a central charge which already exceeds
22.
Thus, we rule out all of these cases.
Finally, we consider the long-root initially regular embeddings for the
$k=6$ case.
As before, there are no embeddings with $G^\ast$ realized at levels one or six;
likewise, all cases for which $G^\ast$ is one of the exceptional groups are
ruled out.
We therefore focus on cases with $G^\ast=SO(n)$, $n\geq 11$, realized at levels
two or three.
There is, however, only one possible initially irregular embedding
of such a group,  namely $SO(n)_2\subset SU(n)_1$, $n\geq 11$.
This leads to the overall embedding
$SO(10)_6\subset SO(n\geq 11)_6\subset SU(n\geq 11)_3$.
Such groups $SU(n\geq 11)_3$ have central charges exceeding $22$,
however, and hence cannot be realized in string theory.

Thus, we have shown that there exist {\it no embeddings for $SO(10)_k$ for
levels $k=5,6,7$}\/ into groups of rank $\leq 22$.
This includes both diagonal and non-diagonal embeddings.
This implies that it is {\it impossible to realize $SO(10)_{5,6,7}$
in free-field string theory}, which in turn implies, for example,
that it is {\it impossible to realize the $\rep{126}$
representation of $SO(10)$ in free-field heterotic string models}.
Note that this result
is wholly independent of the specific free-field string construction employed.
In particular, this conclusion holds not just within, {\it e.g.}, the
free-fermionic
construction (for which a similar conclusion has already been established
\cite{shygut}), but also for all orbifold constructions as well
(regardless of whether they are symmetric or asymmetric, or whether they are
based on abelian or non-abelian groups).

Finally, we discuss the embeddings of the GUT group $E_6$
at levels $k=2,3,4$.
We begin by noting that no maximal Class IA irregular embeddings exist
for {\it any}\/ of the exceptional $E_{6,7,8}$ groups when realized
at levels $k\leq 5$;  moreover,
the embeddings that realize these groups at higher levels involve
groups of rank $\geq 22$.
Thus, the diagonal embeddings are the only initially irregular
embeddings that can realize these groups at higher levels.
Furthermore,
since the exceptional groups cannot be regularly embedded
into any classical groups, this also exhausts the possibilities
for initially regular embeddings.
Thus, we conclude that the only possible ways of
generating the exceptional gauge symmetries
$E_6$, $E_7$, and $E_8$ at higher levels are through diagonal embeddings.
This implies that $E_6$ and $E_7$ cannot be realized
at levels $k\geq 4$, and likewise $E_8$ can be realized only at levels $k=1,2$.
Of course, for string GUT purposes,
we are only interested in realizing $E_6$ (since $E_7$ and $E_8$ do not
have complex representations);  moreover,
it is important to realize $E_6$ at levels $k\geq 2$, since it is only at
such higher levels
that the potentially important $\rep{78}$ representation can appear in
the massless spectrum.
Thus, we see that if we wish to realize the $\rep{78}$ representation of $E_6$,
the diagonal embedding for $E_6$ is the only possibility.
Likewise, we also see that free-field string models can {\it never}\/ give rise
to
any massless representations of $E_6$ other than the $\rep{27}$ and $\rep{78}$
representations, for the highest affine level
at which $E_6$ can ever be realized is $k=3$.

\subsection{Summary:  List of embeddings for string GUT groups}

We now summarize the results of the previous subsection.

In the previous subsection, we
performed a complete classification of {\it all}\/
possible embeddings that can give rise to $SU(5)_k$, $SU(6)_k$, $SO(10)_k$,
and $(E_6)_k$ for the cases $2\leq k\leq 4$.
In the case of $SO(10)$, we were also able to extend these results
to cover {\it all}\/ of the potentially realizable levels,
 {\it i.e.}, $k\leq 7$.
In the table below, we have collected
all of those embeddings
which do not reduce to the trivial diagonal embeddings.
Note that, as required, in each case we have embedded the GUT group
$G'$ at level $k$ into a level-one simply laced group $G$;  for non-maximal
embeddings we have also listed the relevant intermediate group $G^\ast$.
Also note that in this table we have listed only those embeddings
which are {\it distinct}\/ ({\it i.e.}, which have different branching rules).

\beq
\begin{tabular}{rcl|c|c|c|c}
 $G'_{k} $ & $\subset$ &  $G$ & $\xi$ & Maximal? & $G^\ast$ & $\Delta c$\\
\hline
\hline
  $SU(5)_2$  & &  Diagonal only &  &  &  & $8/7$ \\
  $SU(5)_3$  & & $SU(10)_1$ & $+3$ & Yes & --- & $0$ \\
  $SU(5)_4$  & &  $SU(15)_1$ & $+2$ & No & $SU(6)_4$ & $10/3$ \\
  $SU(5)_4$  & &  $SU(16)_1$ & $+1$ & No &  $SO(10)_4$ & $13/3$ \\
\hline
  $ SU(6)_2 $  & &  Diagonal only &  &  &  & $5/4$ \\
  $ SU(6)_3 $  & &  Diagonal only &  &  &  & $10/3$ \\
  $ SU(6)_4 $ & & $SU(15)_1 $  & $+6$ & Yes & --- & $0$  \\
\hline
  $SO(10)_2$  & &   $SU(10)_1$  &  $+1$ & Yes &  --- & $0$ \\
  $ SO(10)_3 $  & &  Diagonal only &  &  &  & $30/11$ \\
  $SO(10)_4$  & &  $SU(16)_1$ & $+5$ & Yes & --- & $0$ \\
  $SO(10)_4$ & & $[SU(10)_1]^2 $ & $+2$ & No &  $[SO(10)_2]^2$ or $SU(10)_2$ &
$3$  \\
  $SO(10)_4$ & & $SU(10)_1 \times [SO(10)_1]^2$ & $+1$ & No &
          $[SO(10)_2]^2$ & $4$ \\
  $ SO(10)_{k>4} $  & &  Impossible &  &  &  \\
\hline
  $ (E_6)_{2} $  & &  Diagonal only &  &  &  & $6/7$\\
  $ (E_6)_{3} $  & &  Diagonal only &  &  &  & $12/5$ \\
  $ (E_6)_{k> 3} $  & &  Impossible &  &  &  \\
\hline
\end{tabular}
\eeq

In the above table, we have defined the quantity $\xi$ as
\beq
       \xi ~\equiv ~  k~{\rm rank}\, G' ~-~ {\rm rank}\, G~.
\eeq
Thus $\xi$ is a measure of the ``compactness''
of the embedding relative to the diagonal embedding $G'_{k}\subset G_1
\times G_1 \times ...\times G_1$, and is positive for embeddings which
require fewer dimensions of the charge
lattice than would be required by the diagonal embedding.
As can be seen from this table, all of the new embeddings
are more compact than the diagonal embedding, and thus can potentially
be implemented more efficiently in string theory.
Furthermore, for each embedding,
we have also listed the quantity
\beq
        \Delta c ~\equiv~ c(G) ~-~ c(G'_{k})
\eeq
where the central charges are
given in (\ref{centralcharge}).
In general, embeddings which are more compact
give rise to smaller values of $\Delta c$.  Embeddings with $\Delta c=0$ are
often called {\it conformal}\/, and have been studied in the
context of the bosonic string \cite{BB,conformalembeddings}.
Thus, for each potential GUT group, this table describes the
most efficient free-field string realization that can possibly exist.

It is an important observation that the majority of these embeddings
have $\Delta c\not= 0$.  For such embeddings, this implies that
the true central-charge ``cost'' of realizing a given GUT group $G'_{k}$
in free-field string constructions
is {\it not}\/ merely the central charge of the GUT group itself,
 {\it but also}\/ the extra cost $\Delta c$
associated with the particular embedding.  Indeed, this extra cost
is the cost of embedding $G'_{k}$ into a level-one simply laced group,
as required in free-field constructions.
Such additional central charges
$\Delta c$ are often overlooked in string GUT model-building analyses,
but have important phenomenological consequences.  For example,
such central charges strictly limit the
sizes of possible accompanying ``hidden'' sector gauge groups.
Indeed, in some cases, we see that this unavoidable extra cost $\Delta c$ can
be
quite substantial, and hence by scanning the above table it is now possible
to determine the {\it minimal}\/ cost $\Delta c$ that must be paid
in order to realize each GUT group at a given level.
Note that it is only because we have done a
complete classification of {\it all}\/ potential GUT-group
free-field embeddings that
such minimum values can now be determined.
This issue will be discussed further in the Conclusions.

Finally, also note from the above table that
$SO(10)$ can be realized only at levels $k\leq 4$,
and that $E_6$ can be realized only at levels $k\leq 3$.
These bounds are stronger than the naive central-charge
constraints (which would have permitted $k\leq 7$ for
$SO(10)$, and $k\leq 4$ for $E_6$), but we found that they
arise because
there are no {\it embeddings}\/ which can realize
such groups and levels.
Thus, for example,
we see that we can never obtain a massless $\rep{126}$ representation
of $SO(10)$ in free-field heterotic string models, nor obtain any $E_6$
representations
other than the $\rep{27}$ or $\rep{78}$ representations.
These results hold for {\it all}\/ free-field string constructions,
and are completely general.

%=====================================================================
\vfill\eject
\setcounter{footnote}{0}
\section{GSO Projections for $SO(10)_2$ and $SU(5)_3$ from $SU(10)_1$}

In the previous section, we classified all methods of realizing
the GUT groups of interest in free-field string models.  We found that
for level-two realizations, the only possible non-diagonal embedding
is $SO(10)_2\subset SU(10)_1$.  Likewise, for level-three realizations,
the only possible non-diagonal embedding is $SU(5)_3\subset SU(10)_1$.
In this section
we shall focus on these two embeddings, and determine which linear combinations
of the nine $SU(10)_1$ Cartan generators must be GSO-projected out of the
spectrum in order to realize either $SO(10)_2$ or $SU(5)_3$.
In so doing,
we will also see explicitly how these embeddings manage to be more efficient
than the diagonal embeddings which have been traditionally employed.

As we have outlined in Sect.~4, the first step in the analysis of
a given embedding $G'_{k'}\subset G$ is the determination of the
corresponding embedding matrix $\calP$,
for this matrix encodes all information concerning the
particular embedding.
In general, such matrices can be fairly complicated to determine,
for one requires independent knowledge of the branching rules for
a number of separate representations of $G$ before the entries
in $\calP$ can be unambiguously determined.
The job can often be made simpler, however,
by making use of various symmetries of the Dynkin diagram
(which ultimately become symmetries of the embedding matrices $\calP$).

Let us begin, however, by considering the simple diagonal embeddings.
In general,
for diagonal embeddings
$G_n\subset G_1\times G_1\times...\times G_1$,
the form that the embedding matrix takes is fairly simple:
\beq
    \calP(G_n\subset G_1\times G_1\times...\times G_1)~=~
    \biggl(~\underbrace{\bone_r ~ \bone_r ~ \bone_r ~
                   \cdots ~\bone_r }_{n~{\rm factors}} ~\biggr)~.
\label{Pdiagonal}
\eeq
Here $\bone_r$ is the $r$-dimensional unit matrix, where $G$ has rank $r$.
Note that the $\calP$-matrix takes this
form for {\it all}\/ groups $G$.
It is easy to check that this matrix, via (\ref{kresult}),
yields the expected subgroup affine level $k=n$.  From
(\ref{Pdiagonal}), we find that the required GSO projections
also take a fairly simple form for diagonal embeddings.
For concreteness, let us consider the diagonal embedding of $SO(10)_2$,
 for which we have
\beq
           \calP~=~
           \pmatrix{
            1 & 0 & 0 & 0 & 0 &     1 & 0 & 0 & 0 & 0 \cr
            0 & 1 & 0 & 0 & 0 &     0 & 1 & 0 & 0 & 0 \cr
            0 & 0 & 1 & 0 & 0 &     0 & 0 & 1 & 0 & 0 \cr
            0 & 0 & 0 & 1 & 0 &     0 & 0 & 0 & 1 & 0 \cr
            0 & 0 & 0 & 0 & 1 &     0 & 0 & 0 & 0 & 1 \cr}~.
\label{Pdiagonalso10}
\eeq
The nullspace of $\calP$ is thus spanned by five
vectors $\vec\beta_i$
with Dynkin indices $a_j=\delta_{ji}-\delta_{j+5,i}$ respectively.
It is then straightforward to
convert these Dynkin indices to the standard Cartesian coordinates
that are appropriate for describing the
embeddings of each $SO(10)$ group factor [as indicated in
(\ref{innerprods})].
If the five Cartesian GSO generators of the first $SO(10)_1$
factor are denoted $U_1$ through $U_5$, and those of the
second factor are denoted
$U_6$ through $U_{10}$, we then find from these vectors $\vec\beta_i$ that
the diagonal subgroup can be realized by GSO-projecting out the following five
generators:\footnote{
  Note that strictly following the procedure outlined in Sect.~4 yields
  linear combinations of the $U_i$ which are more complicated than those listed
in (\ref{Usprojected}).
  However, it is always possible to choose different linear combinations of the
vectors $\vec\beta_i$
  which span the nullspace of ${\cal P}$, and thereby reduce the corresponding
  linear combinations of $U_i$ to the form given in (\ref{Usprojected}).}
\beq
     U_1-U_6~,~~~ U_2-U_7~,~~~ U_3-U_8~,~~~ U_4-U_9~,~~~ U_5-U_{10}~.
\label{Usprojected}
\eeq
Thus, the total rank is reduced by five in this realization.
Of course, this case is somewhat trivial due to the highly symmetric nature
of the diagonal embedding.

By contrast,
let us now turn to the more interesting embedding which also realizes
$SO(10)_2$, namely $SO(10)_2\subset SU(10)_1$.
We have seen in the previous section that this embedding
is actually the $n=10$ case of the general series of embeddings $SO(n)_2\subset
SU(n)_1$.
After some work, we find that the $\calP$-matrix
for this general embedding with even $n$ is given by the following
$(n/2)\times (n-1)$-dimensional matrix:
\beq
\calP ~=~ \pmatrix{
         1 & 0 & \cdots & 0 & 0 & 0 & 0 & 0 & \cdots & 0 & 1 \cr
         0 & 1 & \cdots & 0 & 0 & 0 & 0 & 0 & \cdots & 1 & 0 \cr
         \vdots  & \vdots & \ddots & \vdots  & \vdots  & \vdots  & \vdots &
\vdots
           & {}_\cdot \cdot^\cdot  & \vdots  & \vdots \cr
        0 & 0 & \cdots & 1 & 0 & 0 & 0 & 1 & \cdots & 0 & 0 \cr
        0 & 0 & \cdots & 0 & 1 & 0 & 1 & 0 & \cdots & 0 & 0 \cr
         0 & 0 & \cdots & 0 & 1 & 2 & 1 & 0 & \cdots & 0 & 0 \cr }~.
\eeq
Exchanging the last two rows of this matrix corresponds
to changing the relative chirality of $SO(10)_2$ within $SU(10)$;
in either case, the nullspace of $\calP$ is unaffected,
and the same GSO projections will be produced.
Thus, for the specific case of the
$SO(10)_2\subset SU(10)_1$ embedding, we have
\beq
      \calP ~=~ \pmatrix{
         1 & 0 & 0 & 0 & 0 & 0 & 0 & 0 & 1 \cr
         0 & 1 & 0 & 0 & 0 & 0 & 0 & 1 & 0 \cr
         0 & 0 & 1 & 0 & 0 & 0 & 1 & 0 & 0 \cr
         0 & 0 & 0 & 1 & 0 & 1 & 0 & 0 & 0 \cr
         0 & 0 & 0 & 1 & 2 & 1 & 0 & 0 & 0 \cr }~,
\label{Pnondiagonalso10}
\eeq
which should be contrasted with (\ref{Pdiagonalso10}).
It can be explicitly verified,
inserting (\ref{Pnondiagonalso10}) into (\ref{kresult}), that
the $SO(10)$ subgroup is indeed realized at level $k=2$.
We also find that
the {\it four}\/ vectors $\vec\beta_i$ which span the nullspace of $\calP$ in
this
case have
Dynkin indices $a_j=\delta_{j,i}-\delta_{j,10-i}$.
Thus, at this stage, the results for the $SO(10)_2\subset SU(10)_1$ embedding
seem quite similar to those for the diagonal $SO(10)_2$ embedding.

The conversion of these Dynkin indices to Cartesian coordinates, however,
now depends on the lattice embedding of $SU(10)_1$ rather than
on the embedding of $SO(10)_1\times SO(10)_1$.
Recall from Sect.~4 that the embedding of $SU(10)$ requires
a {\it ten}\/-dimensional lattice, spanned by Cartan generators $U_1$ through
$U_{10}$.
Let us define ${\bf E}\equiv \sum_{i=1}^{10} U_i$, which corresponds
to the $U(1)$ direction that is orthogonal to the $SU(10)$ hyperplane in this
ten-dimensional space.
Following the procedure outlined in Sect.~4, we then find
that the $SO(10)_2\subset SU(10)_1$ embedding is realized
by projecting out any {\it four}\/ of the following five linear combinations
of Cartan generators
\beq
      \hat U_i ~\equiv~ (U_i + U_{11-i})-{\textstyle{1\over 5}}\,{\bf
E}~,~~~~i=1,...,5~.
\label{EU}
\eeq
Note that $\sum_{i=1}^5 \hat U_i=0$, so indeed only four of these
linear combinations are independent.
Also note that
each of these directions $\hat U_i$ is orthogonal
to ${\bf E}$ itself.
Thus, as required, this dimensional truncation
occurs completely within the nine-dimensional $SU(10)$ hyperplane,
so that this embedding truly embeds $SO(10)_2$ into $SU(10)_1$ alone.
Indeed, the orthogonal
tenth direction ${\bf E}$ is unaffected by the dimensional truncation.
In this way, then, we see explicitly how this alternative embedding
for $SO(10)_2$ manages to be more ``compact'' than the diagonal embedding,
and requires the sacrifice of only {\it four}\/ lattice directions.

In fact, the GSO projections (\ref{EU}) that we found
could have been deduced in a rather simple way.
Given the symmetry of
the embedding $\calP$ matrix in (\ref{Pnondiagonalso10})
as well as the symmetry of the matrix $\sum_k G_{jk}(\vec\alpha_k,\hat e_\ell)$
that
describes the conversion from $SU(10)_1$ Dykin indices to Cartan coordinates
as in (\ref{cartesian}),
it is clear that
that we expect our final GSO projections to be symmetric under the interchange
$U_i\leftrightarrow U_{11-i}$.
We also know that our final GSO-projected
linear combinations of the $U_i$ must be orthogonal to ${\bf E}$.
With this information alone, we are immediately led to define
the orthogonal symmetric linear combinations $\hat U_i$ in (\ref{EU}).
Only four of these combinations are independent;  hence a suitable basis
for the GSO projections must consist of any four.
Of course, the $\hat U_i$ are only one basis in which the final result can be
expressed.  For example,
an equivalent alternative set of four GSO projection combinations would be
\beqn
     \hat U_1-\hat U_2 &=& U_1-U_2-U_9+U_{10}~,\nonumber\\
     \hat U_1-\hat U_3 &=& U_1-U_3-U_8+U_{10}~,   \nonumber\\
     \hat U_1-\hat U_4 &=& U_1-U_4-U_7+U_{10}  ~, \nonumber\\
     \hat U_1-\hat U_5 &=& U_1-U_5-U_6+U_{10}  ~.
\label{equivcombos}
\eeqn
We thus see that
in order to realize the $SO(10)_2\subset SU(10)_1$ embedding,
we require GSO projections along only
simple sums and differences of the lattice directions.

We now turn to the $SU(5)_3\subset SU(10)_1$ embedding.
This is one of the other potentially useful GUT-group ``core'' embeddings that
we found in the previous section.
Like the above non-diagonal $SO(10)_1$ embedding,
the starting group here is again
$SU(10)_1$;  thus, the sole difference between this case and the previous
case is the particular set of linear combinations of
$U_i$ which must be GSO-projected out of the spectrum.
As before, we begin the analysis of this embedding
by constructing the corresponding $\calP$-matrix.
In this case, however, we find that there are two
choices for the embedding matrix,
 which we shall denote $\calP^+$ and $\calP^-$.
These two choices ultimately correspond
to the relative chirality of $SU(5)_3$ within $SU(10)_1$.
In other words, it is possible to define the fundamental branching rule
for this embedding to be either
$\rep{10}\to \rep{10}$ (which we shall call the `$+$' chirality embedding)
or $\rep{10}\to \rep{\overline{10}}$ (the `$-$' chirality embedding).
Let us first consider the embedding with `$+$' chirality.
After a detailed derivation of the branching rules that
apply in this case,\footnote{
           Our procedure is as follows.
           Starting with the $\rep{10}\to\rep{10}$ branching rule that defines
             this embedding,  we take repeated tensor products of
            the $\rep{10}$ representation on both sides in order to
              derive the branching rules for the larger representations.
             In this way we find that
              $\rep{45}\to\rep{\overline{45}}$;
              $\rep{120}\to\rep{70}+\rep{50}$;
              $\rep{210}\to\ \rep{\overline{175}} +\rep{\overline{35}}$;
              and
              $\rep{252}\to\rep{126}+\rep{\overline{126}}$.
             Note that in cases of potential ambiguity regarding
               branching rule identifications,
              it often proves useful to compare the indices $\ell_G$ of
              candidate representations.
             Care must also be taken with regards to distinguishing
             representations from their complex conjugates.}
we find the embedding matrix
\beq
    \calP^+~=~
    \pmatrix{ 0 & 1 & 2 & 1 & 2 & 1 & 1 & 0 & 0 \cr
              1 & 0 & 0 & 0 & 0 & 1 & 0 & 1 & 0 \cr
              0 & 1 & 0 & 1 & 1 & 0 & 0 & 0 & 1 \cr
              0 & 0 & 1 & 1 & 0 & 1 & 2 & 1 & 0 \cr  }~.
\label{Pplus}
\eeq
Given this result, it is clear that the embedding with `$-$' chirality
is described by the alternative matrix
\beq
    \calP^-~=~
    \pmatrix{ 0 & 1 & 2 & 1 & 0 & 1 & 1 & 0 & 0 \cr
              1 & 0 & 0 & 0 & 1 & 1 & 0 & 1 & 0 \cr
              0 & 1 & 0 & 1 & 0 & 0 & 0 & 0 & 1 \cr
              0 & 0 & 1 & 1 & 2 & 1 & 2 & 1 & 0 \cr  }~
\label{Pminus}
\eeq
which is the same as $\calP^+$, except turned upside down (which only
affects the middle column, given the symmetries of these matrices for these
groups).

We now
determine the appropriate linear combinations of
Cartan generators $U_i$ that must be GSO projected out of the spectrum
in order to realize these embeddings.
Unlike the previous case, we now expect both symmetric and {\it anti-symmetric}
combinations of the $U_i$ to be involved.
We therefore define, in addition to the symmetric combinations (\ref{EU}),
the anti-symmetric combinations
\beq
        \hat V_i~\equiv~ U_i - U_{11-i}~,~~~~~~ i=1,...,5~.
\eeq
As required, these combinations are already orthogonal to ${\bf E}$.
Following the procedure outlined in Sect.~4,
we then find, after much algebra,
that the $SU(5)_3\subset SU(10)_1$ embeddings
with $\pm $ chiralities respectively
can be realized by GSO-projecting
out of the string spectrum the following {\it five}\/ linear
combinations of Cartan generators:
\beqn
           && \hat U_1~-~\hat U_3~,\nonumber\\
           && \hat U_1~-~\hat U_5~,\nonumber\\
           && \hat U_2~-~\hat U_4 ~\mp~\hat V_5~,  \nonumber\\
           && \hat V_1~-~\hat V_2 ~-~\hat V_4 ~,\nonumber\\
           && \hat V_2~-~\hat V_3 ~-~\hat V_4 ~.
\label{mixed}
\eeqn
As before, each of these linear combinations amounts to simple
sums and differences of lattice directions $U_i$.
We thus see that this embedding
is clearly much more ``compact'' than the diagonal embedding
for $SU(5)_3$
(which would have required that we start
with the rank-{\it twelve}\/ group
$[SU(5)_1]^3$, and then delete {\it eight}\/ Cartan generators).

Thus, starting from the level-one simply laced group
$SU(10)_1$,
we immediately see what distinguishes
the realization of $SO(10)_2$ from that of $SU(5)_3$.
Both constructions begin with projections along the symmetric
$\hat U_1-\hat U_3$ and $\hat U_1-\hat U_5$ lattice directions.
The projections leading to $SO(10)_2$
then continue along the remaining purely symmetric $\hat U_1-\hat U_2$
and $\hat U_1-\hat U_4$ lattice directions, as in (\ref{equivcombos}),
while those leading to $SU(5)_3$ instead follow the anti-symmetric
and mixed directions listed in (\ref{mixed}).

%=====================================================================
\vfill\eject
\setcounter{footnote}{0}
\section{Conclusions and Discussion}

In this paper, we have analyzed the
general methods by which higher-level and/or non-simply laced gauge
symmetries can be realized in free-field heterotic string models.
We found that all such realizations have a common underlying
feature, namely a dimensional truncation of the associated
charge lattice, and this in turn enabled us to obtain a number
of results which have direct bearing on the construction of
such models.

In particular, our main results are as follows:
\begin{itemize}
\item	First, we showed that conformal invariance, together with
        the masslessness constraint for gauge-boson states,
        imply that the only way to realize higher-level
	or non-simply laced gauge symmetries in free-field heterotic string models is
        to start with a level-one simply laced (`L1SL') gauge symmetry, and
        then
	to perform a {\it dimensional truncation}\/ of the charge lattice
	via GSO projection.  This is explained in detail in Sect.~3.
	This is a general result, true for {\it all}\/ free-field heterotic string
	constructions.

\item	Next, we showed that such dimensional truncations correspond uniquely
        to {\it irregular embeddings}\/ of higher-level and/or non-simply laced
        gauge groups within level-one simply laced gauge groups, and that for
        each irregular embedding there exists a corresponding dimensional
        truncation.

\item   This identification then allowed us to derive, in a model-independent
        way,
        the specific GSO constraints that are necessary in order to realize any
        given
        subgroup embedding in free-field string theory.
        This is discussed in Sect.~4.

\item	These general results then enabled us to classify,
        in Sect.~7, {\it all}\/ possible
	embeddings that can give rise to the GUT groups $SU(5)$, $SU(6)$,
	$SO(10)$, and $E_6$ at levels $k=2,3,4$ in free-field string models.
	We found that there exist interesting {\it alternative embeddings}\/
        by which such higher-level groups can be realized.
        These embeddings, in particular, go beyond the commonly employed
diagonal
	embedding $G_k\subset [G_1]^k$,
        and are always more compact and require less central charge than the
        diagonal embeddings.  The results of our
	classification are summarized in Sect.~7.4, while
        Sect.~8 contains
	some examples of the particular GSO projections that are necessary
        in order to realize these alternative embeddings.

\item	We showed that in free-field heterotic string constructions
        it is {\it impossible}\/ to realize $SO(10)$ at levels
        $k>4$, or to realize $E_6$ at levels $k>3$.
        These results are obtained in Sect.~7.3.5, as an extension of our
classification
         for the GUT groups $SO(10)$ and $E_6$.
        Although the simple central-charge constraints for these groups would
        have allowed $SO(10)$ and $E_6$ to be realized up to levels $k=7$ and
$k=4$ respectively,
        we found that $SO(10)_{k>4}$ and $(E_6)_{k>3}$ nevertheless cannot be
realized.
        These stronger constraints ultimately arise due to the lack of suitable
        {\it embeddings}\/ for such groups and levels.

\item	One consequence of the impossibility of realizing $SO(10)_{k>4}$ and
	$(E_6)_{k>3}$ is that there are now significant {\it new constraints
	on the massless matter spectrum}\/ of any effective GUT theory that can
        be derived from a free-field heterotic string model.  For example, we
find
       that the interesting
	$\rep{126}$ representation of $SO(10)$ is {\it disallowed}\/ as
	a massless state in string theory.
        Analogous results hold for $E_6$.

\item	Finally, because we were able to classify {\it all}\/ possible
        realizations
	of the GUT groups $SU(5)$, $SU(6)$, $SO(10)$, and $E_6$ for levels $k=2,3,4$,
	we were also able to
	determine the minimum extra central-charge ``cost''
	$\Delta c$ that each of these groups necessarily requires for its
	realization in free-field string constructions.
        These values of $\Delta c$ are given in the table in Sect.~7.4.
\end{itemize}

\begin{figure}[htb]
\centerline{\epsfxsize 5.5 truein \epsfbox {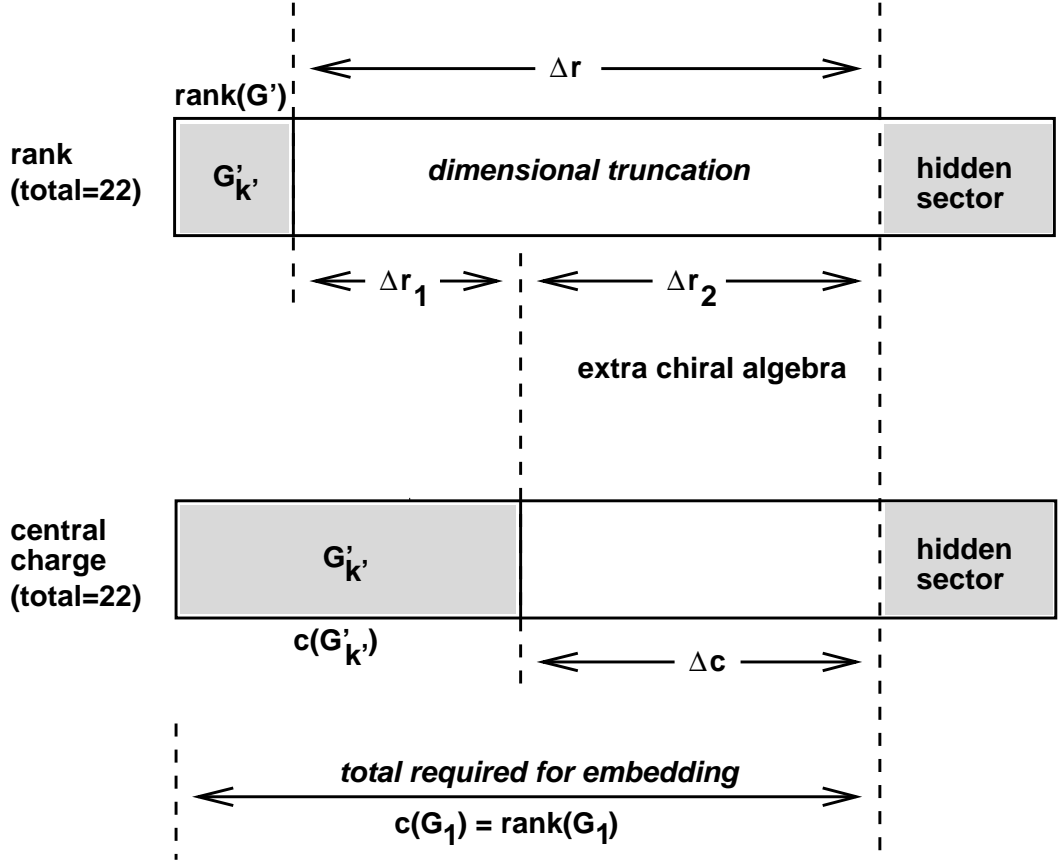}}
\caption{Distribution of rank and central charge for the general embedding
of $G'_{k'}$ within a level-one simply laced group $G_1$.
Here $\Delta r$ and $\Delta c$ are the total rank and central-charge ``costs''
of realizing the embedding.  The degrees of freedom represented by
$\Delta r_2=\Delta c$, in particular, form a (usually unavoidable)
independent extra chiral algebra
which does not contribute to the gauge symmetries of the theory.}
\label{rankchargefig}
\end{figure}

This last issue, concerning the central-charge ``cost'' $\Delta c$,
deserves further comment.
Given a particular embedding $G'_{k'}\subset G_k$, there
are, {\it a priori}\/, two distinct measures of the
compactness or efficiency of the realization:
\beqn
         \Delta c &\equiv& c(G_k)~-~ c(G'_{k'})\nonumber\\
         \Delta r &\equiv& {\rm rank}\,G~-~ {\rm rank}\, G'~.
\eeqn
The first corresponds to the additional {\it central charge}\/ that is required
in
order to achieve the particular embedding, while the second
corresponds to the additional {\it lattice directions}\/ that are required.
These quantities $\Delta c$ and $\Delta r$ are therefore different measures
of the compactness of the embedding.
However, as we have seen above,
in free-field string models
the group $G_k$ must ultimately be simply laced, and realized at level one.
For such groups $G_1$, the central charge and rank are the same.
Therefore, we always have
that ${\rm rank}\,G'+\Delta r= c(G'_{k'})+\Delta c$.
This situation is illustrated in Fig.~\ref{rankchargefig}.

%================== FIGURE INSERTED HERE ==============================
%   If you do not wish to have the figure inserted, just comment
%   out the following lines:
\input epsf
%================== END OF INSERTED FIGURE ============================

Since the subgroup $G'_{k'}$ is either non-simply laced and/or realized
at higher level,
we find that ${\rm rank}\,G'\not= c(G'_{k'})$.
This in turn implies that $\Delta c\not =\Delta r$,
and it is important to understand the consequences of this difference.
As illustrated in Fig.~\ref{rankchargefig},
we may split the total rank-reduction $\Delta r$
into two pieces, $\Delta r_1$ and $\Delta r_2$,
where $\Delta r_1$ is defined as $c(G'_{k'})-{\rm rank}\,G'$.
Note that $\Delta r_1$ and $\Delta r_2$ need not be integers;  indeed,
only their sum is an integer.
The distinction between $\Delta r_1$ and $\Delta r_2$ is meaningful because
these two
contributions ultimately sit within different conformal field theories.
In particular, the contributions $({\rm rank}\, G' + \Delta r_1)$
together comprise the complete affine Lie algebra
corresponding to $G'_{k'}$.  Although the contribution $\Delta r_1$ has
(along with $\Delta r_2$)
been dimensionally truncated out of the charge lattice,
excitations of the degrees of freedom corresponding to $\Delta r_1$ are
nevertheless
part of the affine Lie algebra $G'_{k'}$.
Indeed, as we have seen in Sect.~3,
such excitations into truncated lattice directions are
necessary in order to keep the conformal dimensions
of the gauge-boson vertex operators equal to one.

By contrast, the remaining contribution $\Delta r_2$ is {\it not}\/ part of
the conformal field theory of the affine Lie algebra $G'_{k'}$,
and instead comprises a {\it completely separate chiral algebra}.
This chiral algebra does not form
an affine Lie algebra of its own, nor is it a component of any other
affine Lie algebra.
However, its presence is necessary {\it in string theory}\/
because of the restricted nature of the
embeddings that can be employed in order to {\it realize}\/ the affine Lie
algebra  $G'_{k'}$.
As such, this ``unwanted'' chiral algebra is an unavoidable cost of
realizing a higher-level and/or non-simply laced gauge symmetry
in string theory.

It useful to understand how this distinction between
$\Delta r_1$ and $\Delta r_2$ arises in actual string constructions.
For this purpose, let us recall the ten-fermion example of
Sect.~6.2, and focus on the realization of the $SU(2)_2$ group factor in that
model.
Recall that, as far as the $SU(2)_2$ group factor was concerned,
there were originally four degrees
of freedom prior to the dimensional truncation:
these were represented by the four real fermions $\lambda_{1,2,3,4}$,
and gave rise to the gauge group $SO(4)_1= SU(2)_1\times SU(2)_1$.
By contrast, after the dimensional truncation,
only the $(\lambda_1,\lambda_2)$
fermions combine to form the surviving lattice direction
corresponding to $SU(2)_2$.
Indeed, the lattice direction that had
corresponded to $(\lambda_3, \lambda_4)$ is dimensionally truncated
in this diagonal embedding.
Thus, $(\lambda_3,\lambda_4)$ together correspond to the total $\Delta r$.
However, as can be seen from the GSO constraints given in Sect.~6,
the degree of freedom represented by the fermion $\lambda_3$
contributes to the full $SU(2)_2$ affine Lie algebra, and thus
corresponds to $\Delta r_1$.
Indeed, excitations of $\lambda_3$ are necessary
in order to preserve the conformal
dimension of the gauge-boson vertex operators
(or equivalently, to ensure that the roots
of $SU(2)_2$ had the appropriate ``length''
$\sqrt{2}$ in the {\it pre}\/-truncated lattice).
Thus, in this construction,
$SU(2)_2$ is represented by the three real
fermions $\lambda_{1,2,3}$, with total central charge $3/2$.
By contrast, the remaining degree of freedom $\lambda_4$
contributes to $\Delta r_2$, and
(along with the remaining
degrees of freedom in this model, such as $\lambda_5$)
becomes part of an extra chiral algebra.

Note that for the purposes of string model-building,
one typically seeks to minimize
the size of this extra chiral algebra.  This maximizes
flexibility in constructing the hidden sector, and the
reasons for wanting to maximize this flexibility are two-fold.
On the one hand, we may require the hidden sector to
be sufficiently complicated to allow, \eg, successful gaugino condensation to
occur \cite{gaugino_cond}.
On the other hand, it is also useful
to have enough flexibility in the structure of the
hidden sector in order to be able to arrange for
other phenomenological features that we desire, such
as three chiral generations and an adjoint Higgs with appropriate couplings.
It is therefore important to determine the
minimum value of the ``cost'' $\Delta c$, or equivalently
the minimum size of the extra chiral algebra
that must generated when realizing $G'_{k'}$.

Thanks to our general understanding in Sects.~3 and 4,
we can now address this question in the context of free-field
string constructions.
Indeed, as a result of our complete classification in Sect.~7,
we can now definitively answer this question
in the interesting cases of string GUT's realized at levels $k=2,3,4$.
For example, one might have thought
that in the case of $SU(5)_2$, the minimum ``cost'' would
simply be $\Delta c=1/7$, since this is the smallest addition that
gives an integer total for the central charge, as required for a final
level-one simply laced group.
However, we now see from our classification (as summarized in Sect.~7.4) that
we must in fact have $\Delta c=8/7$ for the case of $SU(5)_2$,
since we found that the diagonal embedding is the {\it only}\/ realization
available
for $SU(5)_2$.
Thus, the central charge available for the hidden sector
in this case is actually smaller by one unit than might
have been thought.  Similar, but often stronger, conclusions hold in many other
cases,
and can be readily determined from the table in Sect.~7.4.
Note that such results are possible only because we
have done a complete classification for these groups and levels.

Given our results, we can now begin to
systematically explore other aspects of the construction
of realistic string GUT models.
As we discussed in the Introduction,
in this paper we have primarily focused on the relation
between the ``GSO Projection'' and ``Group Theory Embedding'' boxes in
Fig.~\ref{flowchartfig}.
However, it may now be possible to address
some questions pertaining to the other connections illustrated in this figure.
For example, one formal question that naturally arises
concerns the relation between
the various possible embeddings that we have identified, and the specific
string
constructions that may be necessary in order to realize such embeddings.
In particular, it is natural to ask what kinds of $\IZ_N$ twists in
an orbifold construction will be required in order to produce a given
type of GSO projection and its associated dimensional truncation.
Such an understanding would be useful in guiding the construction
of alternate string GUT models, and is currently under investigation.

Another more phenomenological question concerns
the relation between a chosen embedding,
and the specific phenomenology that it can accommodate.
In Sect.~7.1, we briefly touched upon some of the
inter-relationships between the level of the
subgroup achieved via a particular embedding,
and the appearance of certain types of representations
({\it e.g.}\/, an adjoint Higgs) in the corresponding massless string spectrum.
However, in general we wish to understand more complicated correlations:
for example, we would like to simultaneously
realize a specific number of chiral generations, an adjoint Higgs,
and a phenomenologically desirable structure for their Yukawa couplings.
It is important to understand how such phenomenological features can be
incorporated in the selection of a particular embedding.
Such issues, however,
necessarily involve details of the hidden sector
as well as the observable sector.
They therefore require an analysis of the {\it total}\/
gauge embeddings
$G'_{k'}\times G_{\rm hidden}\subset {\rm L1SL}$
that are appropriate for the given model.
We intend to perform such an analysis in the future.
The general methods that we have
outlined here should nevertheless
aid in the resolution of these questions,
and should thereby facilitate the construction of
phenomenologically desirable models.

%=============================================================================
%=============================================================================
\bigskip
\medskip
\leftline{\large\bf Acknowledgments}
\medskip

We wish to thank Alon Faraggi for initial collaborations,
and for discussions throughout.
We also thank P. Argyres, K.S. Babu, S. Chaudhuri, S.-W. Chung, A. Font,
J. Lykken, F. Wilczek, and E. Witten for helpful comments.
This work was supported in part by DOE Grant No.\ DE-FG-0290ER40542
and by the W.M. Keck Foundation.

%=============================================================================

%=============================================================================
\vfill\eject
\setcounter{footnote}{0}
\setcounter{section}{0}
\Appendix{Class IIA Embeddings}

In this Appendix we show that all of the Class IIA embeddings
given in (\ref{rule2}) are equivalent to the diagonal embeddings.

First, though, we remark that it is
easy to see why the embeddings, as given in the form
(\ref{rule2}), are unsuitable for our purposes.
These embeddings all have the form
\beq
       (H_1)_{k_1} \times (H_2)_{k_2} ~\subset ~G~,
\eeq
and we know that such embeddings are maximal.
However, the important point to note is that
the subgroup here is not simply $(H_1)_{k_1}$, but
instead always contains
another factor $(H_2)_{k_2}$.
This means that the true sub-embedding that we would
otherwise wish to concentrate on, namely
$(H_1)_{k_1}\subset G$, is {\it not}\/ maximal.
In other words, there may exist a subgroup of $G$
in which $(H_1)_{k_1}$ by itself can be maximally embedded.
As we shall now show, this subgroup is always
$[(H_1)_1]^{k_1}$.

We begin with the general embedding relation of the form
\beq
         H_1\otimes H_2~\subset~ G~
\label{direct}
\eeq
where $H_1$, $H_2$, and $G$ are arbitrary groups.
We shall assume, as appropriate for the Class IIA embeddings,
that such an embedding gives rise to the branching
rule $\rep{mn}\to(\rep{m},\rep{n})$, where $\rep{mn}$, $\rep{m}$,
and $\rep{n}$ are the fundamental representations
of $G$, $H_1$, and $H_2$ respectively.
We then wish to show that this realization of $H_1$, considered
as a subgroup of $G$, is equivalent to the ``diagonal'' realization
\beq
     H_1~\subset~ H_1\otimes H_1\otimes ...  \otimes H_1~\subset~ G
\label{diag}
\eeq
where there are $n$ different $H_1$ factors.  Our procedure will be to compare
the generators of $H_1$ as realized via (\ref{direct}) with those
of $H_1$ as obtained via (\ref{diag}).  The equivalence of the two
sets of generators will demonstrate the equivalence of the two embeddings
of $H_1\subset G$.  For typographical simplicity, we will omit normalization
factors
for the generators.

Given (\ref{direct}), the generators of $H_1$
in the $\rep{mn}$ representation of $G$
are simply $T_a\otimes\bone$, where $T_a$ are the generators
of $H_1$ in the $\rep{m}$ representation and where $\bone$ is the identity
operator in the $\rep{n}$ representation of $H_2$.  Note, in particular, that
\beq
        T_a \otimes \bone ~=~ T_a \oplus T_a \oplus ... \oplus T_a
           ~=~ \mathop{\bigoplus}_{i=1}^{n}\,T_{a}^{(i)}~.
\label{firstway}
\eeq
We now consider the generators of $H_1$ as obtained
via the diagonal embedding (\ref{diag}).
Let $T_{a_i}^{(i)}$ be the
generators of the $i^{\rm th}$ factor of $H_1$ in the tensor product.
Then a ``basis set'' of
generators of $H_1\otimes H_1\otimes...\otimes H_1\equiv [H_1]^n$ is
\beq
       {\cal T}^{(i)}_{a_i}~\equiv ~ \bone\otimes...\otimes
	T_{a_i}^{(i)}\otimes ...\otimes \bone~,
\eeq
since all other generators of $H_1\otimes ... \otimes H_1$ can be
reached by group composition.
As before, we work
in the $\rep{mn}$ representation of $G$.
Now, the $\rep{mn}$ representation of $G$ decomposes into representations of
$H_1\otimes ...\otimes H_1$ according to the branching rule
\beq
    \rep{mn}~\to~ (\rep{m},\bone,...,\bone)\oplus
	(\bone,\rep{m},\bone,...,\bone)\oplus ...
    \oplus(\bone,\bone,...,\bone,\rep{m})~.
\label{decomp}
\eeq
For simplicity, let $\Psi$ denote the $\rep{mn}$ representation of $G$,
and let $\psi_i$ ($i=1,...,n$) respectively denote the representations
of $[H_1]^n$ that appear in (\ref{decomp}).
We see, then, that performing an arbitrary $G$-transformation on $\Psi$
induces the following transformation on $\oplus_{i=1}^n \psi_i$:
\beq
     \biggl\lbrace   T_{a_1}^{(1)}\otimes T_{a_2}^{(2)}\otimes
	T_{a_3}^{(3)}\otimes...\otimes T_{a_n}^{(n)}\biggr\rbrace
	\biggl\lbrace \mathop{\bigoplus}_{i=1}^n \psi_i\biggr\rbrace
	= \mathop{\bigoplus}_{i=1}^n\, \biggl\lbrace {\cal T}_{a_i}^{(i)}
	\psi_i\biggr\rbrace~.
\eeq
This simplification into only the ``basis'' generators ${\cal T}_{a_i}^{(i)}$
arises because all generators $T_{a_j}^{(j)}$ are equivalent to the identity
when acting on their respective singlet representations.
Thus, an arbitrary $[H_1]^n$ group element acting
on the representation $\oplus_i \psi_i$  is given by
\beq
   \exp\biggl\lbrace \mathop{\bigoplus}_{i=1}^n\,
        \epsilon_{a_i}^{(i)} \, {\cal T}_{a_i}^{(i)} \, \psi_i\biggr\rbrace~
\label{intermediate}
\eeq
where $\epsilon_{a_i}^{(i)}$ are the group parameters of the $i^{\rm th}$
factor of $H_1$ within $[H_1]^n$.
Now, by definition, the ``diagonal''
$H_1$ subgroup within the tensor product $[H_1]^n$ is given
by the restriction $\epsilon_{a_i}^{(i)}=\epsilon_a \delta_{a_i,a}$ for
all $i$.  Thus, for transformations in the diagonal $H_1$ subgroup,
(\ref{intermediate}) becomes
\beq
     \exp \biggl \lbrace  \epsilon_a  \left( \mathop{\bigoplus}_{i=1}^n \,
       {\cal T}_a^{(i)} \, \psi_i\right) \biggr\rbrace~,
\eeq
which is equivalent to
\beq
    \exp \biggl \lbrace  \epsilon_a
      \left( \mathop{\bigoplus}_{i=1}^n \,T^{(i)}_{a}\right) \biggr\rbrace  \,
        \left(\mathop{\bigoplus}_{i=1}^n\, \psi_i \right)~.
\label{secondway}
\eeq
However, the generators in (\ref{secondway}) are precisely equivalent
to those in (\ref{firstway}).  Thus we have demonstrated that {\it all}\/
Class IIA embeddings (\ref{direct}) are equivalent to the diagonal embedding
(\ref{diag}).

As an example of such equivalences, let us consider the series of embeddings
\beqn
        SU(5)_n~\times~ SU(n)_5 ~&\subset&~ SU(5n)_1 \nonumber\\
        SO(10)_n~\times~ SO(n)_{10} ~&\subset&~ SO(10n)_1~.
\eeqn
These embeddings
all correspond to the diagonal embeddings of $SU(5)_n$ and $SO(10)_n$
within a tensor product of $n$
factors of $SU(5)_1$ and $SO(10)_1$ respectively.  (Such higher-level
diagonal embeddings can also be realized by iterating these maximal
Class IIA embeddings.)  As  a less trivial example, let us consider
realizing $SO(10)_4$ via the embedding
\beq
          SO(10)_4 \times Sp(2)_{10}~\subset~ Sp(20)_1~.
\eeq
In principle, this is a ``smaller'' embedding of $SO(10)_4$ than the
corresponding four-fold diagonal embedding would be, since the
non-simply laced supergroup $Sp(20)$ has rank 10 rather than rank 20.
However, in string theory, such a non-simply laced group must itself
be realized through an irregular embedding, and indeed the smallest
possible embedding for $Sp(20)_1$ is
\beq
         Sp(20)_1 \times Sp(2)_{10}~\subset~ SO(40)_1~.
\eeq
Thus the total embedding into a level-one simply laced group becomes
\beq
          SO(10)_4 \times Sp(2)_{10}\times
         Sp(2)_{10}~\subset~ SO(40)_1~,
\eeq
and once again this is tantamount to the diagonal embedding for the
$SO(10)_4$ factor.  Thus, although the Class IIA embeddings reproduce the known
diagonal embeddings, we see that they do not give rise to new embeddings.

%=========================================================================
%======================== REFERENCES =====================================
%=========================================================================

\vfill\eject
%  \bigskip
%  \medskip

\bibliographystyle{unsrt}

\begin{thebibliography}{99}

\bibitem{models}
    See, {\it e.g.},\\
    I. Antoniadis, J. Ellis, J. Hagelin, and D.V. Nanopoulos,
                  \PLB{231}{89}{65}; \\
    A.E. Faraggi, D.V. Nanopoulos, and K. Yuan,
      \NPB{335}{90}{347};\\
    J.L. Lopez, D.V. Nanopoulos, and K. Yuan, \NPB{399}{93}{654},
     hep-th/9203025;\\
    I. Antoniadis, G.K. Leontaris, and J. Rizos,
        {\it Phys.\ Lett.}\/ {\bf B245} (1990) {161};\\
     A.E. Faraggi, \PLB{278}{92}{131};
       \NPB{387}{92}{239}, hep-th/9208024;
       \PLB{302}{93}{202}, hep-ph/9301268.
\bibitem{thresholds}
      See, {\it e.g.},\\
      V.S. Kaplunovsky, \NPB{307}{88}{145};
          Erratum:  {\it ibid.}\/ {\bf B382} (1992) 436, hep-th/9205070;\\
              L.J. Dixon, V.S. Kaplunovsky, and J. Louis, \NPB{355}{91}{649};\\
        I. Antoniadis, J. Ellis, R. Lacaze, and D.V. Nanopoulos,
                  \PLB{268}{91}{188};\\
        L. Dolan and J.T. Liu, \NPB{387}{92}{86}, hep-th/9205094;\\
         P. Mayr, H.P. Nilles, and S. Stieberger, \PLB{317}{93}{53},
            hep-th/9307171;\\
         D.M. Pierce, \PRD{50}{94}{6469}, hep-th/9508178;\\
      P. Mayr and S. Stieberger, hep-th/9412196; \PLB{355}{95}{107},
           hep-th/9504129;\\
       M. Chemtob, \PRD{53}{96}{3920}, hep-th/9506178;\\
        H.P. Nilles and S. Stieberger, \PLB{367}{96}{126}, hep-th/9510009.
\bibitem{DF}
  K.R. Dienes and A.E. Faraggi, \PRL{75}{95}{2646}, hep-th/9505018;
           \NPB{457}{95}{409}, hep-th/9505046.
\bibitem{k1}
    J.A. Casas and C. Mu\~noz, \PLB{214}{88}{543}; \\
     L. Ib\'a\~nez, \PLB{318}{93}{73}, hep-ph/9308365;\\
   K.R. Dienes, A.E. Faraggi, and J. March-Russell,
      \NPB{467}{96}{44}, hep-th/9510223.
\bibitem{extramatter}
           I. Antoniadis, J. Ellis, S. Kelley, and D.V. Nanopoulos,
       \PLB{272}{91}{31};\\
        S. Kelley, J.L. Lopez, and D.V. Nanopoulos, \PLB{278}{92}{140};\\
     D. Bailin and A. Love, \PLB{280}{92}{26};  \MODA{7}{92}{1485};
      Erratum: {\it ibid.}\/ {\bf A7} (1992) 2963;\\
    M.K. Gaillard and R. Xiu, {\it Phys.\ Lett.}\/
      {\bf B296} (1992) 71, hep-ph/9206206;\\
      S.P. Martin and P. Ramond, {\it Phys.\ Rev.}\/
      {\bf D51} (1995) 6515, hep-ph/9501244;\\
         B.C. Allanach and S.F. King, hep-ph/9601391.
\bibitem{wittencouplings}
        E. Witten, \NPB{471}{96}{135}, hep-th/9602070.
\bibitem{review}
     K.R. Dienes, hep-th/9602045 (to appear in {\it Physics Reports}\/).
\bibitem{stringguts}
   D.C. Lewellen, \NPB{337}{90}{61};\\
   A. Font, L.E. Ib\'a\~nez, and F. Quevedo, \NPB{345}{90}{389};\\
   J.A. Schwartz, \PRD{42}{90}{1777};\\
   J. Erler, hep-th/9602032;\\
   G. Cleaver,  hep-th/9604183.
\bibitem{shygut}
   S. Chaudhuri, S.-W. Chung, G. Hockney, and J. Lykken,
      \NPB{456}{95}{89}, hep-ph/9501361.
\bibitem{stringguthybrids}
   G. Aldazabal, A. Font, L.E. Ib\'a\~nez, and A.M. Uranga,
    \NPB{452}{95}{3},  hep-th/9410206;  
    \NPB{465}{96}{34}, hep-th/9508033.
\bibitem{shynew}  S. Chaudhuri, G. Hockney, and J. Lykken, 
     \NPB{469}{96}{357}, hep-th/9510241.
\bibitem{GxGmodels}
   A.A. Maslikov, S.M. Sergeev, and G.G. Volkov, \PLB{328}{94}{319};
        \PRD{50}{94}{7440};  \IJMP{9}{94}{5369};\\
   A.A. Maslikov, I. Naumov, and G.G. Volkov,
         \IJMP{11}{96}{1117}, hep-ph/9512429;\\
   D. Finnell, \PRD{53}{95}{5781}, hep-th/9508073.
\bibitem{duality}
      S. Chaudhuri, G. Hockney, and J. Lykken,
        \PRL{75}{95}{2264}, hep-th/9505054;\\
      S. Chaudhuri and J. Polchinski, \PRD{52}{95}{7168}, hep-th/9506048.
\bibitem{KMalgebras} For pedagogical introductions to affine Lie
   algebras, also known as \KM\ algebras, see, {\it e.g.}:\\
    P. Goddard and D. Olive, \IJMP{1}{86}{303};\\
    P. Ginsparg, {\it Applied Conformal Field Theory}, published
     in {\it Fields, Strings, and Critical Phenomena:
     Proceedings of Les Houches, Session XLIX, 1988}, eds.\
     E. Br\'ezin and J. Zinn-Justin (Elsevier, 1989).\\
    A history of these algebras can be found in the Appendix of:\\
     M.B. Halpern, E. Kiritsis, N.A. Obers, and K. Clubok,
     \PRT{265}{96}{1}, hep-th/9501144.\\
    In particular, the affine level was first introduced in:\\
        K. Bardakci and M.B. Halpern, \PRD{3}{71}{2493}.
\bibitem{asymmorbifolds}  K. Narain, M. Sarmadi, and C. Vafa,
        \NPB{288}{87}{551};  \NPB{356}{91}{163}.
\bibitem{KLST}
          H. Kawai, D.C. Lewellen, J.A. Schwartz, and S.-H.H. Tye,
                 \NPB{299}{88}{431}.
\bibitem{KT}  Z. Kakushadze and S.-H.H. Tye, hep-th/9512156.
\bibitem{freefermions}  H. Kawai, D.C. Lewellen, and S.-H.H. Tye,
           {\it Nucl.\ Phys.}\/ {\bf B288} (1987) 1;\\
           I. Antoniadis, C. Bachas, and C. Kounnas,
           {\it Nucl.\ Phys.}\/ {\bf B289} (1987) 87.
\bibitem{witten}
      E. Witten, \NPB{460}{96}{541}, hep-th/9511030.
\bibitem{Eeightmodel}  H. Kawai, D.C. Lewellen, and S.-H.H. Tye,
          \PRD{34}{86}{3794}.
\bibitem{slansky}
        R. Slansky, {\it Phys.\ Rep.}\/ {\bf 79} (1981) 1;\\
        W.G. McKay and J. Patera, {\it Tables of Dimensions, Indices,
               and Branching Rules for Representations of Simple Lie
                Algebras}\/
         (Marcel Dekker, New York, 1981);\\
        M.R. Bremner, R.V. Moody, and J. Patera, {\it Tables of Dominant
            Weight Multiplicities for Representations of Simple Lie Algebras}\/
           (Marcel Dekker, New York, 1985);\\
        H. Georgi, {\it Lie Algebras in Particle Physics}\/ (Benjamin/Cummings,
         Menlo Park, 1982);\\
        L. O'Raifeartaigh, {\it Group Structure of Gauge Theories}\/
           (Cambridge University Press, Cambridge, England, 1986).
\bibitem{dynkin2}
        E.B. Dynkin, {\it Amer.\ Math.\ Soc.\ Trans.}\/ {\bf 2} (1957) 111
        [{\it Mat.\ Sbomik N.S.}\/ {\bf 30} (1952) 349].
\bibitem{BB}
        F.A. Bais and P.G. Bouwknegt, \NPB{279}{87}{561};\\
        A.N. Schellekens and N.P. Warner, \PRD{34}{86}{3092}.
\bibitem{cahn}  R.N. Cahn, {\it Semi-Simple Lie Algebras and their
         Representations}\/ (Benjamin/ Cummings, Menlo Park, 1984);\\
      J. F\"uchs, {\it Affine Lie Algebras and Quantum Groups}\/
   (Cambridge University Press, Cambridge, England, 1992).
\bibitem{dynkin3}
        E.B. Dynkin, {\it Amer.\ Math.\ Soc.\ Trans.}\/ {\bf 2} (1957) 245
         [{\it Trudy Moskov.\ Mat.\ Obsc.}\/ {\bf 1} (1952) 39].
\bibitem{conformalembeddings}
        V.G. Ka\v{c} and M.N. Sanielevici, \PRD{37}{88}{2231};\\
        M.A. Walton, \NPB{322}{89}{775};\\
        D. Verstegen, \CMP{137}{91}{567}.
\bibitem{gaugino_cond}
      See, {\it e.g.},\\
      H.P. Nilles, \PLB{115}{82}{193};\\
      S. Ferrara, L. Girardello, and H.P. Nilles, \PLB{125}{83}{457};\\
      J.P. Derendinger, L. Ib\'a\~nez, and H.P. Nilles, \PLB{155}{85}{65};\\
      M. Dine, R. Rohm, N. Seiberg, and E. Witten, \PLB{156}{85}{55};\\
      C. Kounnas and M. Porrati, \PLB{191}{87}{91};\\
      A. Font, L. Ib\'a\~nez, D. Lust, and
           F. Quevedo, \PLB{245}{90}{401};\\
      T.R. Taylor, \PLB{252}{90}{59}; hep-ph/9510281;\\
      A. de la Macorra and G.G. Ross, \NPB{404}{93}{321},  hep-ph/9210219;\\
      B. de Carlos, J.A. Casas, and C. Mu\~noz, \NPB{399}{93}{623},
hep-th/9204012;\\
      F. Quevedo,  hep-th/9511131.
\end{thebibliography}

\end{document}